\documentclass[%
reprint,superscriptaddress,
nofootinbib,
 amsmath,amssymb,
 aps,
]{revtex4-2}

\usepackage{color}
\usepackage{graphicx}
\usepackage{dcolumn}
\usepackage{bm}
\usepackage{hyperref}
\usepackage{booktabs}

\usepackage{soul} 

\begin{document}

\preprint{APS/123-QED}

\title{Breaking of universal relationships of axial $wI$-modes in hybrid stars:\\rapid and slow hadron-quark conversion scenarios}

\author{Ignacio F. Ranea-Sandoval}
\email{iranea@fcaglp.unlp.edu.ar}
 \affiliation{%
 Grupo de Gravitaci\'on, Astrof\'isica y Cosmolog\'ia, Facultad de Ciencias Astron{\'o}micas y Geof{\'i}sicas, Universidad Nacional de La Plata, Paseo del Bosque S/N, 1900, La Plata, Argentina.}
 \affiliation{CONICET, Godoy Cruz 2290, 1425, CABA, Argentina.}
 
\author{Octavio M. Guilera}
\email{oguilera@fcaglp.unlp.edu.ar}
 \affiliation{%
 Grupo de Gravitaci\'on, Astrof\'isica y Cosmolog\'ia, Facultad de Ciencias Astron{\'o}micas y Geof{\'i}sicas, Universidad Nacional de La Plata, Paseo del Bosque S/N, 1900, La Plata, Argentina.}
 \affiliation{Instituto de Astrofísica de La Plata, CCT-La Plata, CONICET, Argentina.}
 
\author{Mauro Mariani}
\email{mmariani@fcaglp.unlp.edu.ar}
\affiliation{%
 Grupo de Gravitaci\'on, Astrof\'isica y Cosmolog\'ia, Facultad de Ciencias Astron{\'o}micas y Geof{\'i}sicas, Universidad Nacional de La Plata, Paseo del Bosque S/N, 1900, La Plata, Argentina.}
 \affiliation{CONICET, Godoy Cruz 2290, 1425, CABA, Argentina.}
 
\author{Germ\'an Lugones}
\email{german.lugones@ufabc.edu.br}
\affiliation{%
 Centro de Ciências Naturais e Humanas, Universidade Federal do ABC,  Avenida dos Estados 5001, CEP 09210-580, Santo André, SP, Brazil
}%

\date{\today}

\begin{abstract}
Multi-messenger astronomy with gravitational waves is a blooming area whose limits are not clear. After the first detection of binary black hole merger and the famous event GW170817 and its electromagnetic counterpart, the compact-object astrophysical community is starting to grasp the physical implications of such event and trying to improve numerical models to compare with future observations. Moreover, recent detections made by the NICER collaboration increased the tension between several theoretical models used to describe matter in the inner core of compact objects. In this paper, we focus on quadrupolar purely spacetime $wI$-modes of oscillating compact objects described using a wide range of hybrid equations of state able to include several theoretical possibilities of exotic matter in the inner core of such stars. We study the case in which a sharp first order hadron-quark phase transition occurs and explore the scenarios of {\it{rapid}} and {\it{slow}} phase conversions at the interface. We put special attention on the validity of universal relationships for the oscillation frequency and damping time that might help unravel the mysteries hidden at the inner cores of compact objects. We show that, within the slow conversion regime where extended branches of hybrid stars appear, universal relationships for $wI$-modes proposed in the literature do not hold.
\end{abstract}

\maketitle

\section{Introduction}
\label{sec:intro}
Ten years ago, the 2$M_\odot$ binary pulsar PSR J1614-2230 was detected \cite{Demorest2010}. A few years later, PSR J0348+0432 was identified \cite{Antoniadis2013} and, more recently, PSR J0740+6620 \cite{Cromartie2019}. These observations put strong constraints on the \emph{equation of state} (EoS) needed to describe matter inside a \emph{neutron star} (NS). Theoretical models that describe matter in the inner core of such compact objects need to be stiff, since soft EoSs are unable to reproduce this observational constrain (see, for example, Ref.~\cite{lattimer2msol}). In this scenario, the \emph{hybrid star} (HS) hypothesis was strengthen. Later, an analysis of the \emph{gravitational waves} (GWs) emitted by the binary NS merger event GW170817 \cite{GW170817-detection} and its electromagnetic counterpart (see, for example, Ref.~\cite{GW170817-em}) constrained the radius of  $1.4~M_\odot$ NSs to be $R_{1.4} < 13.6$~km (see, for example, Refs.~\cite{Raithel2018,Abbott:2018}). 
More recently, the LIGO Livingston detector observed the event GW190425, a compact binary coalescence with total mass $\sim 3.4 M_{\odot}$ \cite{GW190425-detection}. This was the first confirmed GW detection based on data from a single observatory and no electromagnetic counterpart was found. If
interpreted as a double NS merger, the total gravitational mass is substantially larger than that of the binary systems identified in the Galaxy. This raises the  possibility of GW190425 being a NS-black hole binary merger \cite{Han:2020qmn,Kyutoku:2020xka}.
Important information is also provided by recent NICER observations of the isolated pulsar PSR J0030+0451. Two independent determinations of both its mass and its radius become available \cite{Riley2019,Miller2019}. The first group inferred a mass of $M=1.34^{+0.15}_{-0.14}M_\odot$ and an equatorial radius $R_{\rm eq} = 12.71^{+1.14}_{-1.19}$~km  and the other obtained $M=1.44^{+0.15}_{-0.14}M_\odot$ and $R_{\rm eq} = 13.02^{+1.24}_{-1.06}$~km, respectively. Several additional physical parameters of this object have been published (see, for example Ref.~\cite{Bilous:2019}). More recently, using both NICER and XMM-Newton data, the mass and radius of PSR J0740+6620 has also been inferred \citep{riley2021ApJ-j0740,miller2021ApJ-j0740}. The results show that, despite being almost 50$\%$ more massive than PRS J0030+0451, the radius of PSR J0740+6620 is almost the same. This observation challenges some modern EoSs of NS matter.

More recently, a statistical analysis of GW170817 data using state-of-the art EoSs almost discarded the possibility of not having quark matter in the inner core of very massive NSs \cite{Annala:2019puf}. For this reason, understanding the nature of the hadron-quark phase transition is of paramount importance for the compact object astrophysical community. Several important aspects of the low-temperature and high-density hadron-quark phase transition are not yet properly understood. Firstly, despite several models predict a first-order phase transition in this regime of the QCD phase diagram (see, for example, \citep{tsue2010PThPh,chamel2013A&A,dexheimer2018IJMPE}), others predict a crossover hadron-quark phase transition in the neutron-star equation of state (see, for example, \citep{hatsuda2006PhRvL,baym2019ApJ}). Secondly, if a first-order hadron-quark phase transition is assumed, one of the open questions related to this subject is to understand whether the hadron-quark interface is  a  sharp discontinuity or if a mixed phase, in which hadrons and quarks coexist, is present within the star (see Ref.~\cite{LugGrunf-universe:2021} and references therein). The quantity that would define which behavior is energetically favored is the hadron-quark surface tension (see, for example, Refs.~\cite{Maruyama:2007ey,Lugones:2013ema,LugGrunf.2021PhRvDL} and references therein). If sharp interfaces are favored it is essential to understand whether quark-hadron conversions at the phase-splitting surface are \emph{rapid} or \emph{slow} (for details, see Ref.~\cite{Pereira:2017rmp}) since this would drastically change the dynamic stability of compact objects. On the other hand, if a mixed phase is present, understanding the nature of such region and whether the so-called pasta phase appears or not is still to be understood \cite{Baym-review}.

Another exciting branch of GW astronomy is related to non-radial oscillations of compact objects (see, for example, the modern review \cite{andersson2021Univ} and the references mentioned in it). NS oscillation modes are complex quantities whose real part describes the oscillation frequency and the inverse of its imaginary part, the damping time. Such modes, known as \emph{quasi-normal modes} (QNMs), are highly dependent on the EoS used to describe matter inside a NS. For this reason, studying and analyzing the spectrum of QNMs can be used to extract information about the structure and the internal composition of such compact objects and shed some light on the behavior of matter at densities several times larger that the nuclear saturation one. These oscillations produce GW emission that might be detected by ground base detectors in the future (see, for example, Ref.~\cite{Punturo:2010zz}). Even though it is known that their characteristic frequencies are strongly dependent of the EoS, it has been shown that some {\emph{universal relationships}}\footnote{Smart parametrizations of the frequency and damping time that are almost insensitive to the EoS used to describe matter inside the compact object.} exist for some particular modes. The most known relationship is the one for the polar $f-$mode \cite{AK,Rosofsky:2018,Flores:2018pnn}. More recently, a universal relationship has been proposed for the $g-$mode for compact objects with a sharp hadron-quark phase transition \cite{RSetalJCAP,Rodriguez-etalg2}. Moreover, it has been proposed that detection of a few $wI$-modes would be enough to reconstruct the EoS of matter in the high-density regime \cite{Mena-Fernandez2019} (also see, Ref.~\cite{chirenti2020-wmode}). The $r-$modes and their stability window have been studied since the CFS instability was discovered \cite{chandra-CFS,FS-CFS}. Such $r-$modes are known to appear in rotating compact objects and they are completely absent (degenerated to null frequency) in non-rotating stars.  In this historical context of fast evolving technology and data processing, having new theoretical calculations to compare observations with is highly relevant. 

In this work, we present calculations related to the so-called $wI$-modes, putting special attention on the fundamental, $wI^{(0)}$, and first overtone, $wI^{(1)}$, of this family. We will perform such calculations using a broad variety of modern EoSs considering the possible occurrence of a hadron-quark phase transition. To model such phase transition, we will assume that the hadron-quark surface tension is high enough to favor a sharp first order transition in which no mixed phase is formed \cite{LugGrunf.2021PhRvDL}. Moreover, we will consider two different scenarios for such phase transition: rapid and slow conversion regimes (for a review on this subject, see Ref.~\cite{LugGrunf-universe:2021}). 
Our main focus is to overview universal relationships proposed to be valid for these modes \cite{w-modes_universal,Benhar2004PRD,Tsui2005MNRAS}, and check whether they remain valid or not for hybrid stars when the quark-hadron conversion speed at a perturbed interface is taken into account. Our analysis will give special attention to the slow hadron-quark conversion regime that gives rise to extended branches of stable stellar configurations. We will show that already known universal relationships valid both for $wI^{(0)}$ and $wI^{(1)}$ do not hold if one considers the slow conversion regime.

Throughout this paper, we use geometrical units ($G=c=1$) except to present the results, which are presented using more physically convenient units. The work is structured in the following manner. In Section~\ref{EoS} we summarize some aspects of the Quantum Chromodynamics (QCD) phase diagram with emphasis in some of the open issues that are relevant for astrophysics.  We describe some basic aspects of hadron-quark phase transitions and describe the hybrid EoSs we work with in this paper. In Section~\ref{axial-modes} we present the basic equations that describe axial modes and explain the numerical scheme used to calculate QNMs. In Section~\ref{results} we show our results for the oscillation frequencies and damping times of the fundamental and first overtone of $wI$-modes obtained using modern EoSs including some hybrid ones, considering the rapid and slow conversion scenarios. We show that in the case of slow conversions there are significant deviations from already known universal relationships described in the literature. A summary of the work, a discussion about the astrophysical implications of our results, and our main conclusions are provided in Section~\ref{conc}.

\section{Describing matter inside a compact object} \label{EoS}

Although the great efforts done over several decades, the understanding of the behavior of matter at densities several times greater than the saturation one ($n_0 = 0.16~{\rm fm}^{-3}$), is still incomplete. In fact, strong interactions are ruled by QCD, but the QCD phase diagram is not fully understood, particularly the region in which matter inside NSs lies (see, for example, Refs.~\cite{qcd01,qcd02,BraunMunzinger:2009}). This is one of the main reasons why the era of multi-messenger NS astrophysics with GWs could be extremely helpful to shed some light into some of these mysteries, and help us understand how matter behaves at densities beyond $n_0$. 

Performing lattice QCD calculations at finite density is not an easy task given, among other technical issues, the sign problem of QCD \cite{2017PTEP.2017c1D01G}. For this reason, several phenomenological models that reproduce some aspects of the theory have been proposed over the last decades (see, for example, Refs.~ \cite{mariani-nestor,Chodos1974,Nefediev2009,Dexheimer:2009,Contrera2010,Orsaria2013,benic2014,Ranea-Sandoval:2017,Malfatti:2019}). 

The structure of a NS can be simplified to be composed of three layers: a crust, an outer core, and an inner core. Physical description of matter in the crust is properly understood since almost fifty years ago \cite{BPS1971,BBP1971}. Conditions in the crust are similar to those that can be found in White Dwarfs \cite{kipp}. The description of matter in the outer core is usually done using the Relativistic Mean Field approximation \cite{walecka1974,Lalazissis:1996}. Several parametrizations are available, including models that use density dependent coupling constants \cite{spinellaTH,DD2,swl4}. The astrophysical community has not yet achieved a total consensus and questions regarding the zoo of particles that might appear in this layer need still to be answered \cite{Baym-review,orsaria-rev}. Despite these uncertainties, most of the theoretical open questions are related to the description of matter in the inner core where densities might be several times larger that $n_0$ \cite{Baym-review,orsaria-rev,Lattimer:2012ARNPS,Burgio:2021PrPNP}. One theoretical possibility that has been exploited is that of a phase transition in which hadrons, at extremely large densities, dissolve to form a soup of free quarks (and gluons, if finite temperature effects are taken into account) (see, for example, Refs.~\cite{Alcock:1986hz,Lattimer:2004,Weber:2004,Weissenborn:2011}). The nature of such exotic phase is not clear, some models predict the formation of diquarks in a color superconducting phase \cite{Ranea-Sandoval:2017,Baldo:2002,Alford:2007}. 

\subsection{Hadron-quark phase transition}

The nature of the phase hadron-quark transition is believed to be mainly determined by the value of the surface tension between these two states of matter, $\sigma _{{\rm hq}}$. There is a large range of theoretical values for such parameter showing that it is highly model dependent. Several works found that, for low values of \mbox{$\sigma _{{\rm hq}} \sim 5 - 30\,{\rm MeV / fm}^2$}, the appearance of a mixed phase is favored \cite{Alford:2001zr}. On the other hand, others found larger values for the surface tension, $\sigma _{{\rm hq}} \sim 50 - 300\,{\rm MeV / fm}^2$ (see, for example, Ref.~\cite{LugGrunf-universe:2021}, and references therein). In the latter case, a sharp hadron-quark phase transition would take place. 

If the phase transition is sharp, no mixed phase of quarks coexisting with hadrons is formed; on the contrary, each phase is constrained to a particular volume of the star. In this scenario, the so-called \emph{Maxwell formalism} is used to describe the phase transition. One aspect to put attention in this case is related to the conversion speed {-compared to the characteristic time of radial modes-} between hadrons and quarks that are in contact at the interface. {Regarding this timescale, we must state that phase transitions are non-linear, highly collective phenomena. For the particular case of the hadron–quark phase transition, its conversion timescale it is not expected to be simply the consequence of particles confining or deconfining independently. A possible mechanism in which a first-order hadron-quark phase transition can proceed at high densities is nucleation \citep{Olesen:1993ek, Lugones:1997gg,Iida:1998pi,Bombaci:2004mt}. This mechanism has a large activation barrier in NSs because a direct conversion of hadronic matter into quark matter in equilibrium under weak interactions is in general a high-order weak process which is strongly suppressed. Thus, nucleation should proceed through an intermediate (flavor conserving) state that is not easily accessible in the neighborhood of the interface (see Figs.~$1$ and $2$ of  Ref.~\cite{Lugones:2015bya}). Calculations of the quantum and thermal nucleation timescales give results larger than the age of the universe for temperatures below a few MeV (this timescale drops dramatically at higher temperatures) \cite{Bombaci:2016xuj}. For the reverse reaction, a similar argument is expected to be valid: a high-order weak interaction process would be necessary within quark matter in order to produce beta-equilibrated hadronic matter with a lower free energy, making the process extremely unlikely. Thus, although the conversion timescale is uncertain, there are good arguments to believe that it could be slow. }


In Ref.~\cite{Pereira:2017rmp}, the authors proved that the conversion speed at the interface might have drastic consequences on the stability of HSs. 
Indeed, if the hadron-quark conversion timescale  around the sharp interface  is {\it slow} compared to the characteristic time of radial modes, some stellar configurations may be stable even if $\partial M / \partial {\epsilon _c} < 0$, being $M$ the stellar mass and ${\epsilon _c}$ its central energy density. These stellar configurations are dynamically stable because all radial eigenfrequencies turn out to be real. The last stable object of an extended branch of slow stable configurations is the one for which the frequency of the fundamental radial eigenmode is zero. The mass of the last stable object will be called \textit{terminal mass} (see, Refs.~\cite{Pereira:2017rmp,LugGrunf-universe:2021,VasquezFlores:2012vf,Mariani:2019vve}, for a detailed description of the hadron-quark conversion process).
This fact is known to lead to the so called  {\it slow-stable hybrid stars} and allows the existence of high-mass stellar twin configurations \cite{Pereira:2017rmp,Mariani:2019vve,Malfatti2020PRD}.

\begin{figure*}[t]
\centering
\includegraphics[width=0.7\linewidth,angle=0]{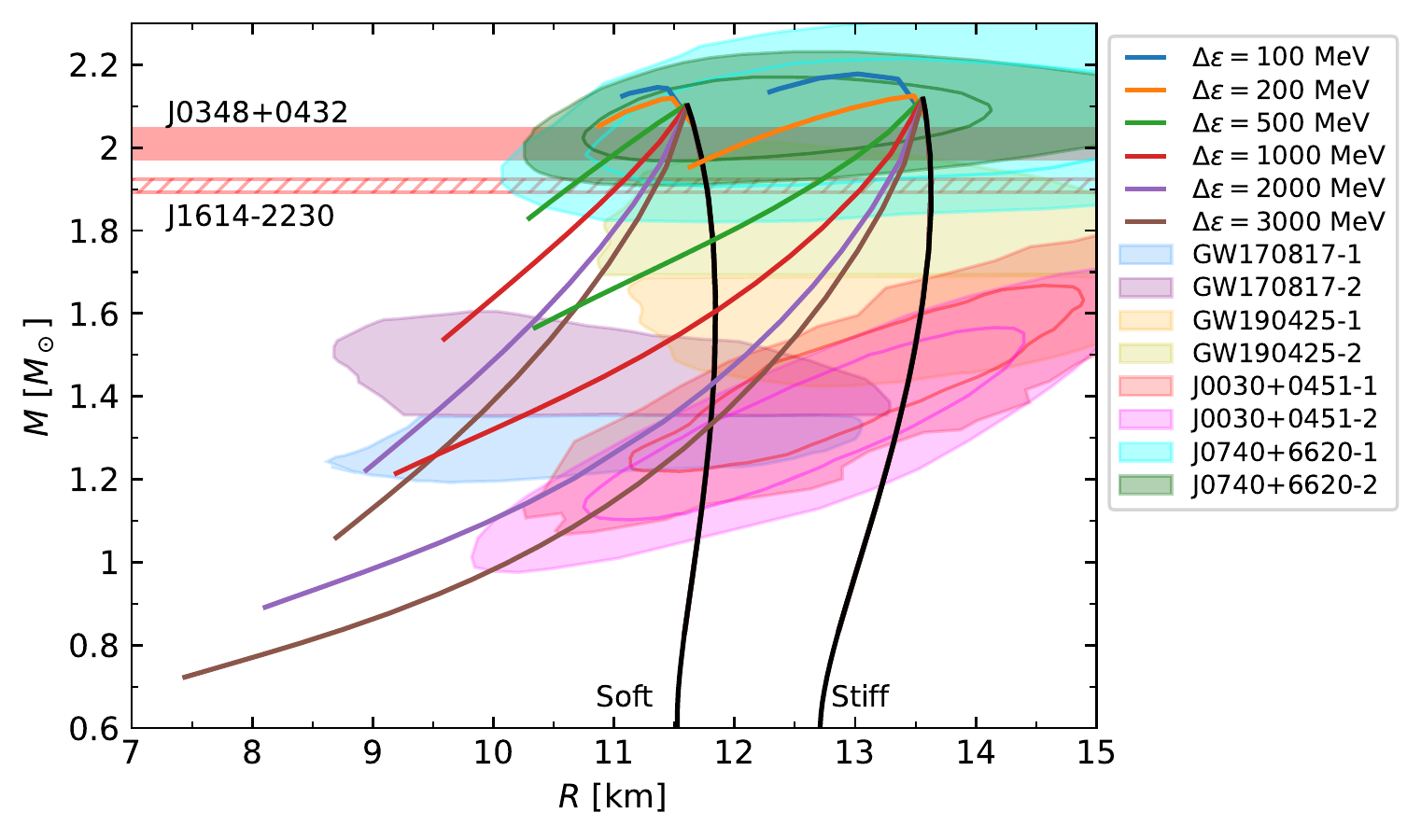}
\caption{Mass-Radius relationships constructed using the selected hybrid EoS. With black lines we present hadronic configurations and with different colors hybrid configurations for different values of $\Delta \epsilon$. All displayed configurations are stable in the slow conversion regime. If rapid conversions are assumed, the only stable configurations are those up to the mass peak. Modern astronomical constraints are also shown in the figure. Notice that, within our model, such objects can be explained as hadronic NSs or as slow stable HSs.} 
\label{fig:MR}
\end{figure*}

A possible scenario where the \textit{slow-stable hybrid star} branch can be populated has been discussed recently \cite{LugGrunf-universe:2021}. Notice that most hadronic stars in  Fig. \ref{fig:MR} that have a hybrid twin, are in metastable states because it is energetically convenient for them to convert into a more compact \textit{slow-stable hybrid star} with same baryonic mass.  However, such conversion appears to be strongly suppressed in practice because the transition pressure $P_{t}$ is attained only at the center of the maximum mass object. Therefore, any further accretion onto such a star would not be able to grow a quark matter core but would instead produce a collapse into a black hole.

However, for hot hadronic stars  the  picture is different. First, the QCD phase diagram suggests that the hadron-quark transition density becomes smaller as the temperature increases.  Second, typical transition mechanisms such as nucleation and spinodal decomposition get easier as the temperature increases (see e.g. Ref. \cite{Bombaci:2016xuj} for the case of quark matter nucleation in dense hadronic matter). Therefore, the critical stellar mass above which a metastable hadronic star could undergo a phase transition would be significantly reduced when the object is hot enough (see  Ref. \cite{LugGrunf-universe:2021} for more details and  for some numerical estimates). As a consequence, a significant portion of the upper part of the hadronic branches in Fig. \ref{fig:MR}  would be able to decay to the \textit{slow-stable hybrid star} branch in the hot protoneutron star phase coming after a core collapse suopernova or in a hot hadronic object resulting from a binary compact star merger.  Below the `hot' critical mass, a proto-hadronic star would survive the early stages of its evolution without decaying to a \textit{slow-stable hybrid star}. When such hadronic object cools down, the critical mass rises and the star can accrete additional mass from a companion keeping its hadronic nature. This qualitative analysis deserves further investigation, but it suggests that there are feasible astrophysical channels for populating both branches, the hadronic and the slow-stable hybrid one.

\subsection{Hybrid EoS model}

In this paper, we construct our hybrid EoSs using for the hadronic sector two different sets of values following the prescription of Ref.~\cite{OBoyle-etal-2020} that ensures continuity in the pressure, the energy density, and the sound speed. We construct a \emph{stiff} and a \emph{soft} EoS using the parameters of Table~\ref{tabla:param_selec}. To keep our treatment as general as possible, we use the CSS parametrization for the quark sector \cite{css-original} and span the three parameters --the hadron-quark transition pressure, $P_t$, the discontinuity in energy density, $\Delta \epsilon$, and the speed of sound of the quark matter, $c_\mathrm{s}$-- in a wide range of physically acceptable values. In this way, we construct over 5000 different hybrid EoSs, considering both the stiff and soft hadronic EoSs and using the CSS parameters in the following ranges:
\begin{align*}
    10~ \mathrm{MeV/fm^3} &\le P_{t} \le 300~\mathrm{MeV/fm^3} \, , \\
    100~\mathrm{MeV/fm^3} &\le \Delta \epsilon \le 3000~\mathrm{MeV/fm^3} \, , \\
    0.2 &\le c_\mathrm{s}^2 \le 1 \, ,
\end{align*}
We select only those EoSs whose stellar models verify $2M_{\odot} < M_{\max}  < 2.3M_{\odot}$ and that are representative of the general and qualitative behavior of our model and, also, sensitive to the effect over the $w$-modes that we aim to study. The upper bound for $M_{\max}$ is taken to be consistent with results from Ref. \cite{Rezzolla:2017aly}.


\begin{table}
\centering
\begin{tabular}{cccccccc}
\toprule
 & $\log_{10}\rho_0$ & $\log_{10}\rho_1$ & $\log_{10}\rho_2$  & $\Gamma_1$ & $\Gamma_2$ & $\Gamma_3$ & $\log_{10}K_1$ \vspace{0.1cm}  \\
\toprule
\emph{soft} & 13.902 & 14.45 & 14.58 & 2.752 & 4.5 & 3.5 & -27.22 \\ \midrule
\emph{stiff} & 13.902 & 14.45 & 14.58 & 2.764 & 8.5 & 3.2 & -27.22 \\
\bottomrule
\end{tabular}
\caption{{Parameters of the selected hybrid EoSs constructed with the prescription of Ref.~\cite{OBoyle-etal-2020}. The values of the first piece of the hadronic EoS, $\Gamma_1$ and $\log_{10}K_1$, were selected  to match the upper limit predicted by cEFT EoSs.}}
\label{tabla:param_selec}
\end{table}

\begin{figure}[t]
\centering
\includegraphics[width=0.95\linewidth,angle=0]{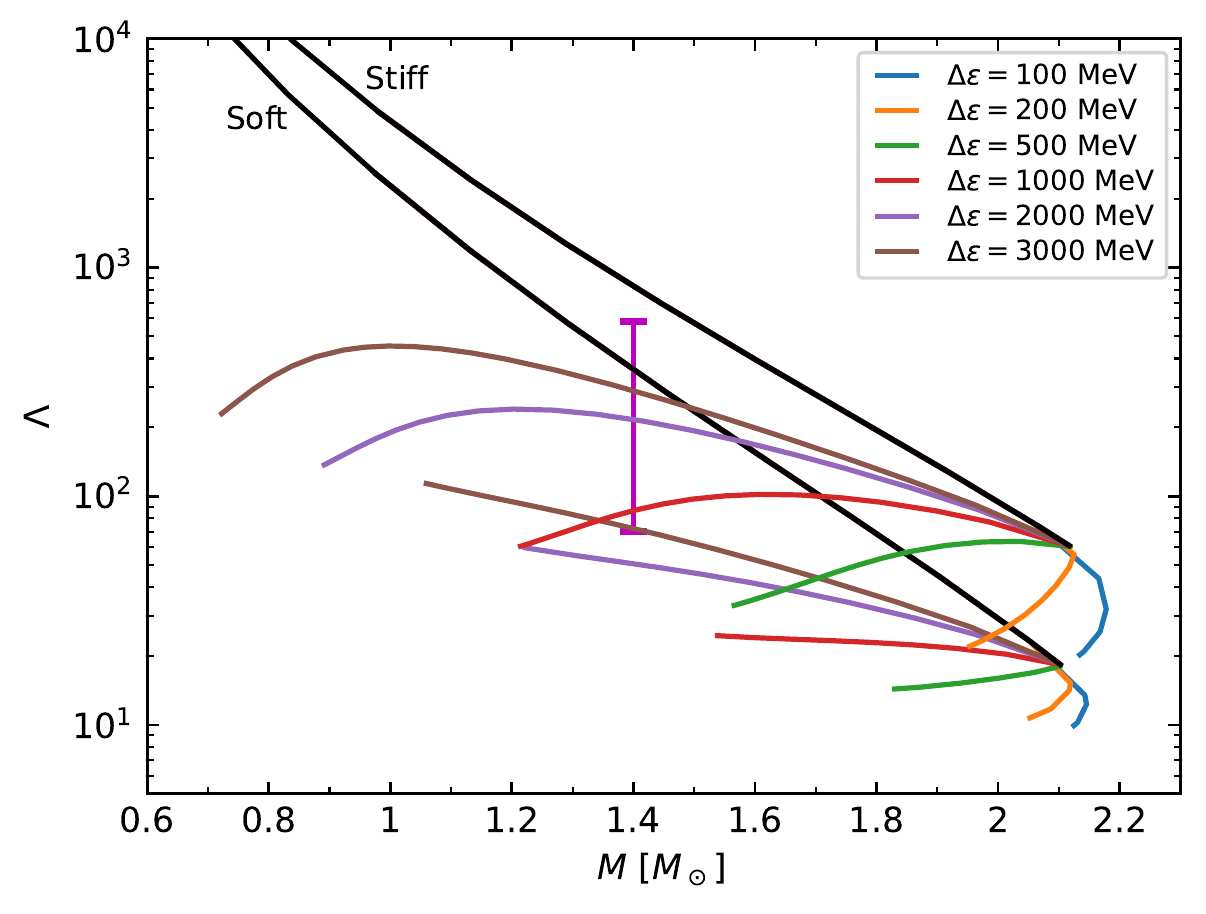}
\caption{Dimensionless tidal deformability, $\Lambda$, as a function of the mass for the stellar configurations constructed using the selected hybrid EoS. With black lines we present hadronic configurations and with different colors hybrid configurations for different values of $\Delta \epsilon$. All displayed configurations are stable in the slow conversion regime. If rapid conversions are assumed, the only stable configurations are those up to the mass peak. The vertical magenta line indicates the inferred value for the 1.4$M_{\odot}$ compact object, $70 < \Lambda_{1.4} < 580$ \citep{Abbott:2018}.}
\label{fig:tidal}
\end{figure}

\section{Axial modes}
\label{axial-modes}

In this section, we will present a short introduction to axial modes described through the Regge-Wheeler equation~\cite{regge1957}. A particular family of such modes are the $w$-modes, which are purely spacetime modes, directly related to curvature changes in compact objects. These modes were first obtained in Ref.~\cite{kokkotas:1992}. As occurs with other QNMs, their frequencies and damping times are EoS-dependent quantities \cite{Benhar:1998}. Working within the Regge-Wheeler gauge, the quadrupolar ($\ell =2$) axial metric perturbations are governed by the following differential equation:
\begin{eqnarray}\label{Regge-Wheeler}
Z_{,tt}&=&\frac{e^{\nu (r)}}{e^{-\lambda (r)}}\Bigg( Z_{,rr}-\bigg(\left[\frac{1}{r}\left(1-e^{\lambda (r)} \right)+  Q(r)\right]Z_{,r} \nonumber \\ && + \left[\frac{3}{r^2}\left(1+e^{\lambda (r)} \right)+\frac{Q(r)}{r}\right]Z \bigg) \Bigg),
\end{eqnarray}
where the metric coefficients $\nu (r)$ and $\lambda (r)$, the pressure $P(r)$, and the energy density $\epsilon (r)$ profiles are obtained after solving the TOV equations and \mbox{$Q(r)=4\pi r e^{\lambda(r)}(\epsilon (r) - P(r))$}. This is the Regge-Wheler equation, although it is more commonly presented using the tortoise coordinate \cite{kokkotas:1999}.

To obtain the QNMs we are interested in, we must solve Eq.~\eqref{Regge-Wheeler} with the following boundary conditions: the solution must be regular at the center of the star ($r=0$), it must behave like a purely outgoing wave at infinity $(r \to \infty$), and  has to be continuous at the surface of the star ($r=R$).

The approach we follow to obtain the QNMs is the classical variable separation method. We will assume that the function $Z(r,t)$ can be written as:
\begin{equation}
Z(r,t) = \zeta (r) e^{i\omega t},
\end{equation}
where the QNM frequency $\omega = 2\pi \nu + i/\tau$ is a complex number.

Introducing this ansatz, Eq.~\eqref{Regge-Wheeler} can be written (inside the star) in the following form:
\begin{eqnarray}
\label{radialRWin}
\zeta_{,rr} &&+ \left(1 - \frac{2m(r)}{r}\right)^{-1}\bigg[\left(\frac{2m(r)}{r^2} - \frac{Q(r)}{e^{\lambda(r)}} \right)\zeta_{,r}  \\
&& - \left[\frac{6}{r^2}\left(1 - \frac{m(r)}{r}\right)^{-1} + \frac{Q(r)}{r e^{\lambda(r)}}  \right] \zeta + \omega ^ 2 e^{-\nu (r)} \zeta \bigg] = 0, \nonumber
\end{eqnarray}
where $m(r)$ describes the gravitational mass profile inside the compact object.

Outside the star, Eq.~\eqref{radialRWin} simplifies to
\begin{eqnarray}
\label{radialRWout}
\zeta_{,rr} &&+ \left(1 - \frac{2M}{r}\right)^{-1} \\ \times && \left[\frac{2M}{r^2}\zeta_{,r}  - \frac{6}{r^2}\left(1 - \frac{M}{r}\right)^{-1} \zeta + \omega ^ 2 \frac{\zeta}{1 - \frac{2M}{r}} \right] = 0, \nonumber
\end{eqnarray}
where $M$ is the total gravitational mass of the stellar configuration.

Following the ideas presented in Ref.~\cite{Mena-Fernandez2019}, we introduce the following function
\begin{equation}
	\kappa=\zeta ^{-1}\zeta_{,r},
\end{equation}
with which Eq.~\eqref{radialRWout} becomes
\begin{eqnarray}\label{eq10}
	\kappa_{,r} &&+ \kappa ^ 2 + \left(1-\frac{2M}{r} \right)^{-1} \\
	&&\times \left[\frac{2M}{r^2}\kappa - \frac{6}{r^2}\left(1-\frac{M}{r} \right) + \omega ^2 \left(1-\frac{2M}{r} \right)^{-1} \right] = 0. \nonumber
\end{eqnarray}

The metric function $\zeta$ needs to be regular at $r=0$. Studying its Frobenius expansion, we obtain that the first terms of the well behaved solution read (up to an arbitrary amplitude with no physical meaning)
\begin{equation} \label{origin}
	\zeta \sim r^3+\frac{16\pi \left(\epsilon_0-P_0\right)-\omega^2e^{-\nu_0}}{14}r^5,
\end{equation}
where quantities with a $0$ subscript are evaluated at $r=0$. We will use Eq.~\eqref{origin} to start the numerical integration of Eq.~\eqref{radialRWin}.

To perform the integration outside the star we will use a {\it complexification} of the radial coordinate. The main numerical advantages of such process are exposed in Refs.~\cite{w-modes_universal,Mena-Fernandez2019}. We will briefly discuss the most important aspects related to the numerical implementation of the method  within our particular problem.

We define a new variable $\rho$ by means of:
\begin{equation}
	r=R+ \rho e^{i\alpha},\quad \rho \in [0,\infty),
\end{equation}
where $R$ is the radius of the configuration and the free parameter $\alpha$ must satisfy
\begin{equation}
	\Re\,(\omega)\sin\alpha+\Im \,(\omega)\cos\alpha<0.
\end{equation}
$\Re$ and $\Im$ stand respectively for the real and the imaginary parts of a given complex number.

To impose the outgoing wave behavior condition, we compactify the variable $\rho$. The major benefit of this procedure is that no {\it numerical definition} of infinity needs to be done. The compactification is performed using new coordinate $0 \le x < 1$ defined by
\begin{equation}
	\rho=\frac{1-x}{x}.
\end{equation}

The first terms of the Frobenius solution compatible with a purely outgoing wave is given by
\begin{equation}
	\kappa \sim -i\omega\left[1+2Me^{-i\alpha}x \right].
\end{equation}
This allow us to integrate Eq.~\eqref{radialRWout} from infinity to the surface of the star.

The last condition that has to be imposed is the continuity of the interior ($in$) and exterior ($ext$) solutions at the surface of the star. Such condition is guaranteed demanding that the function $\kappa$ is continuous at the surface of the stellar configuration, 
\begin{equation}\label{junction}
	\kappa^{(\omega)}_{in}\Big \rvert _{{\rm surf}}=\kappa^{(\omega)}_{ext}\Big \rvert _{{\rm surf}} .
\end{equation}


\begin{figure}[t]
\centering
\includegraphics[width=0.95\linewidth,angle=0]{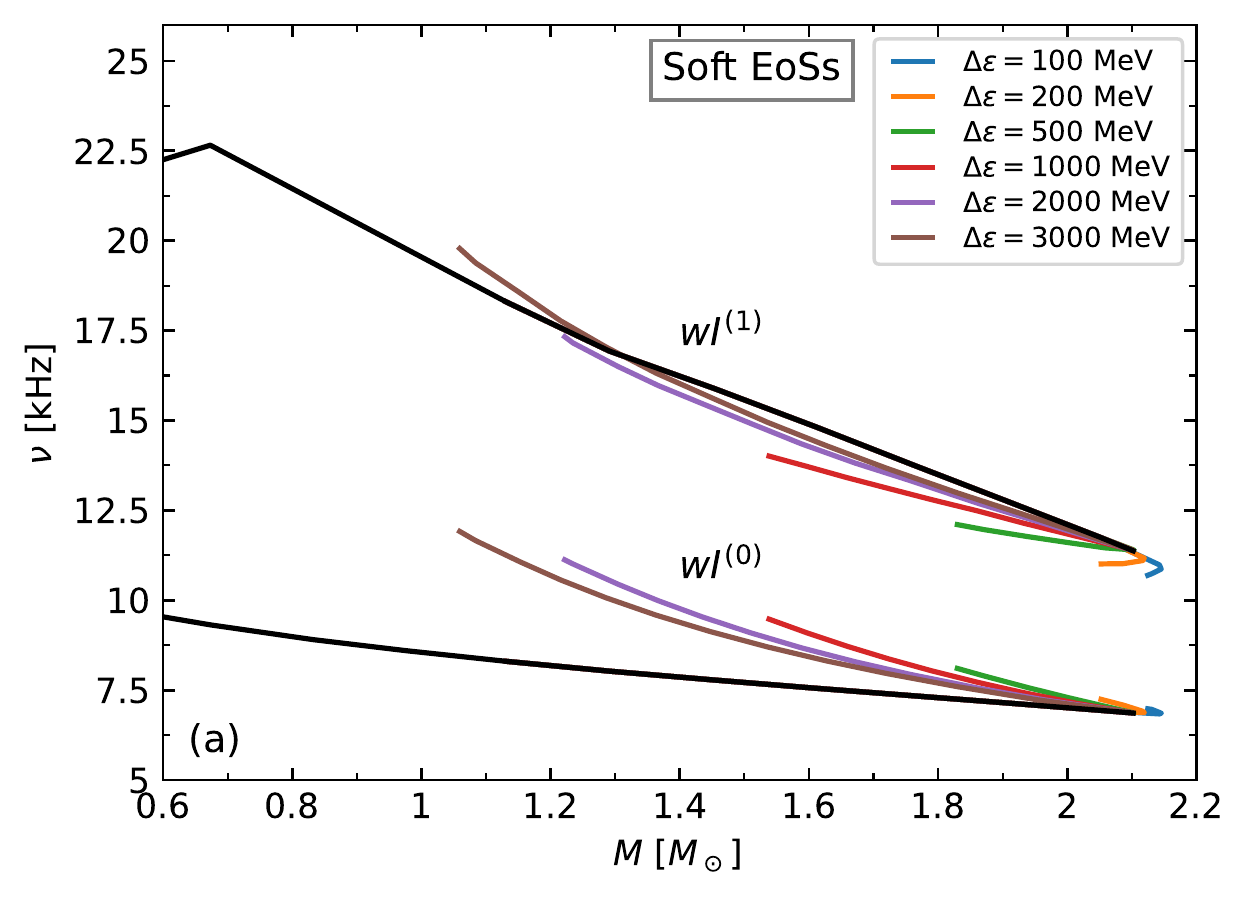}
\includegraphics[width=0.95\linewidth,angle=0]{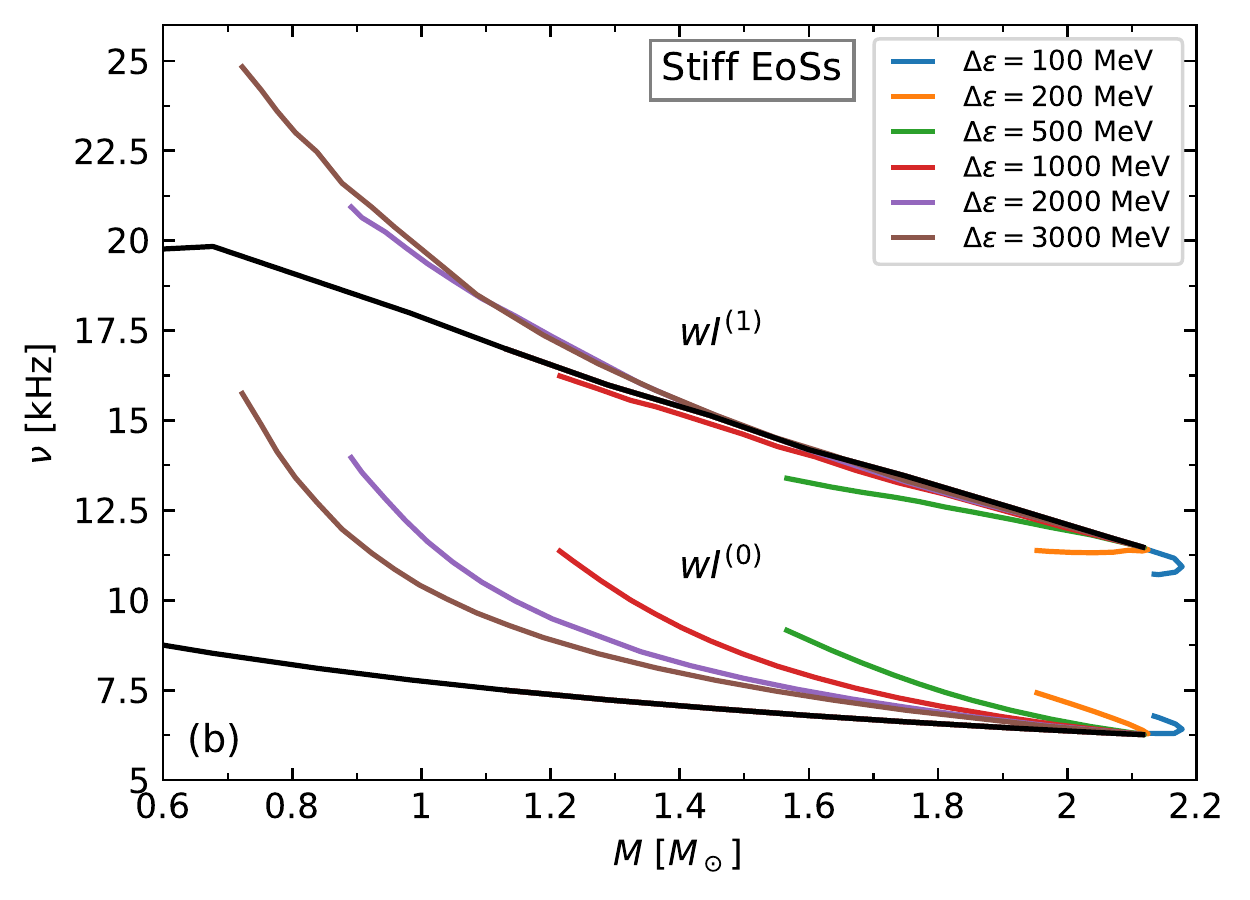}
\caption{Frequency, $\nu$, for both the fundamental and first overtone as a function of the stellar mass, $M$. In the top (bottom) panel, results were obtained using the soft (stiff) hadronic EoS. Color selection is the same used in Fig.~\ref{fig:MR}.} 
\label{fig:num}
\end{figure}

The integration of the Regee-Wheler equation and the calculation of the $wI$-modes is performed, after obtaining the TOV solutions with the \emph{Neutron Star Object Research} (NeStOR) code \cite{mariani-nestor}, using the \emph{Frequency Identificator and Extractor Library} (FIdEL\footnote{FIdEL is a modular code written in {\sc{Fortran 90}} that,  given the structure of NSs calculated previously, computes the $wI$-modes. FIdEL is available for the community upon reasonable request.}). Integration of differential equations is performed using a Runge-Kutta-Fehlberg integrator. A similar procedure has been used to obtain polar QNMs of HSs using {\it Cowling Frequency Key} (CFK) \cite{RSetalJCAP,Rodriguez-etalg2}. In order to obtain the $wI$-modes given an EoS and a family of stars, FIdEL computes the roots of the function 
\begin{equation}
    f(\omega)= \kappa^{(\omega)}_{in}\Big \rvert _{{\rm surf}} - \kappa^{(\omega)}_{ext}\Big \rvert _{{\rm surf}}, 
    \label{eq_zero_w}
\end{equation}
using M\"uller's method. To do that, we start with the less massive star of the family and choose three initial conditions, $\omega_0$ and $\omega_0 \pm \delta \omega_0$. We apply the M\"uller's  method to obtain a correction $w^*$ of $w_0$, and repeat the procedure using, now, $\omega^*$ and $\omega^*\pm \delta \omega^*$ until the frequency of the desired QNM is obtained to a given precision. Finally, when the $wI$-mode is obtained for the first stellar configuration, we use it as the initial condition, $\omega_0$, for the next one and repeat the computations until the $wI$-modes of the whole family of stellar configurations are obtained.

\section{Results} \label{results}

\begin{figure}[t]
\centering
\includegraphics[width=0.93\linewidth,angle=0]{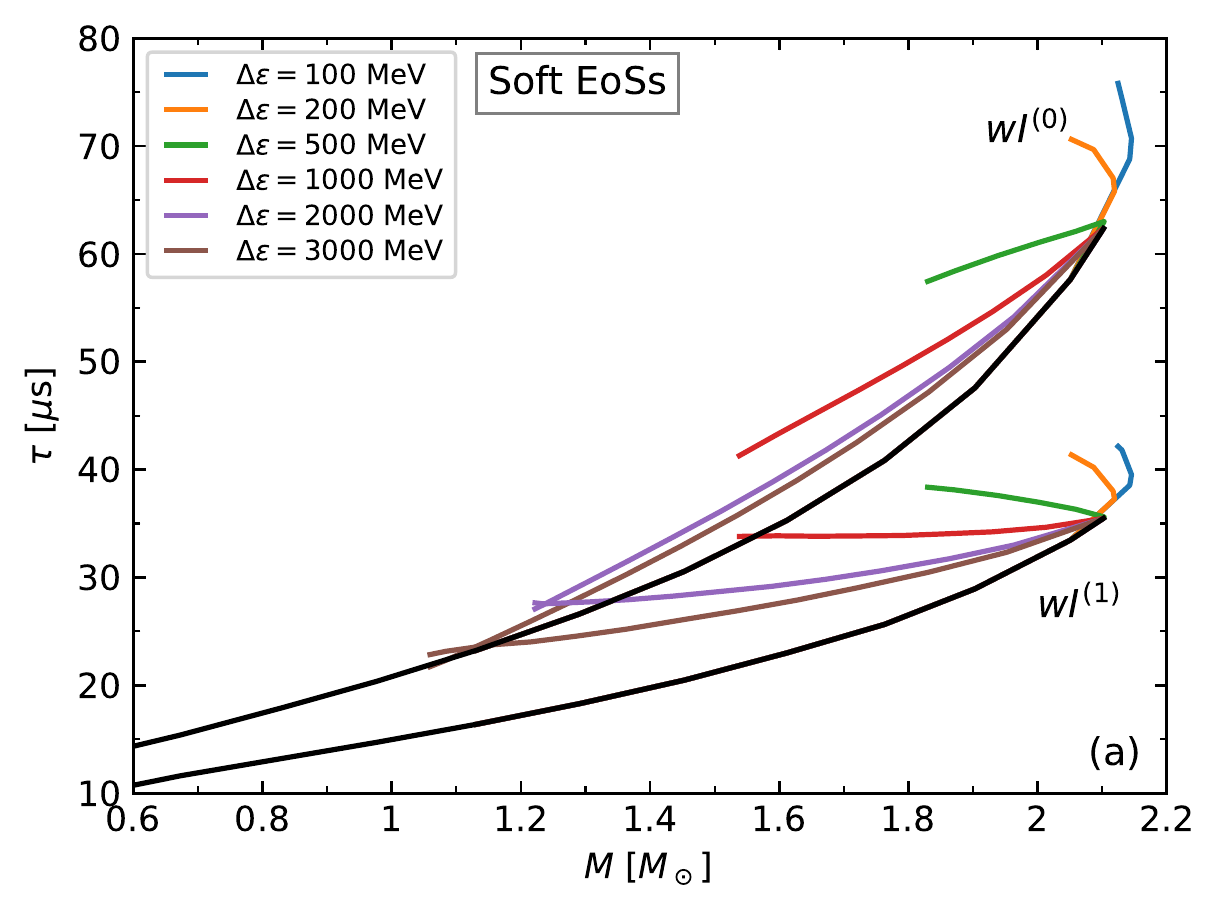}
\includegraphics[width=0.93\linewidth,angle=0]{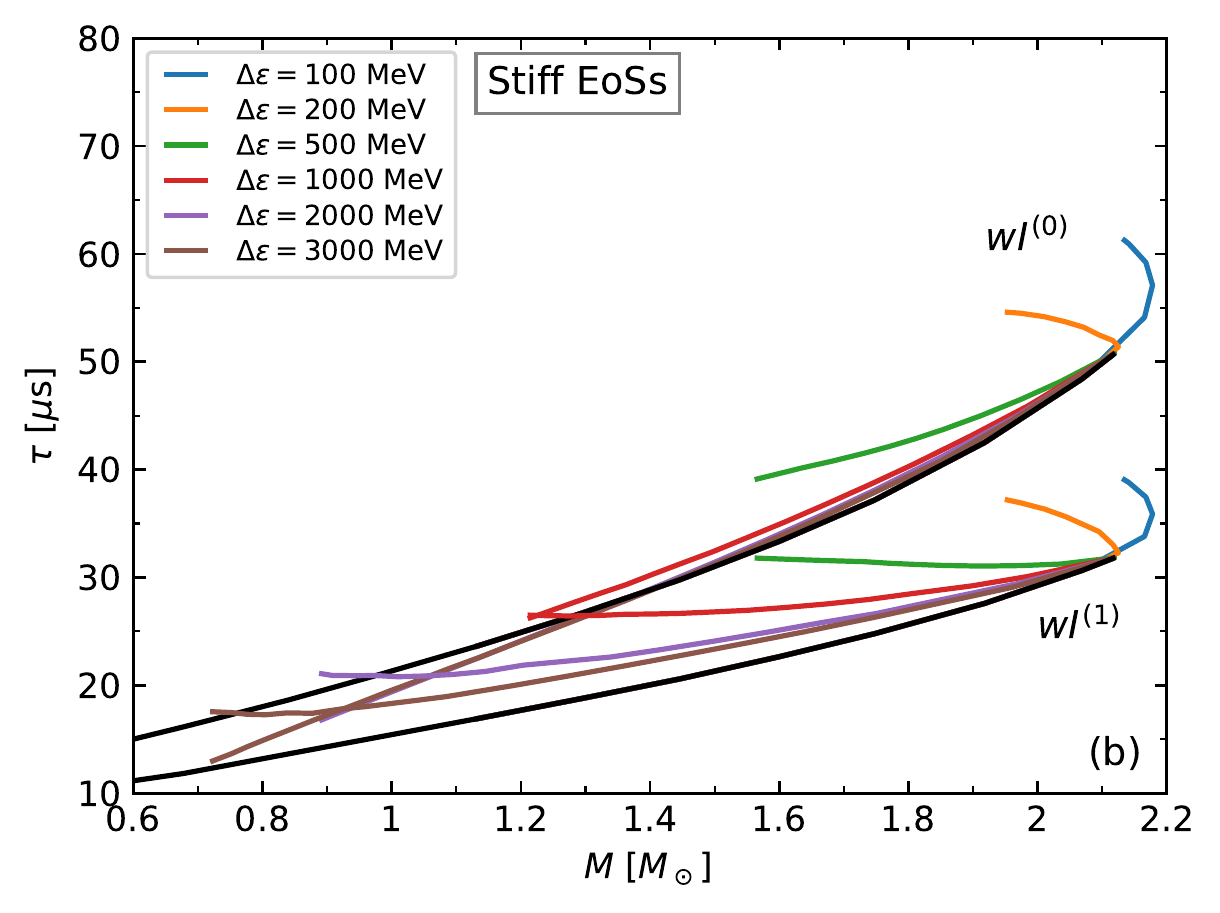}
\caption{Damping time, $\tau$, for both the fundamental and the first overtone as a function of the stellar mass, $M$. In the top (bottom) panel, results were obtained using the soft (stiff) hadronic EoS. Colors have the same meaning as in Fig.~\ref{fig:MR}.} 
\label{fig:taum}
\end{figure}

\begin{figure}[t]
\centering
\includegraphics[width=0.95\linewidth,angle=0]{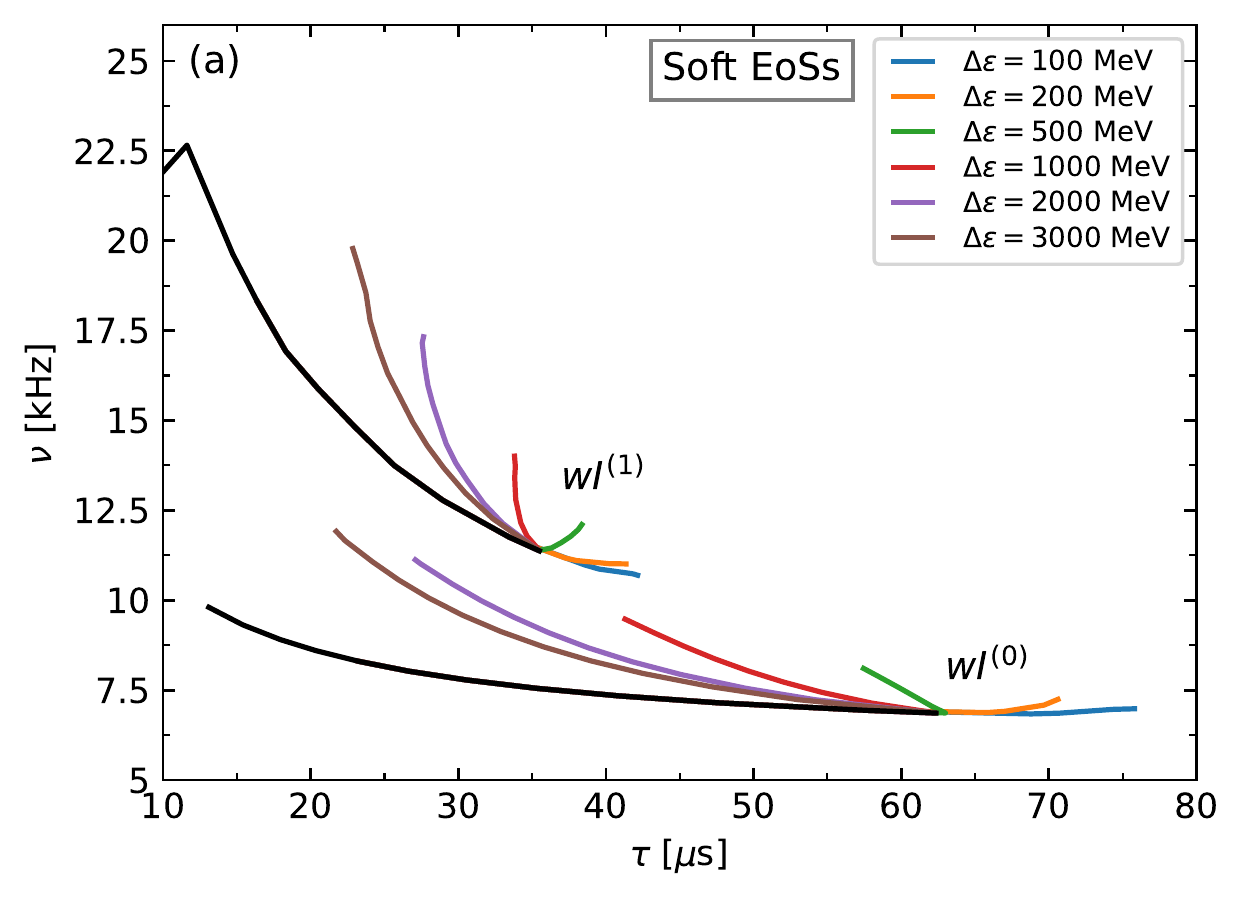}
\includegraphics[width=0.95\linewidth,angle=0]{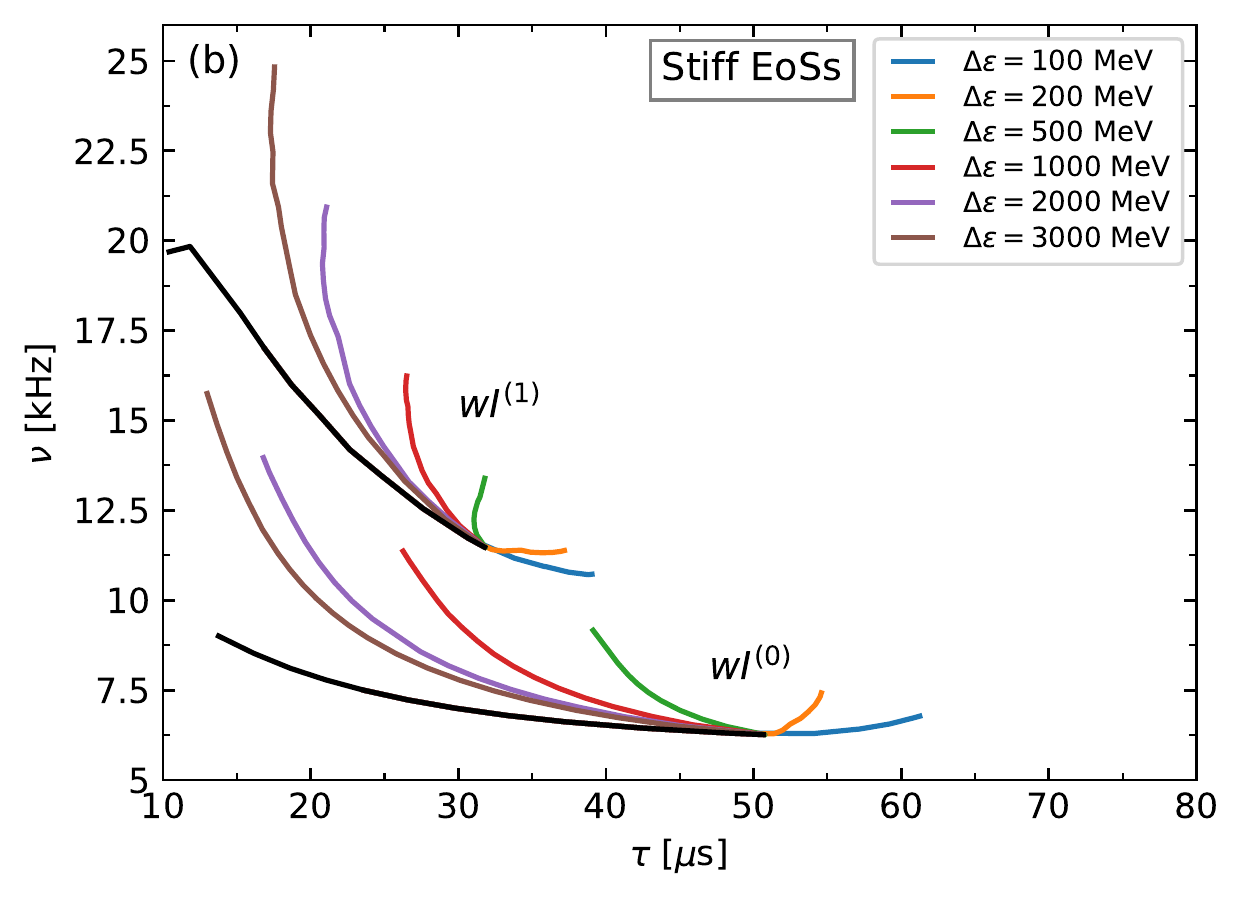}
\caption{Oscillation frequency, $\nu$, and damping time, $\tau$, for the fundamental mode $wI^{(0)}$, and the first overtone $wI^{(1)}$, of hybrid compact objects constructed with soft (top panel) and stiff (bottom panel) hadronic EoSs. When the extended stable branch is present, a drastic change in the behavior of the curves is found as the value of $\Delta \epsilon$ grows. Color selection is the same used in Fig.~\ref{fig:MR}.} 
\label{fig:nutau}
\end{figure}

We have constructed a broad family of hybrid EoSs with a sharp hadron-quark phase transition. The analysis of all the hybrid EoS we have constructed allowed us to reach the conclusion that the CSS parameter that has the most relevant effect on the length of the extended hybrid branch of stable configurations is $\Delta \epsilon$. Based on this fact, we studied six different values for this parameter: 100, 200, 500, 1000, 2000 and 3000 MeV/fm$^3$. In these cases, we have fixed $c_\mathrm{s}^2=0.33$. In addition, to show the impact on our results of variations in $c_\mathrm{s}^2$, we fix $\Delta \epsilon = 1000$ MeV/fm$^3$ and study three values: 0.33, 0.50, 0.70 and 1.00.  (We leave for Appendix \ref{app1} the analysis of the impact on the $wI$ axial QNMs of changes in the speed of sound, $c_{\rm s}$.) For each hadronic EoS, we have fixed the value of $P_t$. We selected the value of the transition pressure $P_t = 250$~MeV/fm$^3$ and $P_t = 110$~Mev/fm$^3$ when using the soft and stiff hadronic EoSs, respectively. We do not study the impact of varying this parameter of the CSS parametrization EoS as selecting low values of it and demanding consistency with the astronomical data leads to the appearance of long \emph{traditional} (\textit{i.e} having $\partial M/ \partial \epsilon_c >0$) hybrid branches and, in some cases, hybrid twin branches (examples of this situation can be found in Refs. \citep{Alvarez-Castillo:2018pve,jakobus2021EPJC,2022EPJC}). We are not interested in this situations as $wI$-modes of these stellar configurations are known to fulfill the universal relationships \citep{w-modes_universal}. Moreover, larger values also leads to mass-radius relationships with maximum masses above 2.3$M_\odot$ and are discarded.

With these EoSs, we have solved TOV equations to construct spherically symmetric stellar configurations and studied their stability under linearized radial perturbations, following the ideas presented in Ref.~\cite{Pereira:2017rmp} (also see the recent review on this subject \cite{LugGrunf-universe:2021}). As can be seen, in the mass-radius relationship presented (for the hybrid EoSes with $c_{\rm s}^2=0.33$) in Fig.~\ref{fig:MR}, several of the hybrid EoSs with sharp hadron-quark phase transition produce (in the slow conversion regime) long extended branches of stable stellar configurations. Moreover, we see that the extended branches of stellar configurations are compatible with astronomical data and, for that reason, should, in principle not be discarded as a viable theoretical scenario (see, Ref.~\cite{lugones2021arXiv} for a more detailed discussion on this subject). In the rapid conversion regime, the constructed stellar configurations are stable only up to the maximum mass, and the appearance of quark matter almost immediately destabilizes the compact object. In addition to the mass-radius relationship, we present, in Fig.~\ref{fig:tidal}, the dimensionless tidal deformability, $\Lambda$, as a function of the mass. As previously shown in Refs. \citep{Mariani:2019vve,curin:2021hsw,lugones2021arXiv,mariani2022MNRAS}, this quantity is not a monotonic decreasing function of the mass, and a rather peculiar behavior can be seen for the extended stability branches. For the stiff hadronic EoS, some of the selected hybrid EoSs (those with $\Delta \epsilon \lesssim 1000$~MeV/fm$^3$) are not compatible with restrictions from GW170817 and its electromagnetic counterpart, as it can be seen both in Fig.~\ref{fig:MR} and \ref{fig:tidal}. Despite this fact, as we are interested in studying universal relationships for $wI$-modes, we also present the results obtained using these EoSs.

\begin{figure*}[t]
\centering
\includegraphics[width=0.49\linewidth,angle=0]{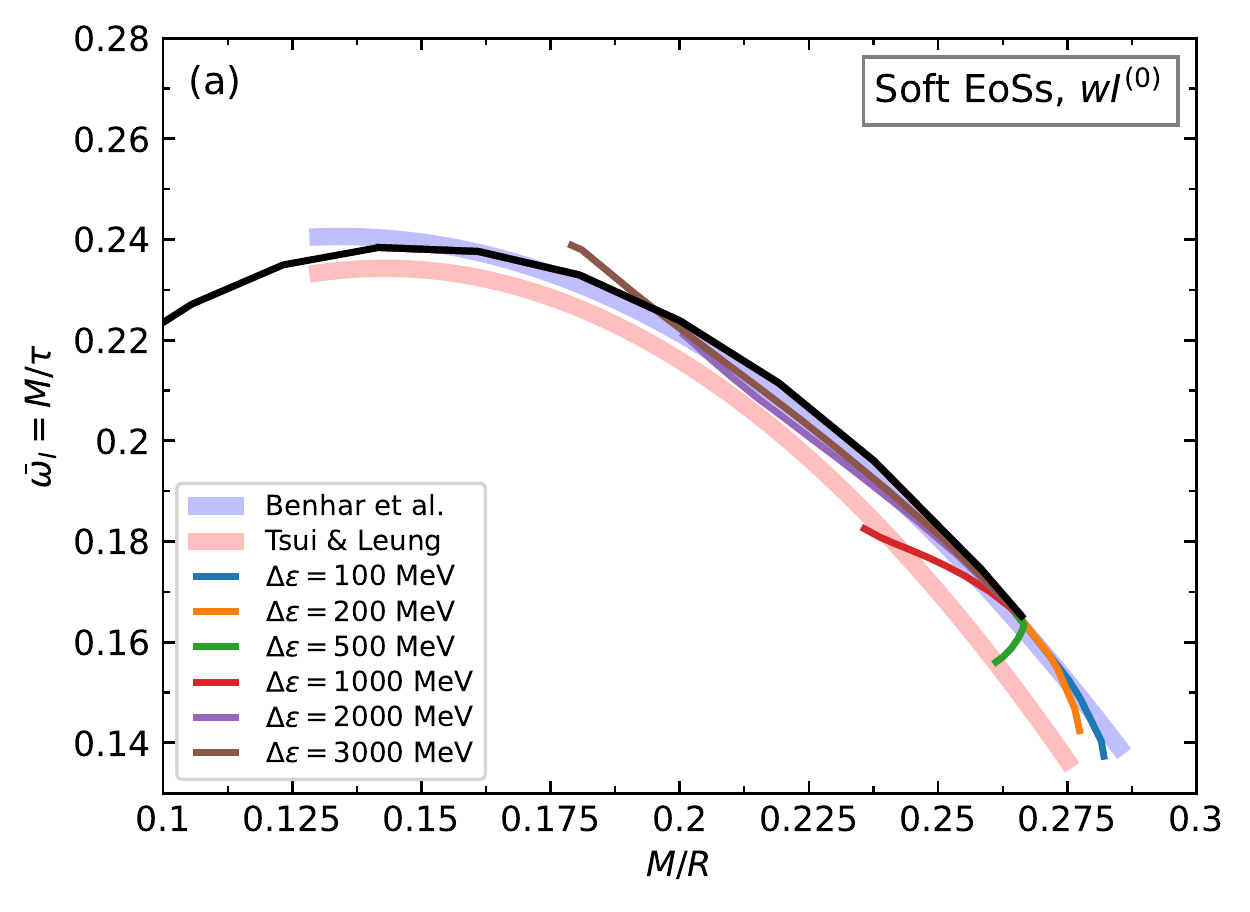}
\includegraphics[width=0.49\linewidth,angle=0]{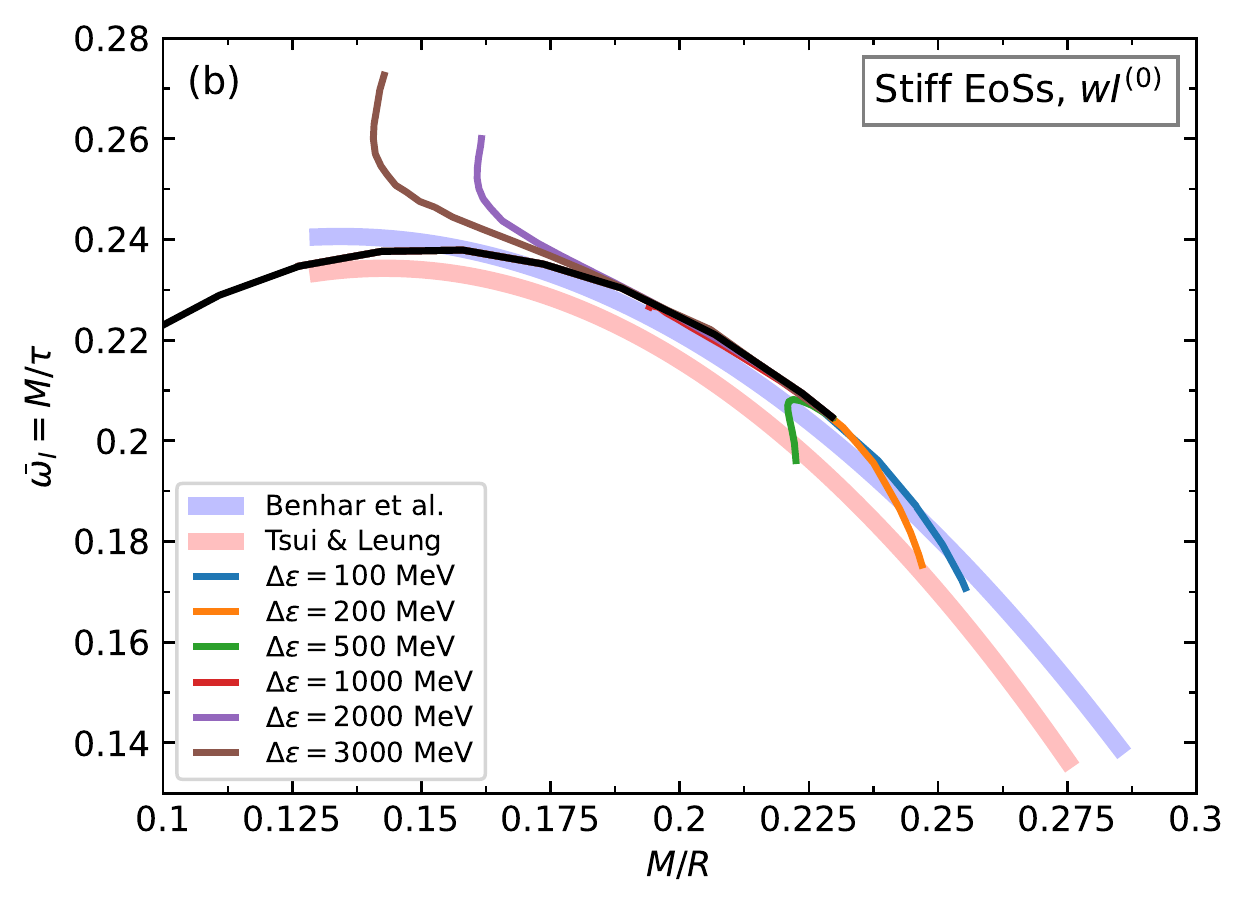}
\includegraphics[width=0.49\linewidth,angle=0]{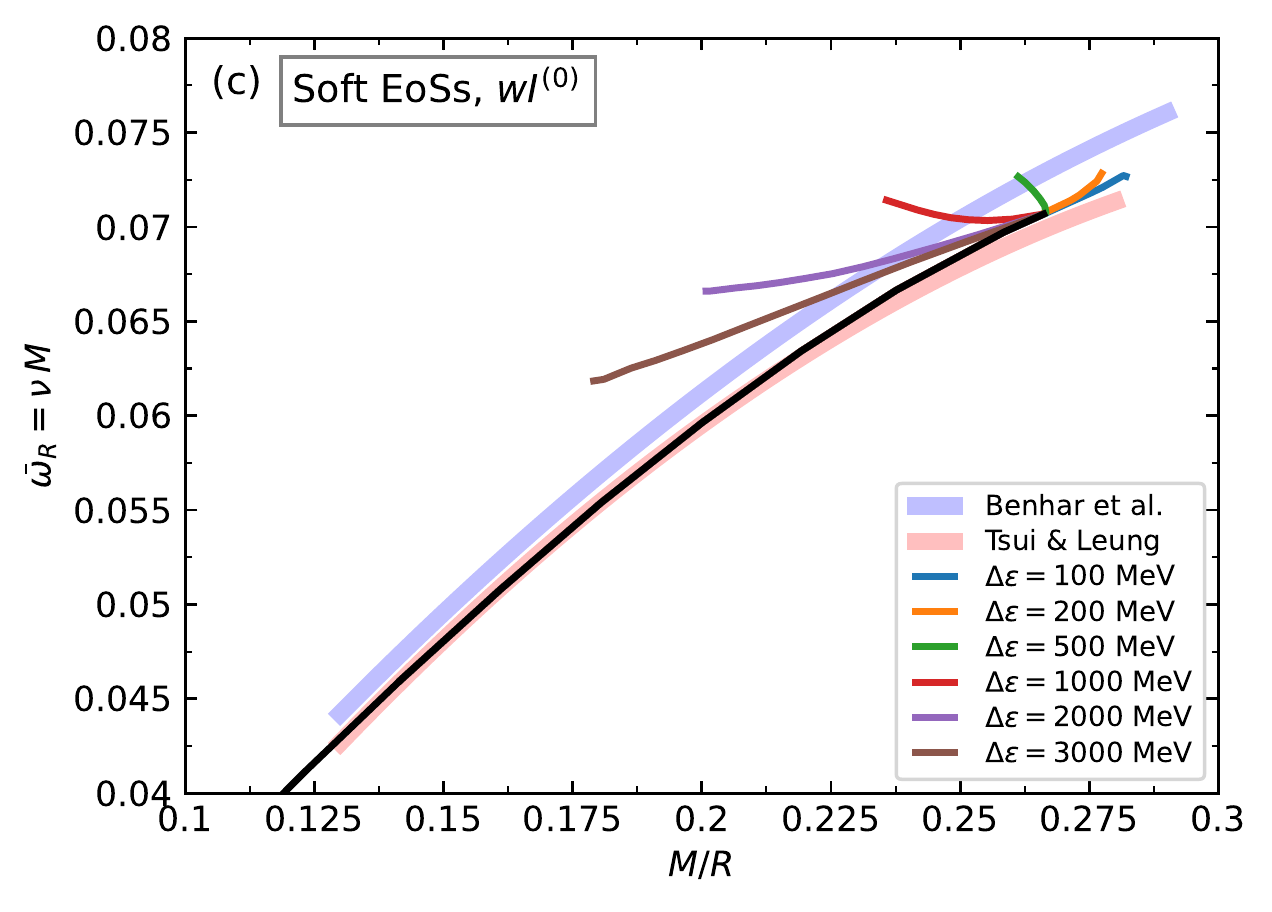}
\includegraphics[width=0.49\linewidth,angle=0]{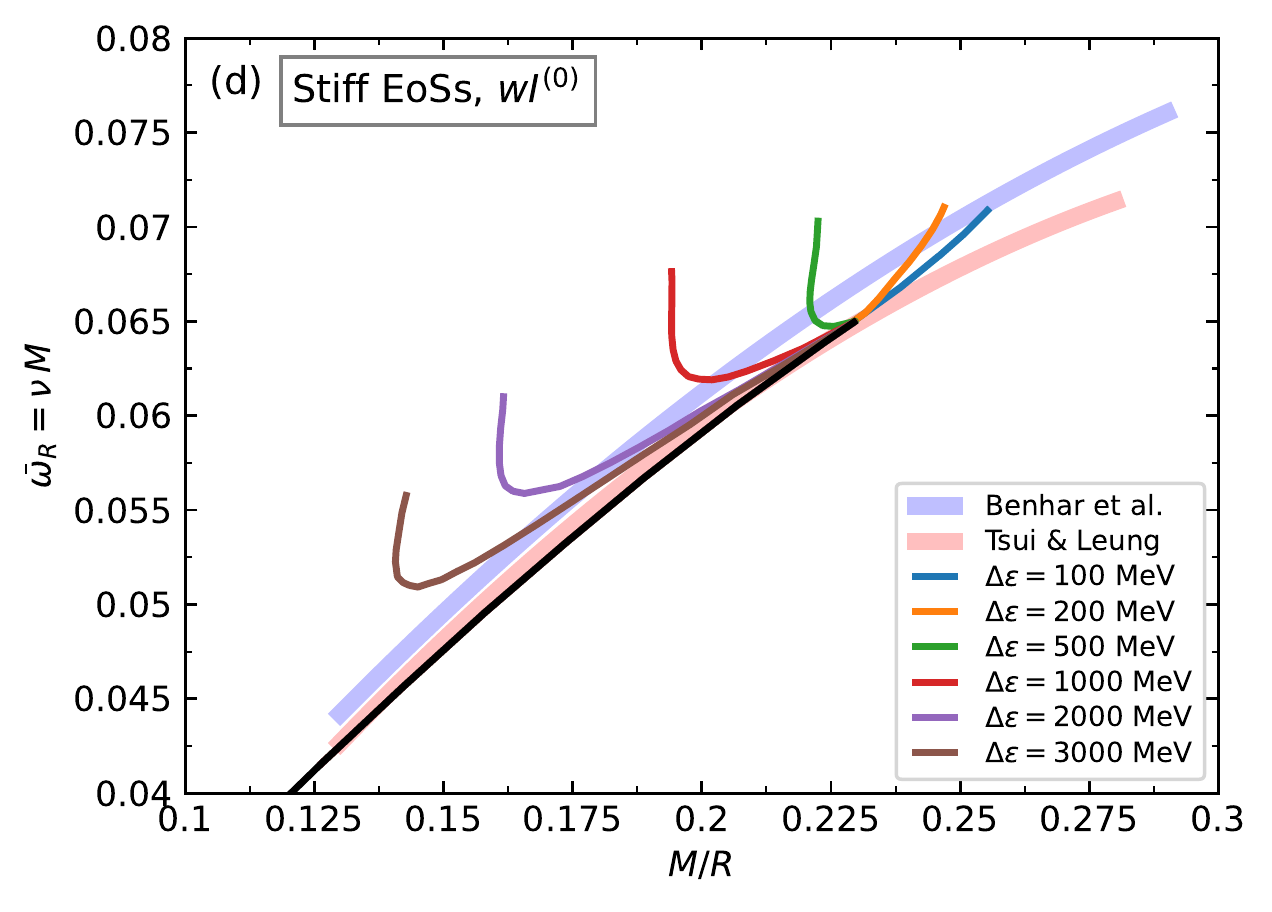}
\caption{Comparison of our results for the $wI^{(0)}$ mode as a function of stellar compactness and universal relationships proposed in Refs.~\cite{Tsui2005MNRAS,Benhar2004PRD}, shown using thick lines. In the upper panels we show results related to $\bar{\omega}_R = M/\tau$ and in the bottom ones those for $\bar{\omega}_R = M \nu$.} 
\label{fig:univ3_w0}
\end{figure*}

After obtaining stellar configurations we have calculated $wI$ axial QNMs of two representative families of hybrid objects presenting (long) extended branches of slow stable hybrid objects. In Fig.~\ref{fig:num} (upper panel for the soft hadronic EoS and lower panel for the stiff one) we present the frequency, $\nu$, of both the $wI^{(0)}$ and $wI^{(1)}$ modes for these hybrid stars. We see that for both modes, the frequency of hadronic stars decreases as the mass of the object becomes larger. For the fundamental mode, this situation is the same for slow stable hybrid objects and, for the first overtone, the behavior depends mainly on the value of $\Delta \epsilon$, but also on the hadronic EoS used. We see that for the soft EoS, if $\Delta \epsilon < 3000$~MeV/fm${^3}$, the frequency of the slow stable hybrid twin is larger that the one corresponding to the hadronic sibling up to the terminal mass, but that for the case in which $\Delta \epsilon = 3000$~MeV/fm${^3}$ this situation is inverted near the terminal mass. When the stiff EoS is used, similar results are obtained for $\Delta \epsilon < 1000$~MeV/fm${^3}$ and, for values above, the situation is inverted for almost all the objects in the slow stable hybrid branch of stellar configurations.

In Fig.~\ref{fig:taum}, we show the damping time, $\tau$, of the first two $wI$-modes as a function of the mass of the star. We find that for hadronic stars, $\tau$ increases monotonically with the mass of the object. For the fundamental mode, this trend is similar for slow stable hybrid configurations when $\Delta \epsilon \gtrsim 2000$~MeV/fm${^3}$. For smaller values, this situation is altered. It is important to note that in the hadronic branch, $\tau$ for the fundamental mode is larger than the one of the first overtone, but this situation can change for some slow stable hybrid configurations for large enough values of $\Delta \epsilon$. For the first overtone, a similar behavior can be seen but we can notice that, for some cases, $\tau$ becomes almost insensitive to the mass ($\Delta \epsilon = 1000$~MeV/fm${^3}$ for the soft EoS and $\Delta \epsilon = 500$~MeV/fm${^3}$ for the stiff one).

\begin{figure*}[t]
\centering
\includegraphics[width=0.49\linewidth,angle=0]{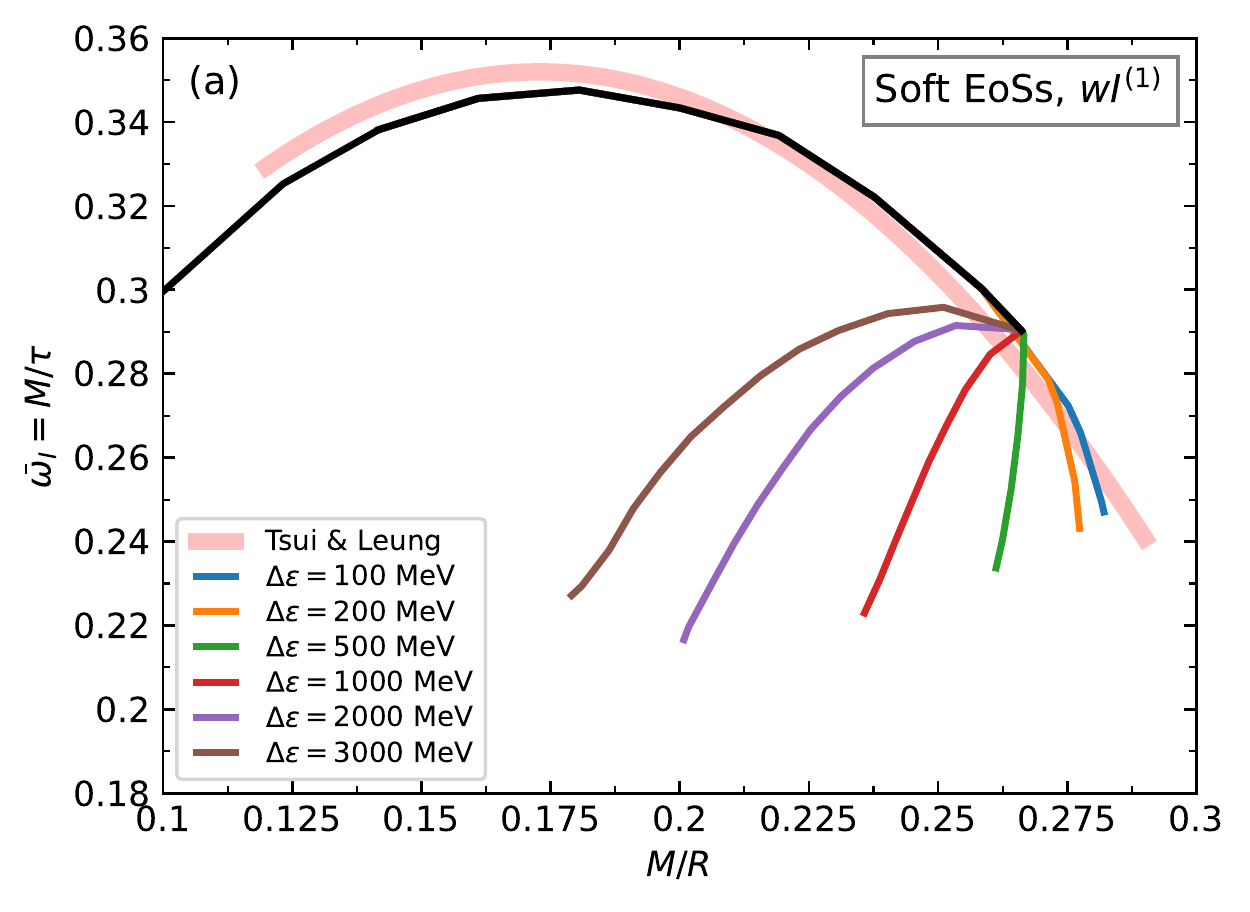}
\includegraphics[width=0.49\linewidth,angle=0]{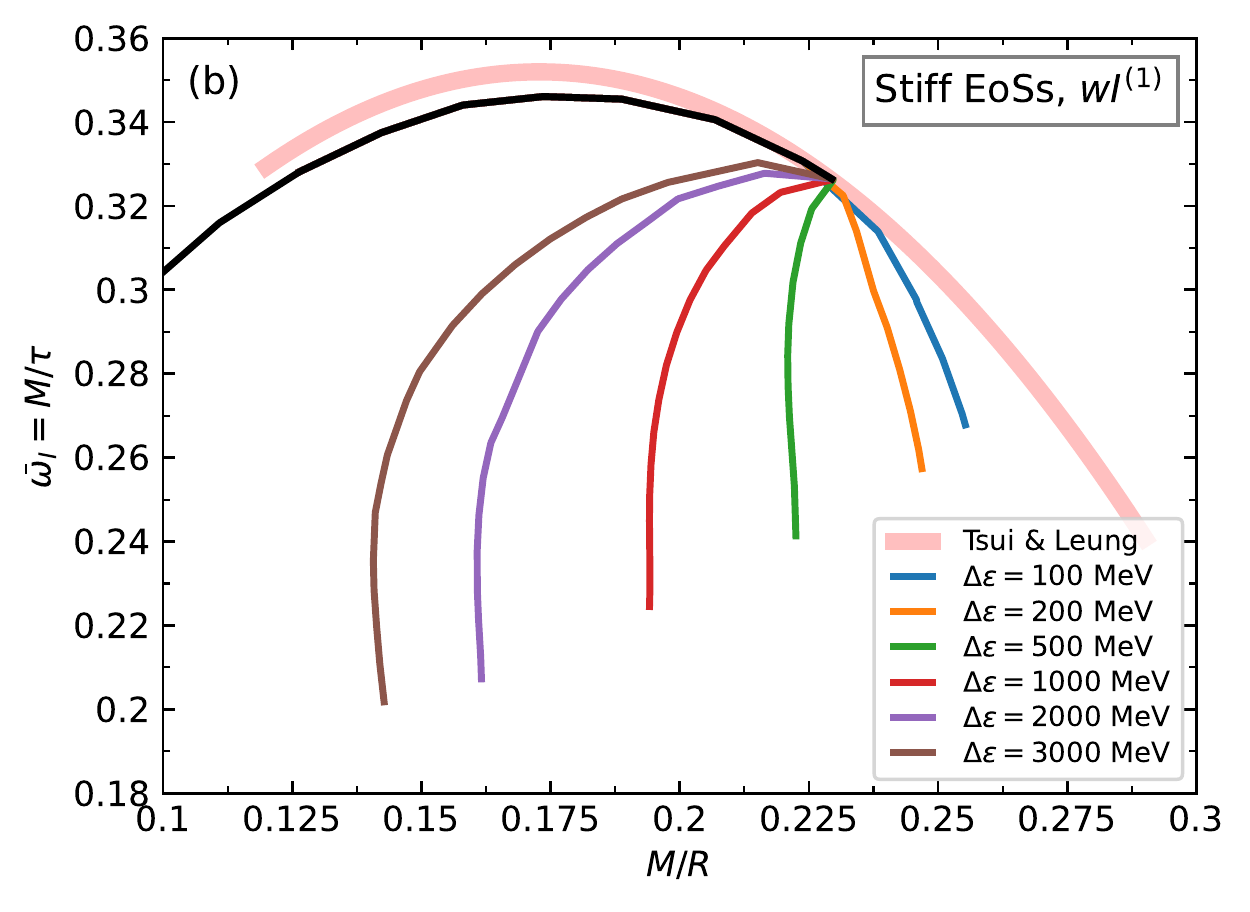}
\includegraphics[width=0.49\linewidth,angle=0]{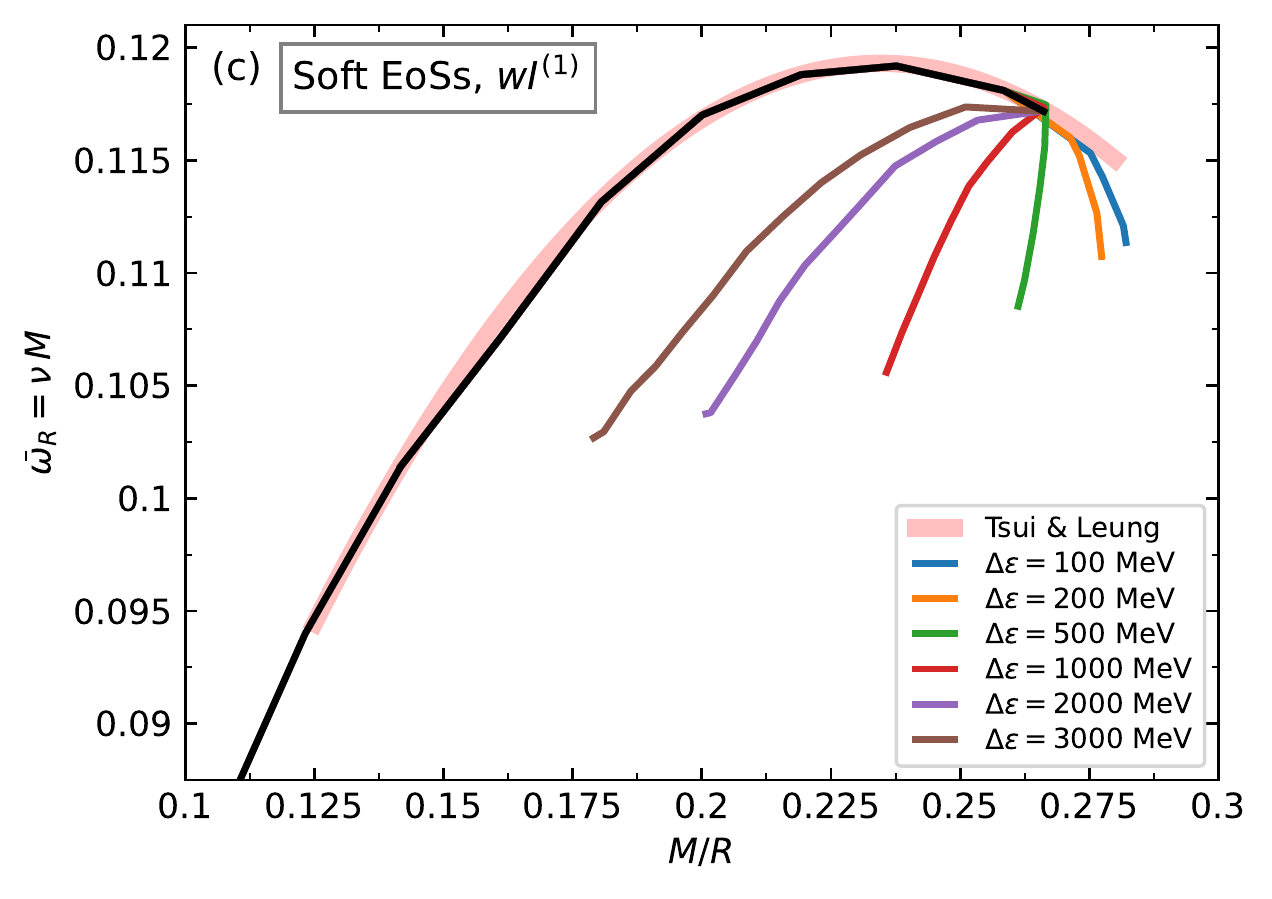}
\includegraphics[width=0.49\linewidth,angle=0]{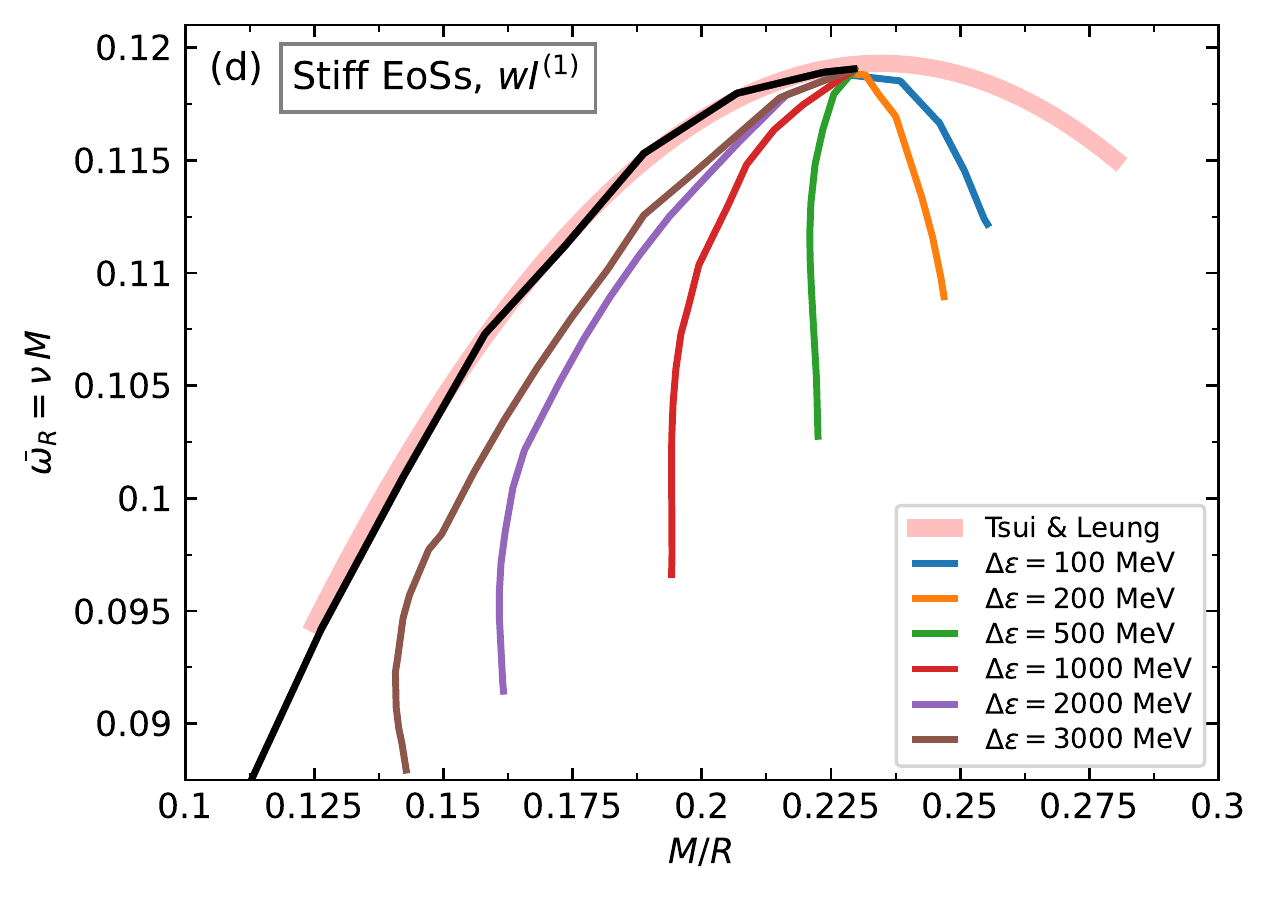}
\caption{Same as in Fig.~\ref{fig:univ3_w0}, but for the first overtone, $wI^{(1)}$. For this mode, Ref.~\cite{Benhar2004PRD} does not present a universal relationship. In all panels a clear deviation from the relationship of Ref.~\cite{Tsui2005MNRAS} can be seen for $\Delta \epsilon \gtrsim$ 500 MeV/fm${}^3$.}  
\label{fig:univ3_w1}
\end{figure*}

In Fig.~\ref{fig:nutau} we present $\nu$ as a function of $\tau$ for the $wI^{(0)}$ and $wI^{(1)}$ modes of compact objects constructed using hybrid EoSs with the soft (left panel) and stiff (right panel) hadronic EoS. 

For the hadronic branch, the frequency of the fundamental mode is almost constant with a value $\nu \sim 7.5$~kHz, despite it is a little larger for low mass objects and, in general, larger when the soft hadronic EoS is used, compared to the stiff one. The damping time increases from $\tau \sim 20$~ms to $\tau \sim 60 \, (50)$~ms if the soft (stiff) hadronic EoS is used. The general behavior for the first overtone is similar, but the frequency changes more clearly along the mass-radius curve and the damping times are in general shorter that those of the fundamental mode (in the range from $15$ to $35$~ms).

\begin{figure*}[t]
\centering
\includegraphics[width=0.49\linewidth,angle=0]{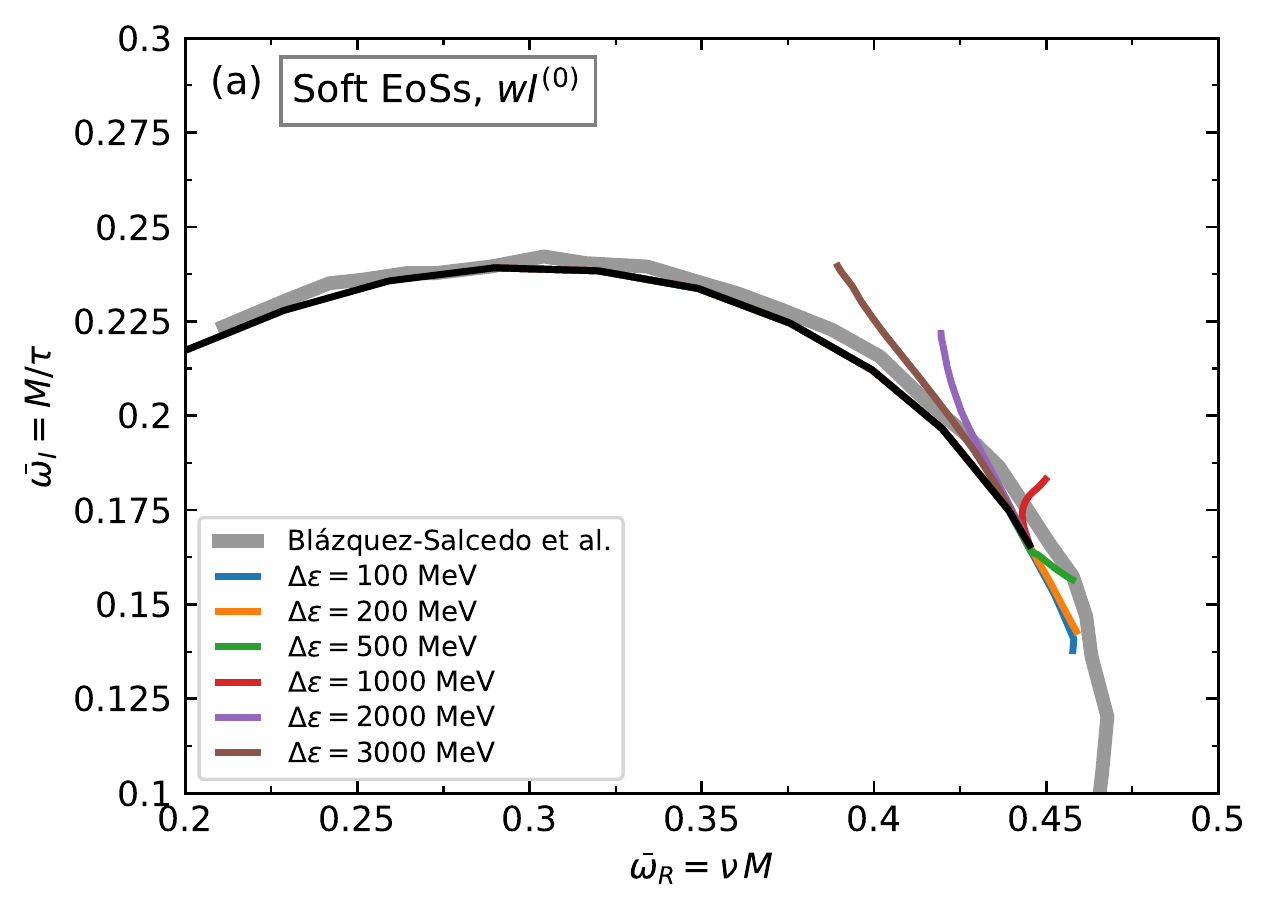}
\includegraphics[width=0.49\linewidth,angle=0]{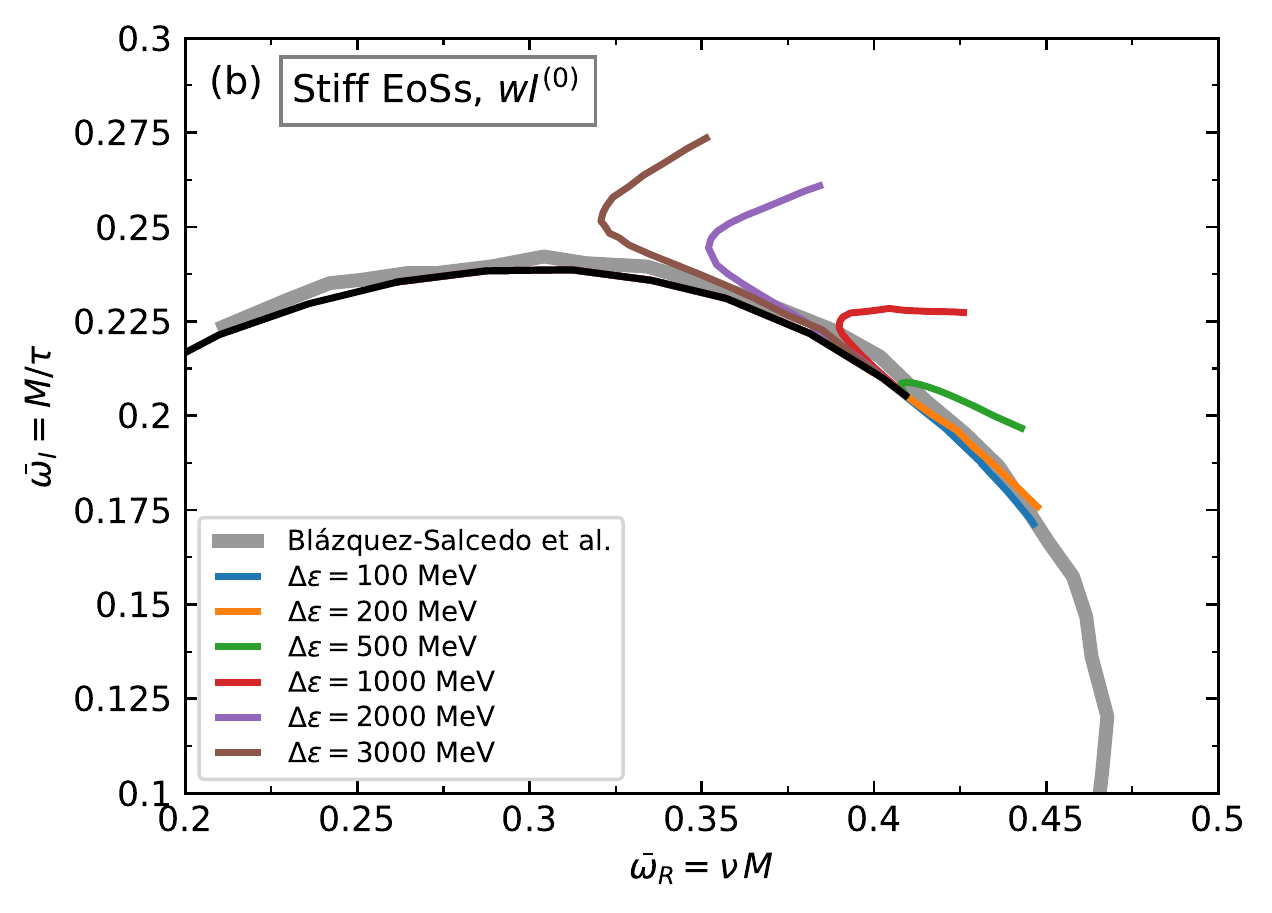}
\includegraphics[width=0.49\linewidth,angle=0]{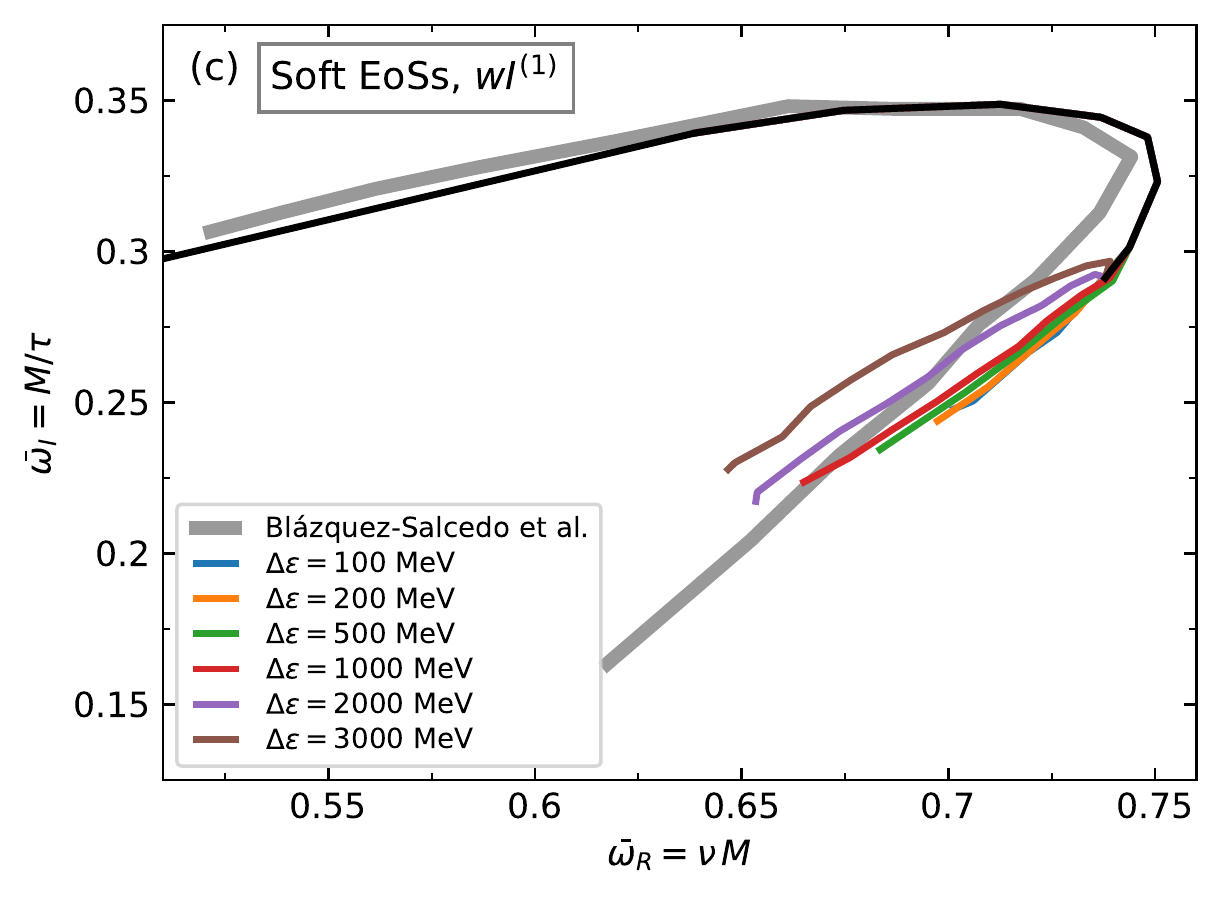}
\includegraphics[width=0.49\linewidth,angle=0]{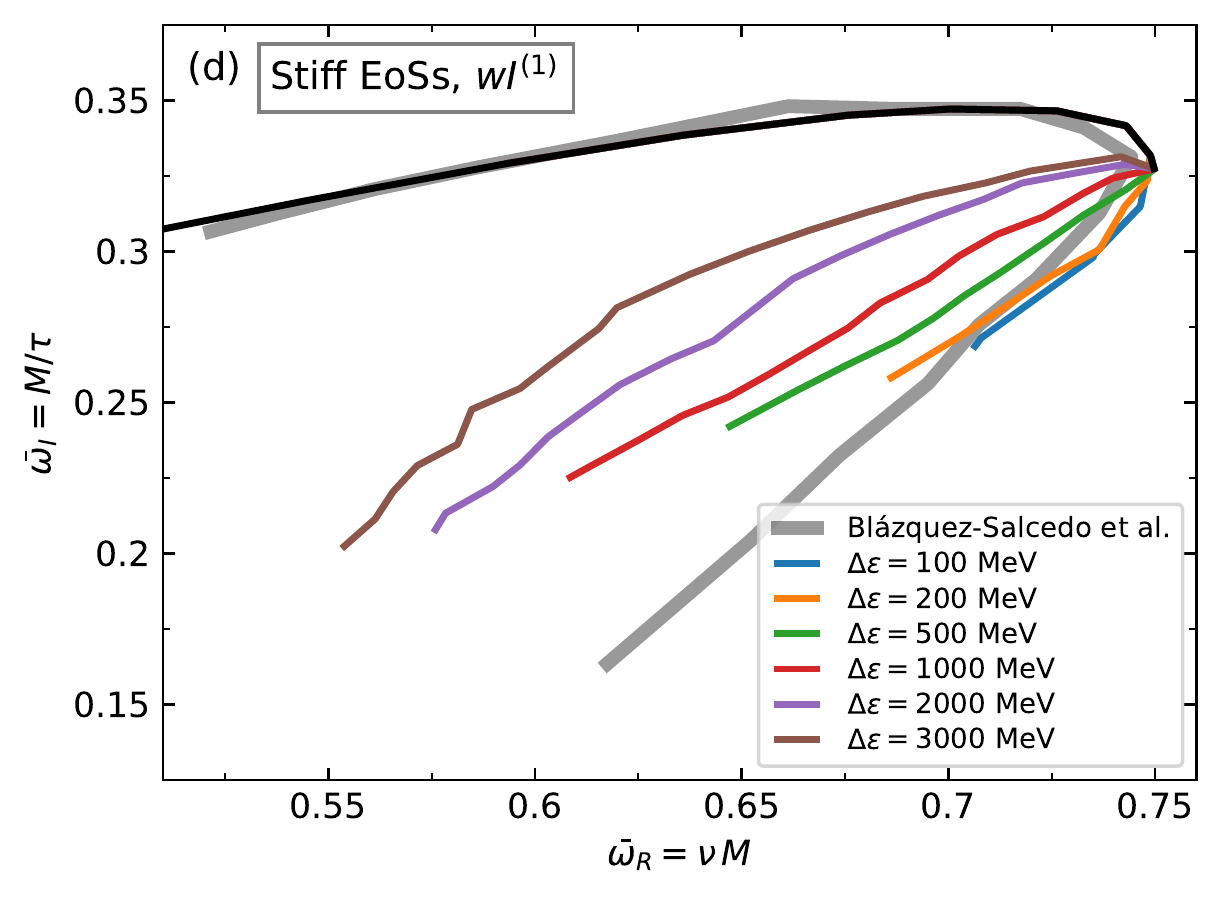}
\caption{In gray, universal relationships for the fundamental mode, $wI^{(0)}$, (upper panels) and first overtone, $wI^{(1)}$, (lower panels) presented in Ref~\cite{w-modes_universal} for the variables $\bar{\omega}_R$ and $\bar{\omega}_I$. With colors, results obtained for compact objects constructed using soft (left panels) and stiff (right panels) hadronic EoSs. The change in the behavior for different values $\Delta \epsilon$ is clear for $\Delta \epsilon \gtrsim 2000$~MeV/fm$^3$. An exception is found in panel~(c), where the curves for the $wI^{(1)}$ mode with a soft hadronic EoS are in reasonable agreement with the proposed fit.
} 
\label{fig:univ1}
\end{figure*}

For these two modes we see that, as the mass increases along the hadronic branch,  $\nu$ decreases and  $\tau$ becomes larger. When extended branches of stable HSs are possible, and mass decreases up to the terminal mass, results depend strongly on the value of $\Delta \epsilon$. For $\Delta \epsilon \lesssim 200$~MeV/fm$^3$, the damping time continues to decrease despite the mass is decreasing, and the frequency (despite some counter examples) always increases as the mass decreases.

The extremely different $M$-$R$ relationship of objects in the extended branch, compared to their hadronic twins, led us to investigate whether known universal relationships for $wI$-modes hold when these objects are considered. Universal relationships for axial (and polar) $w$-modes of NSs, strange quark stars and also HSs, have been proposed in the literature (see, for example Refs.~\cite{AK,chirenti2020-wmode,w-modes_universal,Benhar2004PRD,Tsui2005MNRAS}). In these works, it has been shown that for the fundamental and first overtone of the $wI$-modes, the quantities {$\bar{\omega}_I = M/\tau$ and $\bar{\omega}_R = M \nu$ are simple polynomial functions of the compactness of the stellar object, $M/R$}. In Fig.~\ref{fig:univ3_w0}, we present results for our hybrid stellar configurations and the universal relationship presented in Refs.~\cite{Benhar2004PRD,Tsui2005MNRAS}\footnote{It is worth mentioning that, in Ref.~\cite{Benhar2004PRD}, the authors present a universal relationship for $R \nu$ as a function of the compactness, $M/R$. In our work, we adapt this relationship in order to study it in the $\bar{\omega}_R$-$M/R$ plane.}. In the upper panels, we show results for $\bar{\omega}_I$, panel~(a) for hybrid EoSs constructed with the soft hadronic EoS and panel~(b) with the stiff one. The same is presented for $\bar{\omega}_R$ in the bottom panels, (c) and (d). We find a general agreement between the universal fits and our hadronic configurations. This is not the case for slow stable HSs for which a (clear) departure from the universal relationships can be seen as the parameter $\Delta \epsilon$ grows and the slow-stable branch becomes longer.  In particular, for the hybrid EoSs constructed using the soft hadronic EoS, we find that after the phase transition, the behavior of the curve changes and bends over itself for $\Delta \epsilon \gtrsim 500$~MeV/fm${}^3$. For these cases, as compactness decreases after the peak mass, $\bar{\omega}_R$ is, for a given value of $M/R$, always larger for the slow stable twin than for its hadronic sibling. The results for hybrid EoSs constructed with the stiff hadronic EoS are qualitatively similar, but a clearer deviation can be seen for the low-mass objects of the slow-stable branch, where $\bar{\omega}_R$ shows an increase despite the compactness remains almost constant (see panel (d) of Fig.~\ref{fig:univ3_w0}).

\begin{figure*}[t]
\centering
\includegraphics[width=0.49\linewidth,angle=0]{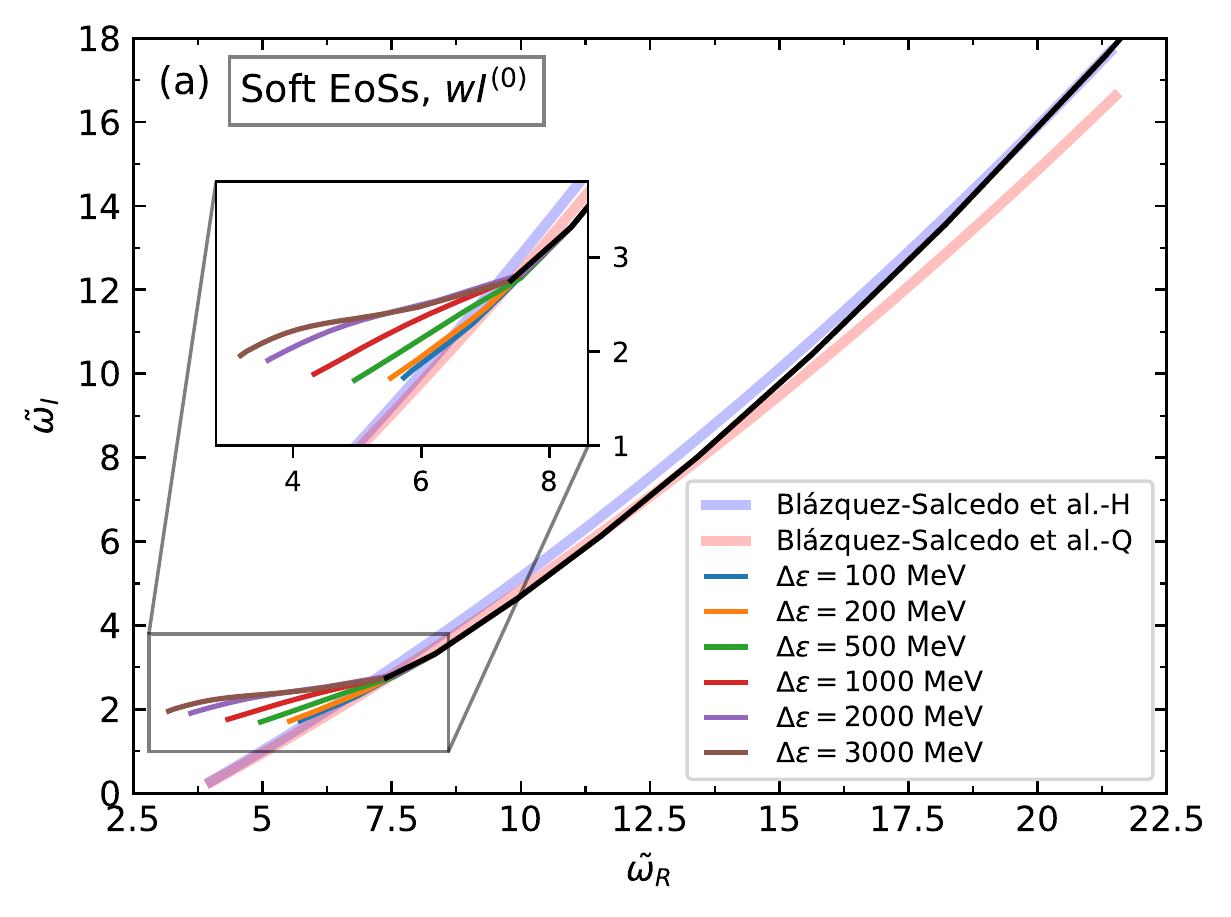}
\includegraphics[width=0.475\linewidth,angle=0]{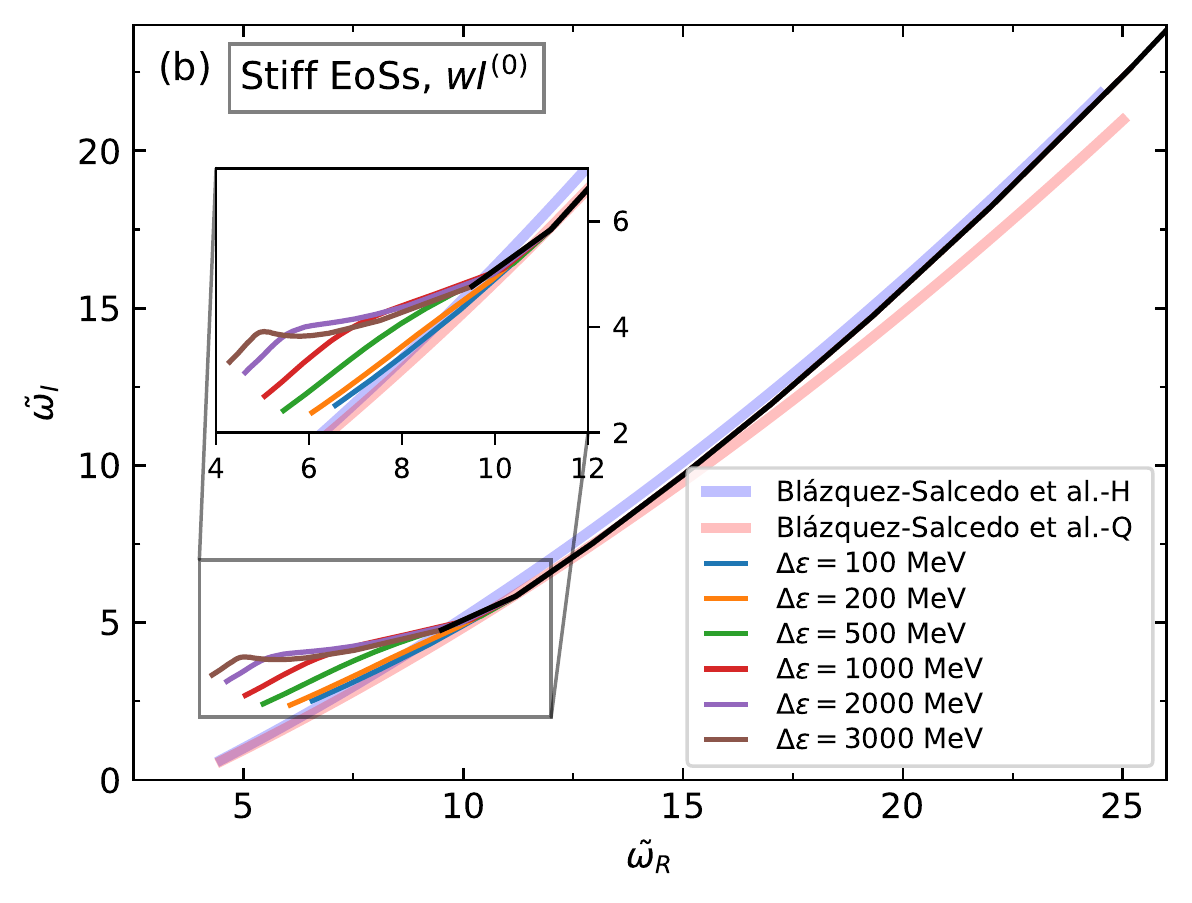}
\includegraphics[width=0.49\linewidth,angle=0]{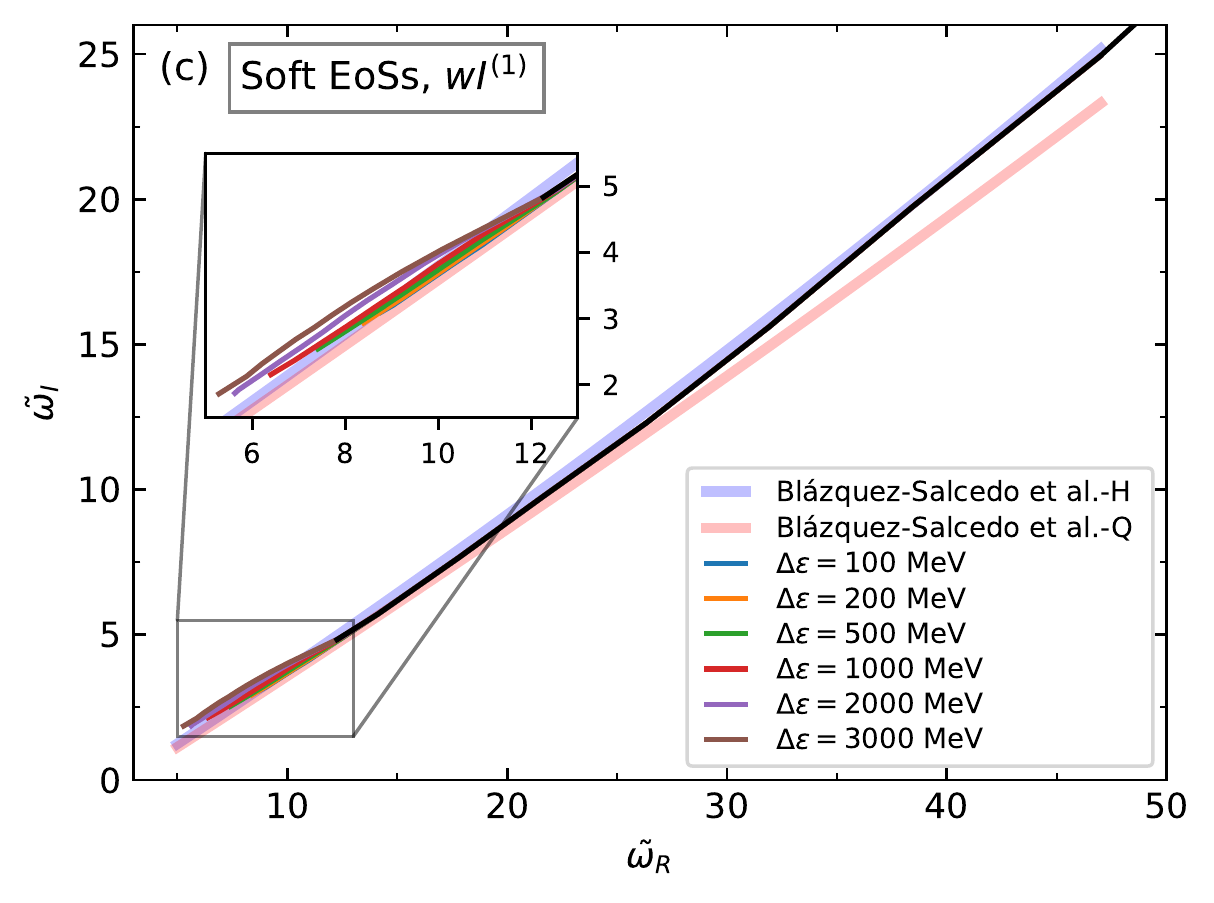}
\includegraphics[width=0.48\linewidth,angle=0]{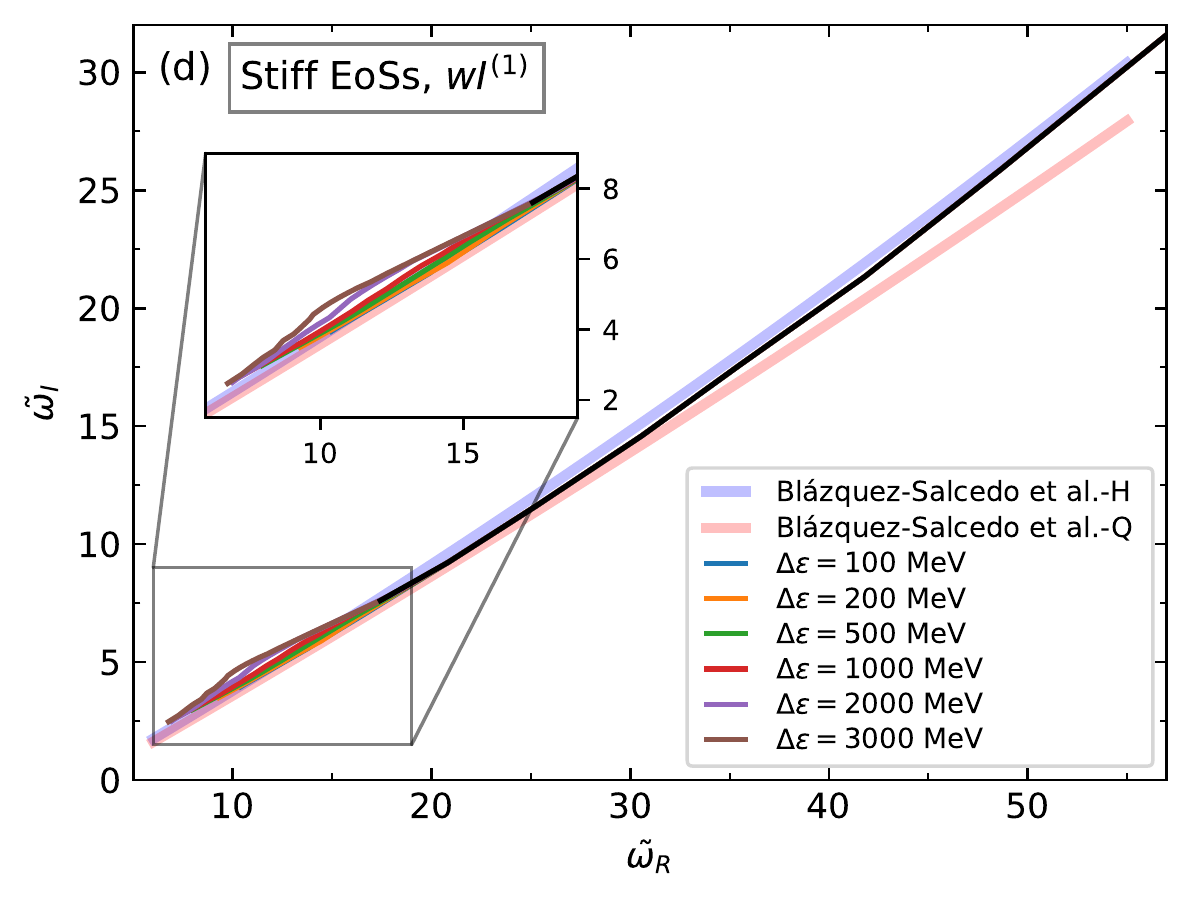}
\caption{In light blue (pink), universal relationship for the fundamental mode, $wI^{(0)}$, and first overtone, $wI^{(1)}$, obtained for hadronic (quark) stellar configurations in \cite{w-modes_universal} for the variables defined in Eq.~\eqref{univ2}. With colors, in the right (left) panel, results obtained for compact objects constructed using soft (stiff) hadronic EoS. A zoom with the region corresponding to the extended branch of stability is presented. For the fundamental mode, deviation from the proposed universal relationships can clearly be seen in this region for values $\Delta \epsilon \gtrsim 200$~MeV/fm$^3$. For the first overtone, the deviation is not significant and results are in general agreement with the proposed fit.} 
\label{fig:univ2}
\end{figure*}

For $\bar{\omega}_I$ we find, despite not that clear, a deviation from universal relationships for $\Delta \epsilon \gtrsim 500$~MeV/fm${}^3$. When a soft EoS is used, $\bar{\omega}_I$ is almost always smaller for objects of the slow-stable branch (the only exception being the case with $\Delta \epsilon = 3000$~MeV/fm${}^3$, where at the region close to the terminal mass the slow-stable branch is placed above). In these cases, as seen in panel~(a) of Fig.~\ref{fig:univ3_w0}, despite presenting clearly different behaviors, all curves lie close to the universal relationship. The situation is similar when the stiff hadronic EoS is used,  with differences at the end of the slow-stable branch, where $\bar{\omega}_I$ increases for an almost constant value of the compactness.

For the first overtone, $wI^{(1)}$, we studied the validity of the EoS-insensitive relationship proposed in Ref.~\cite{Tsui2005MNRAS} for $\bar{\omega}_I = M/\tau$ and $\bar{\omega}_R = M \nu$. As seen in Fig.~\ref{fig:univ3_w1}, where our results are shown together with the fit proposed in Ref.~\cite{Tsui2005MNRAS}, the general conclusion is that the relationships for both the oscillation frequency and the damping times do not hold valid when hadron-quark conversions at the interface are slow and extended stable stellar branches arise. As mentioned for the fundamental mode, deviations from the universal behavior are larger as $\Delta \epsilon$ grows.

Additionally, we analyzed the validity of the universal relationships proposed in Ref.~\cite{w-modes_universal} for $\bar{\omega}_R$ and $\bar{\omega}_I$. In Fig.~\ref{fig:univ1}, we show our results for the fundamental -panels~(a) and (b)- and first overtone -panels~(c) and (d)- of the $wI$ family of QNMs of hybrid compact stars constructed using the soft -panels~(a) and (c)- and stiff -panels~(b) and (d)- hadronic EoSs together with the universal relationships presented in Ref.~\cite{w-modes_universal}. The impact of changing the energy density gap $\Delta \epsilon$ between phases is clear. If we focus on $wI^{(0)}$ we see that, for both hadronic models, configurations in the extended branch clearly deviate from the universal relationship proposed in the literature. This can be seen as an abrupt change in the behavior of the curve that bends over itself. In general, such deviation is more evident for large values of  $\Delta \epsilon$. In particular, for $wI^{(0)}$, if we use the soft hadronic EoS, the deviation is more evident for larger values of $\Delta \epsilon$ that for the stiff case. The other two CSS parameters do not have any qualitative impact on the results.

Results for $wI^{(1)}$ are qualitatively similar to those obtained for the fundamental mode with the exception of the case when a soft hadronic EoS is used, see panel~(c). A clear spreading in the curves (but without any systematic deviation) can be seen for different values of $\Delta \epsilon$, but results can be understood to be in general agreement with the universal relationship presented in the literature. This is not the case for the stiff EoS, panel~(d), where a strong systematic deviation from the universal relationship is seen for  $\Delta \epsilon \gtrsim 200$~MeV/fm$^3$.

Other interesting universal relationships have also been presented in Ref.~\cite{w-modes_universal} for $\tilde{\omega}_R$ and $\tilde{\omega}_I$, which are defined by: 
\begin{equation}
\label{univ2}
    \tilde{\omega}_R = \frac{2\pi  \nu ({\rm{Hz}})}{c\sqrt{P_c({\rm{cm}}^{-2})}}, \qquad \tilde{\omega}_I = \frac{1}{c \sqrt{P_c({\rm{cm}}^{-2} )}} \frac{1}{\tau ({\rm{s}})},
\end{equation}
where $P_c$ is the central pressure of each configuration. Such relationship has been argued to be useful to determine the central pressure of compact objects from a detection of a $wI$-mode. In the upper panels of Fig.~\ref{fig:univ2}, we show our results for $wI^{(0)}$ of hybrid stars constructed using the soft and stiff EoSs. In both cases, a deviation from universal relationships occurs when slow-stable stellar configurations are considered. As seen in the zoom, this deviation is more apparent and qualitatively independent of the hadronic EoS for  $\Delta \epsilon \gtrsim 200$~MeV/fm$^3$. Moreover, we find that the deviation from the universal relationship does not appear to follow any clear EoS insensitive relation. 
Results for the first overtone $wI^{(1)}$ are shown in  the bottom panels of Fig.~\ref{fig:univ2}. In this case, the deviation from the proposed universal relationships is barely visible, see zoomed areas in panels~(c) and (d). As occurs for $wI^{(0)}$, in the case of hybrid EoSs constructed with the soft hadronic EoS, the deviation from the universal relationship increases with $\Delta \epsilon$.

\section{Summary and Discussion} \label{conc}

In this work we studied $wI$-modes of compact stars constructed with hybrid EoSs containing a sharp hadron-quark phase transition. We analyzed the impact of assuming that the hadron-quark conversion speed at the interface is rapid or slow. The later situation is known to alter drastically the dynamic stability of compact objects and gives rise to extended branches of slow stable twin configurations. We have used a stiff and a soft hadronic EoS and a wide range of values of the three parameters of the CSS parametrization for the quark matter EoS. In this way, we have obtained results of great generality. We focused our attention into the fundamental and first overtones for which universal relationships have been proposed \cite{w-modes_universal,Benhar2004PRD,Tsui2005MNRAS}.
Universal relationships for QNMs are important because they would help, due to their EoS-insensitive nature, to obtain physical information related to compact objects from their detection.

\begin{figure*}[t]
\centering
\includegraphics[width=0.49\linewidth,angle=0]{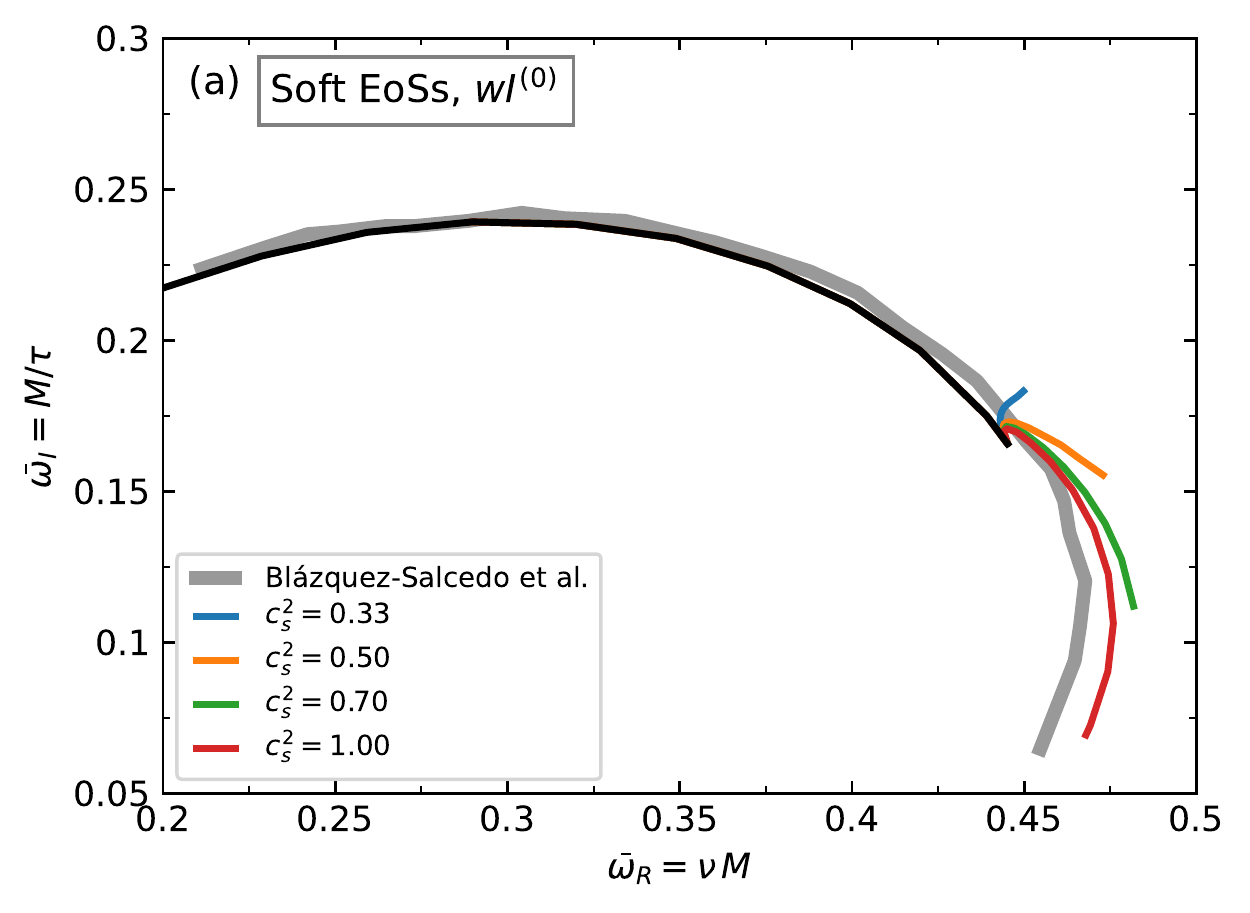}
\includegraphics[width=0.49\linewidth,angle=0]{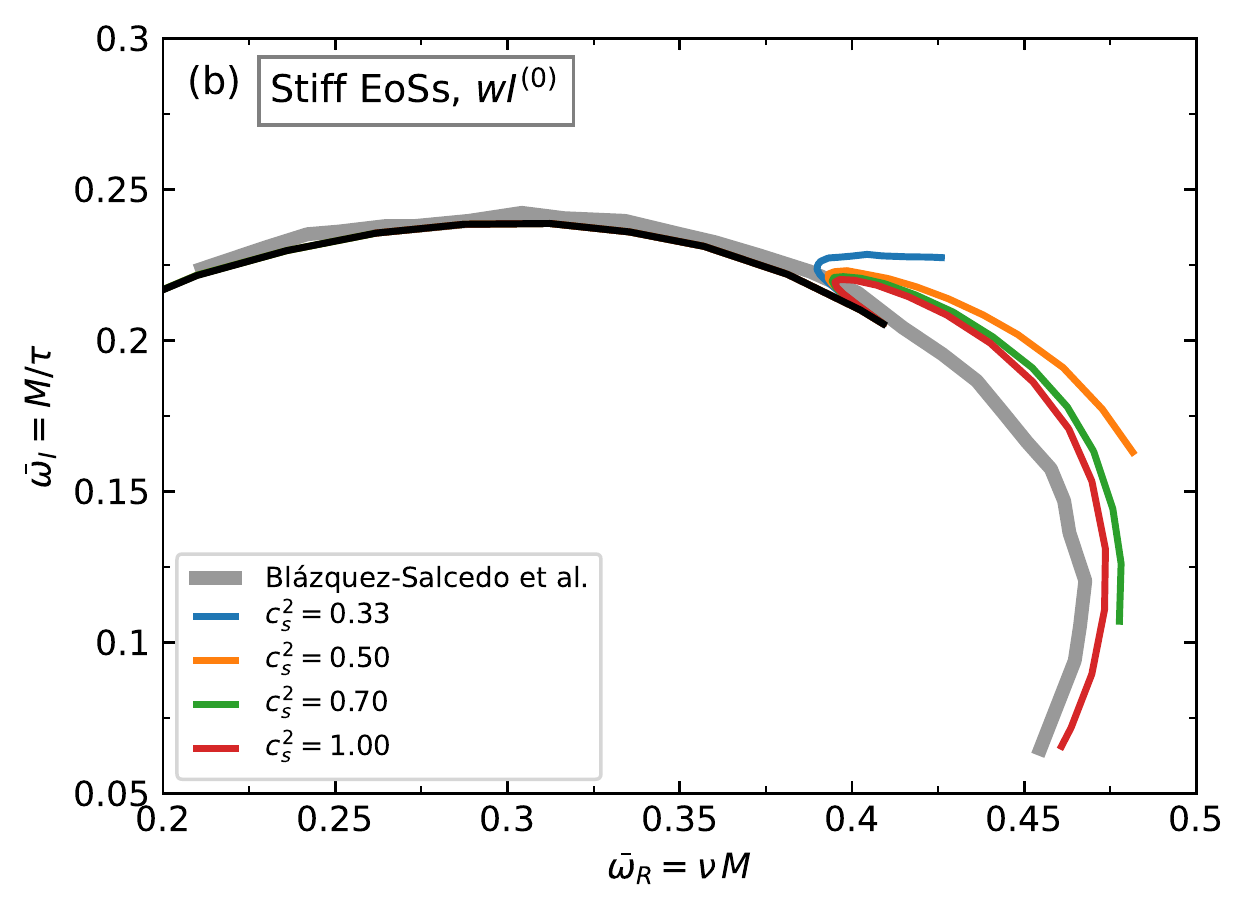}
\includegraphics[width=0.49\linewidth,angle=0]{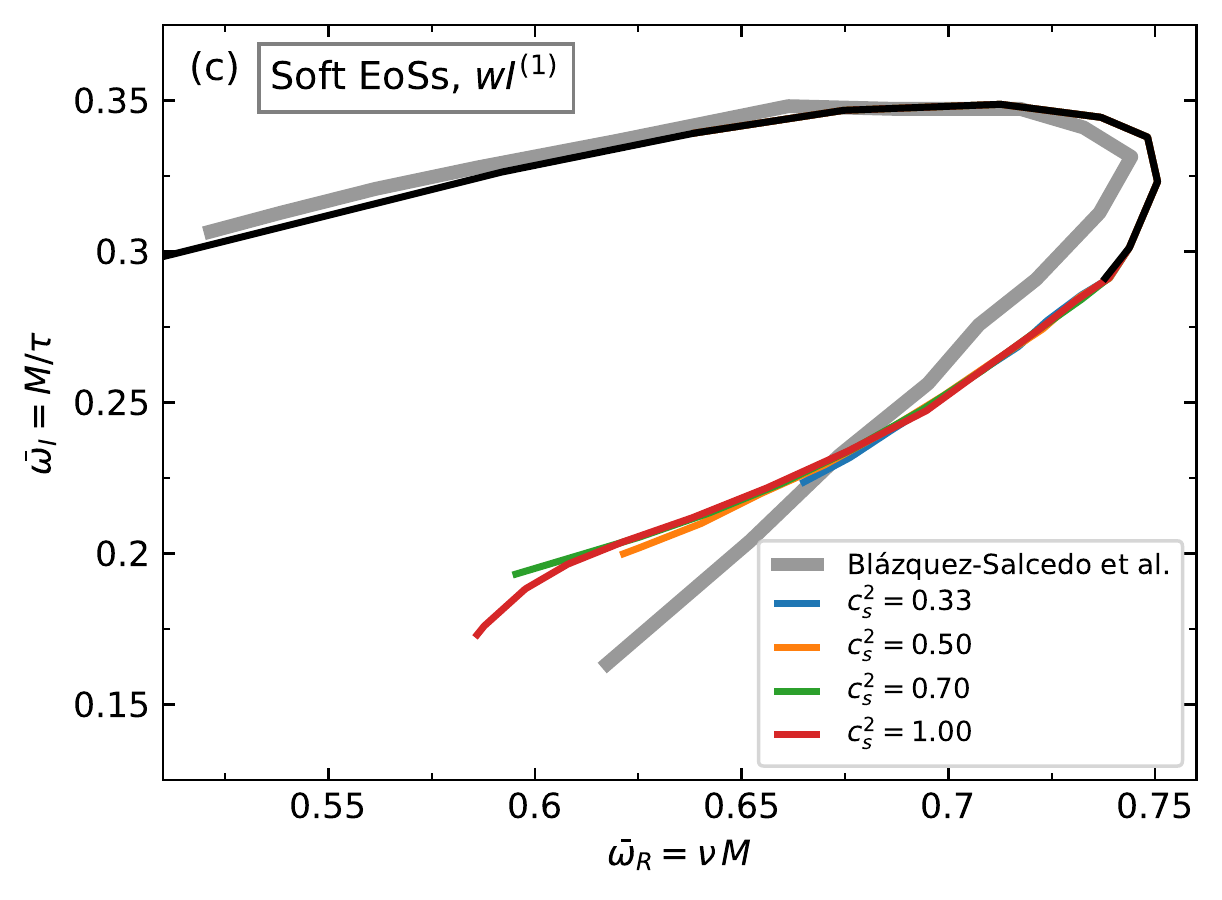}
\includegraphics[width=0.49\linewidth,angle=0]{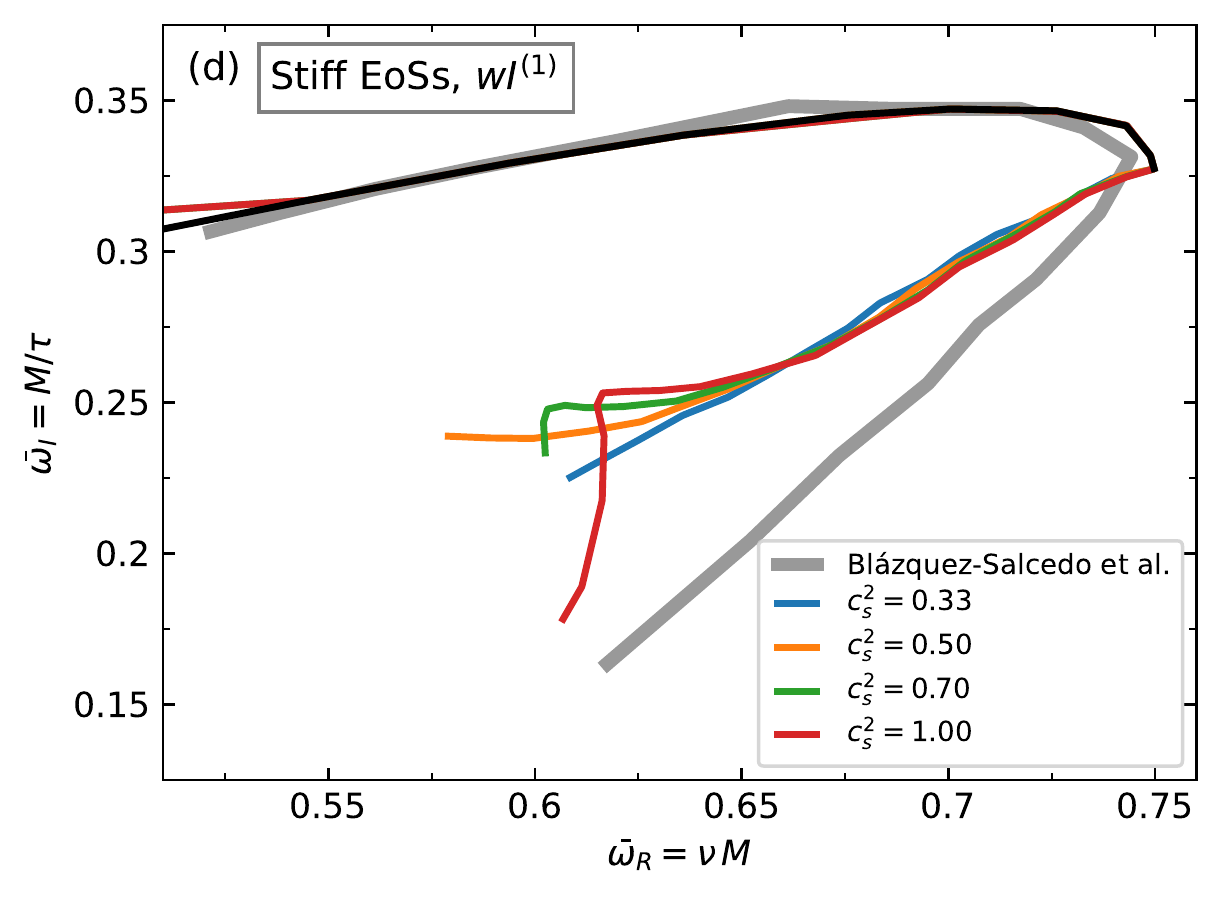}
\caption{In gray, universal relationships for the fundamental mode, $wI^{(0)}$, (upper panels) and first overtone, $wI^{(1)}$, (lower panels) presented in Ref~\cite{w-modes_universal} for the variables $\bar{\omega}_R$ and $\bar{\omega}_I$. With colors, results obtained for compact objects constructed using soft (left panels) and stiff (right panels) hadronic EoSs, and fixed value for $\Delta \epsilon = 1000$~MeV/fm$^3$.} 
\label{fig:univ1cs}
\end{figure*}

We have shown that, for slow stable hybrid objects, already known universal relationships associated with $wI$-modes do not hold in general. In particular, for the fundamental mode, we have studied the behavior of $M \nu$ and $M/\tau$ with the compactness of stellar configurations. We have shown that objects of the extended branch of stellar configurations do not follow, in general, the universal behavior presented in Refs.~\cite{Benhar2004PRD,Tsui2005MNRAS}. Such deviation is more apparent for larger values of the energy density jump $\Delta \epsilon$ at the quark-hadron interface. Moreover, we have found that no universal relationship is valid between $\tilde{w}_R$ and $\tilde{w}_I$. The same occurs for the variables $\bar{w}_R$ and $\bar{w}_I$. For this reason, the applicability of such relationships to determine physical parameters of hybrid stars using measurements of both frequency and damping times must be re-examined. 

We have studied the same for the first $wI$ overtone, $wI^{(1)}$, and obtained similar results when the stiff hadronic EoS is used. Results are not that conclusive when the hadronic part of the hybrid EoS is soft.

This degeneracy might be overcome if fluid modes of the same source can also be detected. Particularly interesting for this quest is to have in mind that the frequency of the $f$-mode for slow-stable twins are significantly larger than the one of its hadronic sibling \cite{lucasygerman,mariani2022MNRAS}. Another distinctive feature of slow-stable HSs is the existence of a $g$-mode associated to the sharp phase transition in the core of the compact object \cite{Rodriguez-etalg2,lucasygerman}. A detection of such mode might be a clear indication of the existence of a sharp hadron-quark phase transition and would also provide some insights about the hadron-quark conversion process. Moreover, since the nature of the phase transition (mixed or sharp \cite{LugGrunf-universe:2021}) is mainly determined by the value of the hadron-quark surface tension, it will provide information about this quantity.

We have put special attention to the impact on the frequencies and damping times of the fundamental and first overtone. The general result is that hybrid stars of the extended branch of stellar configuration break universal relationships for $wI$-modes. The deviation from universal relationships is clear for larger values of the energy density jump, $\Delta \epsilon$. The impact of changes in $c_{\rm s}^2$ shows that deviation from universality is larger for intermediate values. As $c_{\rm s}$ approaches the speed of light, universality seems to be regained for objects near the terminal mass. The reason of this might be that this objects contain a large core of quark matter and so their properties might resemble those of quark stars that are known to fulfill universal relationships (see, Refs. \citep{Tsui2005MNRAS,w-modes_universal} for more details).

Third generation GW detectors are expected to be sensitive enough to detect QNMs of isolated pulsating compact objects with errors not larger that a few tens of Hz \cite{Pratten:2019sed}. Such precision in the measurement of their oscillatory properties would provide key information that might be useful to unravel the mysteries hidden in the inner core of compact objects. In this context, the existence (or not) of objects in the extended branches could, in principle, be falsifiable through GW astronomy.

\begin{figure*}[t]
\centering
\includegraphics[width=0.49\linewidth,angle=0]{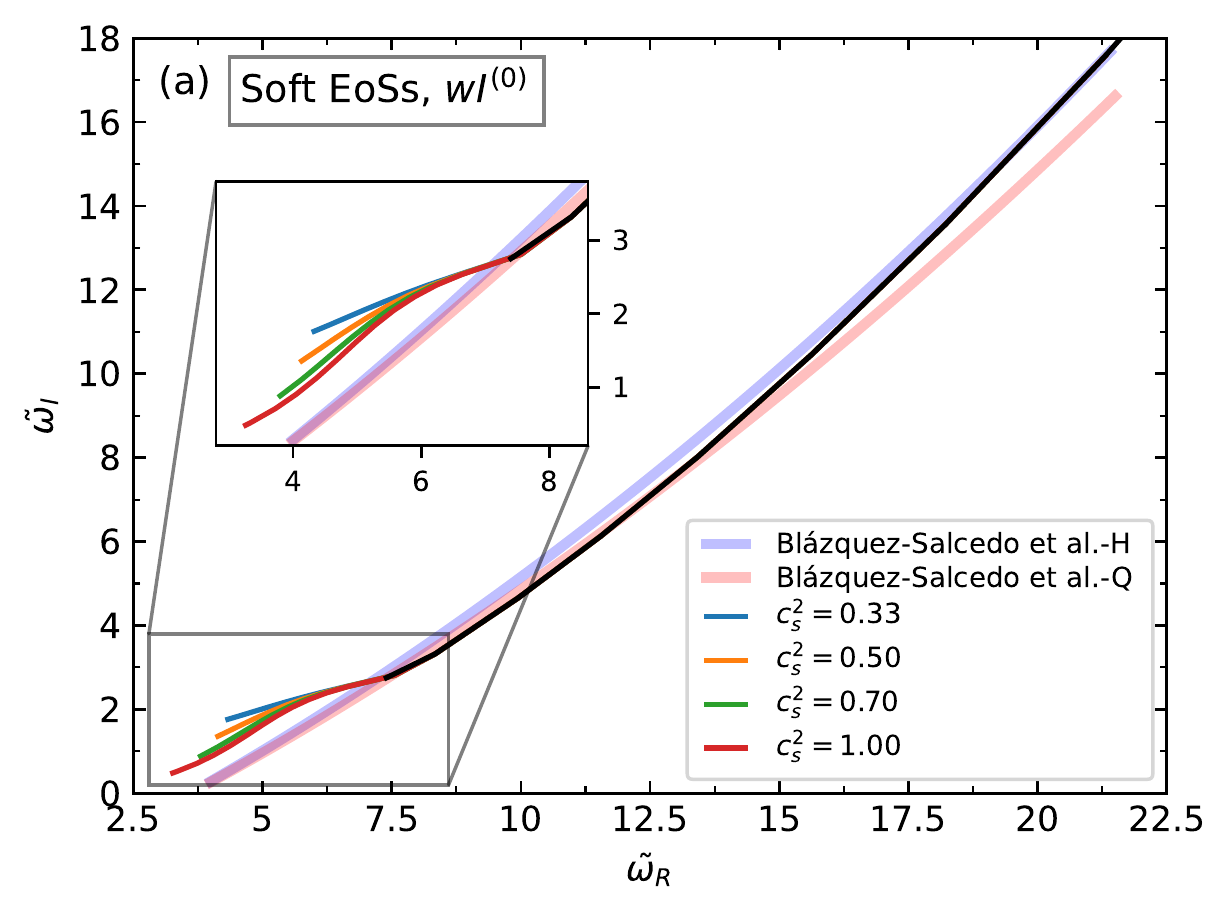}
\includegraphics[width=0.475\linewidth,angle=0]{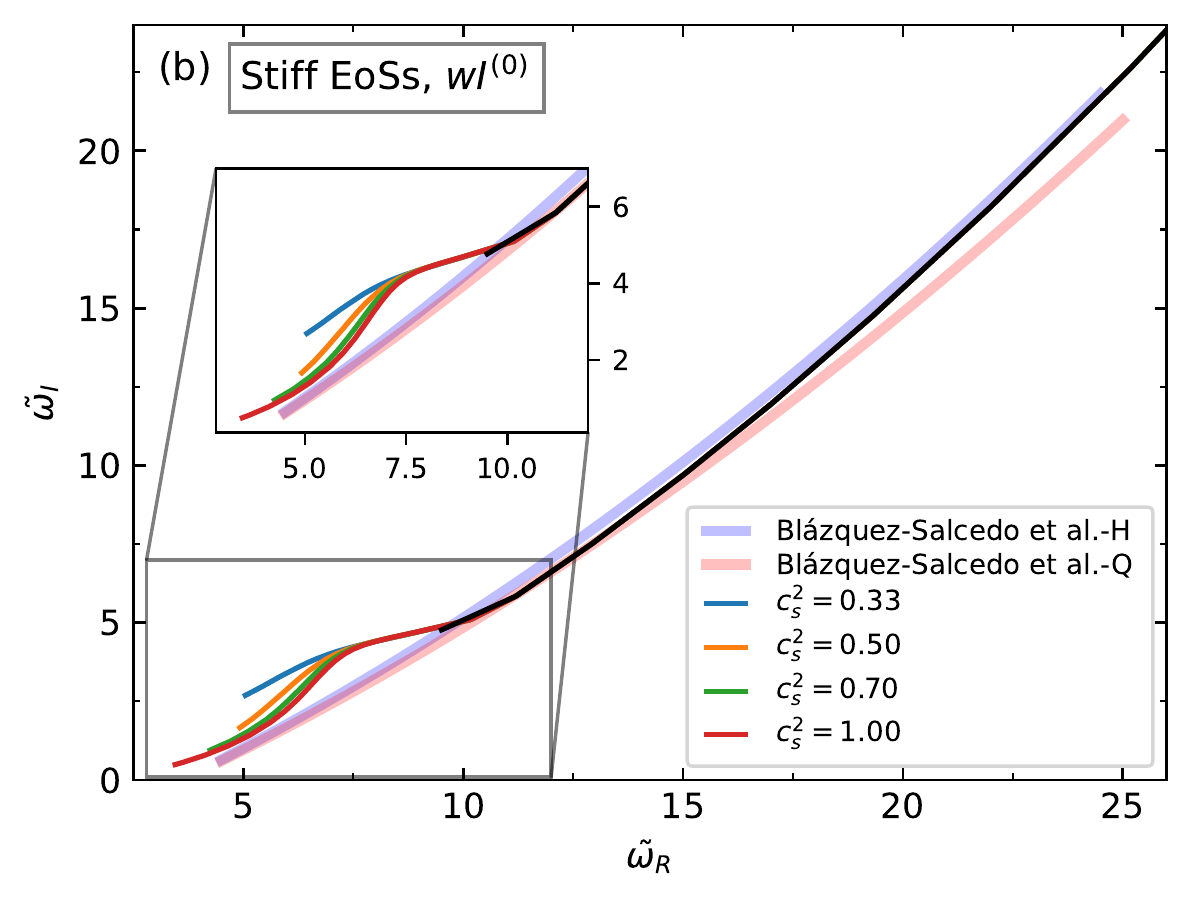}
\includegraphics[width=0.49\linewidth,angle=0]{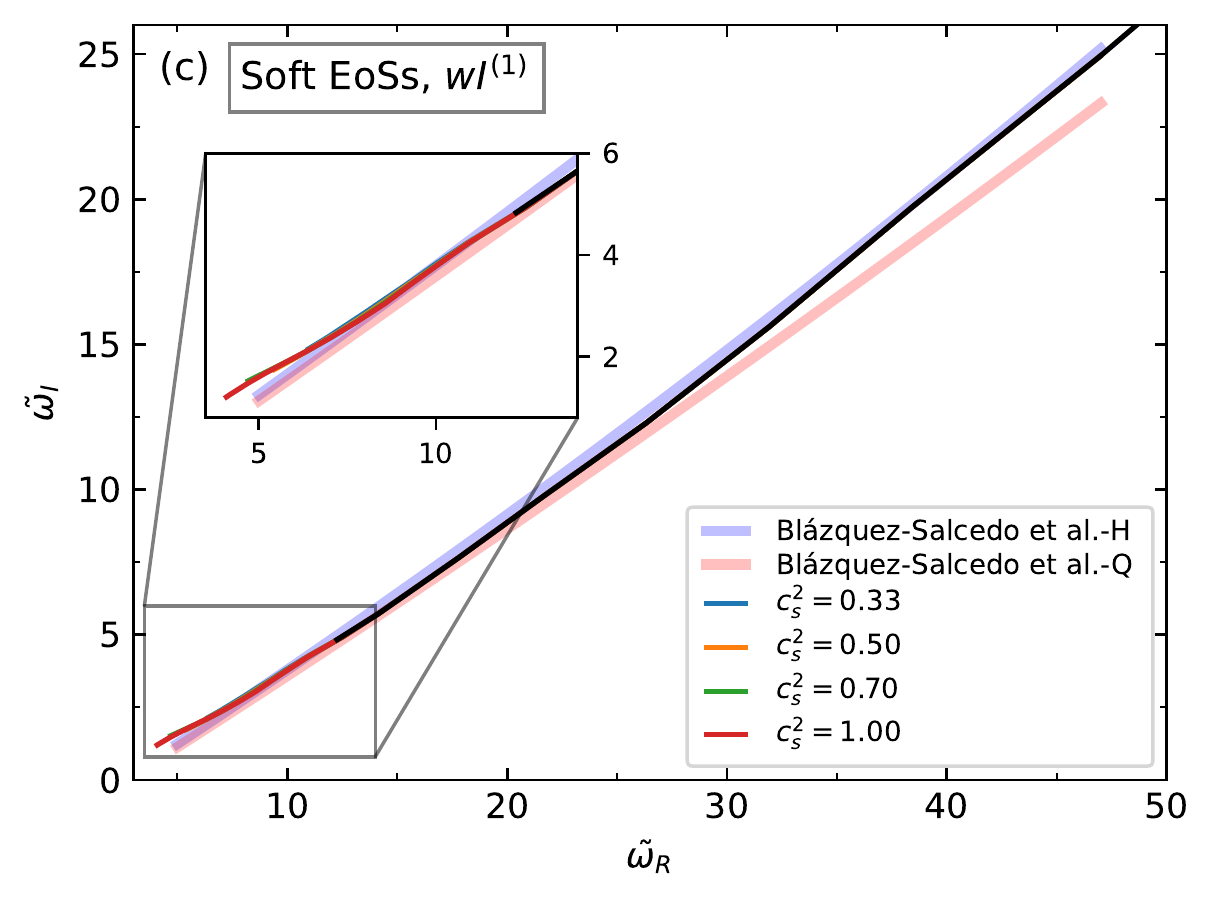}
\includegraphics[width=0.48\linewidth,angle=0]{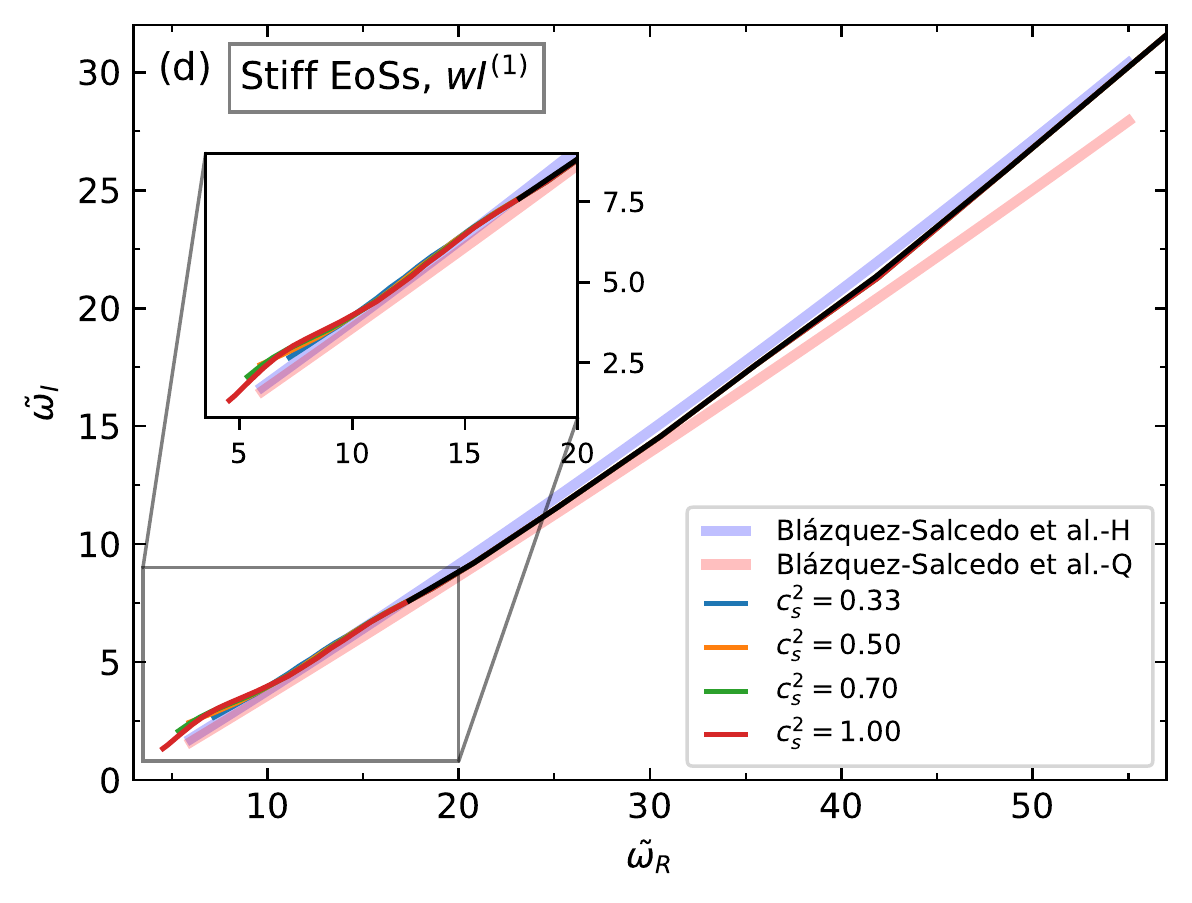}
\caption{In light blue (pink), universal relationship for the fundamental mode, $wI^{(0)}$, and first overtone, $wI^{(1)}$, obtained for hadronic (quark) stellar configurations in Ref.~\cite{w-modes_universal} for the variables defined in Eq.~\eqref{univ2}. With colors, in the right (left) panel, results obtained for compact objects constructed using soft (stiff) hadronic EoS, and fixed value for $\Delta \epsilon = 1000$~MeV/fm$^3$. A zoom with the region corresponding to the extended branch of stability is presented.} 
\label{fig:univ2cs}
\end{figure*}

\section*{Acknowledgments}
The authors are grateful to J. Mena-Fernández and V. Guedes for useful discussions at the early stages of this work. I.F.R.-S.,  O.M.G. and M.M. thank  CONICET  and UNLP for financial support under grants PIP-0714 and G140, G157, G007, X824. I.F.R.-S. is also partially supported by PICT 2019-0366 from ANPCyT (Argentina) and by the National Science Foundation (USA) under Grant PHY-2012152. M.M is a postdoctoral fellow of CONICET. G. L. acknowledges the support of the Brazilian agencies CNPq (grant 316844/2021-7) and FAPESP (grants 2013/10559-5 and 2022/02341-9). 

\appendix

\section{Dependence of the results with $c_{\rm s}^2$} \label{app1}

In this Appendix we explore the dependence of the $wI^{(0)}$ and $wI^{(1)}$ QNMs with the speed of sound of quark matter.

We plot the results that show the general behavior of the frequencies of the fundamental and first overtone $wI$-modes. In particular, we fix the value of the  energy density jump, $\Delta \epsilon = 1000$MeV/fm${}^3$, and study the cases $c_{\rm s}^2 = 0.33$, $0.50$, $0.70$ and $1.00$. Comparisons with the following universal relationships existing in the literature are presented: for $\bar{\omega_R}$ as a function of $\bar{\omega_I}$ in Fig. \ref{fig:univ1cs} and for $\tilde{\omega_R}$ as a function of $\tilde{\omega_I}$ in Fig. \ref{fig:univ2cs}.

In the top panels of Fig. \ref{fig:univ1cs}, we present our results for the fundamental mode. In the left (right) panel results using soft (stiff) hadronic EoS. We can see that deviation from the universal relationships is larger for intermediate values of $c_{\rm s}^2$. As it is increased up to $c_{\rm s}^2=1.00$, results approach (gradually) to the universal behavior (this is especially visible for the stiff case in which the extended branch is the longest). In the bottom panels, we show the same for the first overtone. We find that results are almost insensitive to changes in $c_{\rm s}^2$ for the soft hadronic case (left panel) and that this situation changes for the stiff hadronic EoS (right panel) where deviation increases as $c_s$ approaches the speed of light.

The results shown in Fig. \ref{fig:univ2cs} show that for the fundamental mode (upper panels) deviation is larger for the smaller values of $c_{\rm s}^2$. On the contrary, results are almost insensitive in the case of the first overtone independently of the hadronic EoS used (lower panels).

\renewcommand{\href}[2]{#2}

\newcommand{\apjl}{Astrophys. J. Lett.\ }
\newcommand{\mnras}{Mon. Not. R. Astron. Soc.\ }
\newcommand{\aap}{Astron. Astrophys.\ }
\newcommand{\jcap}{Journal of Cosmology and Astroparticle Physics\ }

\bibliography{ifrs-bib}

\begin{thebibliography}{97}%
\makeatletter
\providecommand \@ifxundefined [1]{%
 \@ifx{#1\undefined}
}%
\providecommand \@ifnum [1]{%
 \ifnum #1\expandafter \@firstoftwo
 \else \expandafter \@secondoftwo
 \fi
}%
\providecommand \@ifx [1]{%
 \ifx #1\expandafter \@firstoftwo
 \else \expandafter \@secondoftwo
 \fi
}%
\providecommand \natexlab [1]{#1}%
\providecommand \enquote  [1]{``#1''}%
\providecommand \bibnamefont  [1]{#1}%
\providecommand \bibfnamefont [1]{#1}%
\providecommand \citenamefont [1]{#1}%
\providecommand \href@noop [0]{\@secondoftwo}%
\providecommand \href [0]{\begingroup \@sanitize@url \@href}%
\providecommand \@href[1]{\@@startlink{#1}\@@href}%
\providecommand \@@href[1]{\endgroup#1\@@endlink}%
\providecommand \@sanitize@url [0]{\catcode `\\12\catcode `\$12\catcode
  `\&12\catcode `\#12\catcode `\^12\catcode `\_12\catcode `\%12\relax}%
\providecommand \@@startlink[1]{}%
\providecommand \@@endlink[0]{}%
\providecommand \url  [0]{\begingroup\@sanitize@url \@url }%
\providecommand \@url [1]{\endgroup\@href {#1}{\urlprefix }}%
\providecommand \urlprefix  [0]{URL }%
\providecommand \Eprint [0]{\href }%
\providecommand \doibase [0]{https://doi.org/}%
\providecommand \selectlanguage [0]{\@gobble}%
\providecommand \bibinfo  [0]{\@secondoftwo}%
\providecommand \bibfield  [0]{\@secondoftwo}%
\providecommand \translation [1]{[#1]}%
\providecommand \BibitemOpen [0]{}%
\providecommand \bibitemStop [0]{}%
\providecommand \bibitemNoStop [0]{.\EOS\space}%
\providecommand \EOS [0]{\spacefactor3000\relax}%
\providecommand \BibitemShut  [1]{\csname bibitem#1\endcsname}%
\let\auto@bib@innerbib\@empty
\bibitem [{\citenamefont {Demorest}\ \emph {et~al.}(2010)\citenamefont
  {Demorest}, \citenamefont {Pennucci}, \citenamefont {Ransom}, \citenamefont
  {Roberts},\ and\ \citenamefont {Hessels}}]{Demorest2010}%
  \BibitemOpen
  \bibfield  {author} {\bibinfo {author} {\bibfnamefont {P.}~\bibnamefont
  {Demorest}}, \bibinfo {author} {\bibfnamefont {T.}~\bibnamefont {Pennucci}},
  \bibinfo {author} {\bibfnamefont {S.}~\bibnamefont {Ransom}}, \bibinfo
  {author} {\bibfnamefont {M.}~\bibnamefont {Roberts}},\ and\ \bibinfo {author}
  {\bibfnamefont {J.}~\bibnamefont {Hessels}},\ }\bibfield  {title} {\bibinfo
  {title} {{Shapiro Delay Measurement of A Two Solar Mass Neutron Star}},\
  }\href {https://doi.org/10.1038/nature09466} {\bibfield  {journal} {\bibinfo
  {journal} {Nature}\ }\textbf {\bibinfo {volume} {467}},\ \bibinfo {pages}
  {1081} (\bibinfo {year} {2010})},\ \Eprint {https://arxiv.org/abs/1010.5788}
  {arXiv:1010.5788 [astro-ph.HE]} \BibitemShut {NoStop}%
\bibitem [{\citenamefont {Antoniadis}\ \emph {et~al.}(2013)\citenamefont
  {Antoniadis} \emph {et~al.}}]{Antoniadis2013}%
  \BibitemOpen
  \bibfield  {author} {\bibinfo {author} {\bibfnamefont {J.}~\bibnamefont
  {Antoniadis}} \emph {et~al.},\ }\bibfield  {title} {\bibinfo {title} {{A
  Massive Pulsar in a Compact Relativistic Binary}},\ }\href
  {https://doi.org/10.1126/science.1233232} {\bibfield  {journal} {\bibinfo
  {journal} {Science}\ }\textbf {\bibinfo {volume} {340}},\ \bibinfo {pages}
  {6131} (\bibinfo {year} {2013})},\ \Eprint {https://arxiv.org/abs/1304.6875}
  {arXiv:1304.6875 [astro-ph.HE]} \BibitemShut {NoStop}%
\bibitem [{\citenamefont {{Cromartie}}\ and\ \citenamefont
  {et~al.}(2020)}]{Cromartie2019}%
  \BibitemOpen
  \bibfield  {author} {\bibinfo {author} {\bibfnamefont {H.~T.}\ \bibnamefont
  {{Cromartie}}}\ and\ \bibinfo {author} {\bibnamefont {et~al.}},\ }\bibfield
  {title} {\bibinfo {title} {{Relativistic Shapiro delay measurements of an
  extremely massive millisecond pulsar}},\ }\href
  {https://doi.org/10.1038/s41550-019-0880-2} {\bibfield  {journal} {\bibinfo
  {journal} {Nature Astronomy}\ }\textbf {\bibinfo {volume} {4}},\ \bibinfo
  {pages} {72} (\bibinfo {year} {2020})},\ \Eprint
  {https://arxiv.org/abs/1904.06759} {arXiv:1904.06759 [astro-ph.HE]}
  \BibitemShut {NoStop}%
\bibitem [{\citenamefont {Lattimer}\ and\ \citenamefont
  {Prakash}(2011)}]{lattimer2msol}%
  \BibitemOpen
  \bibfield  {author} {\bibinfo {author} {\bibfnamefont {J.~M.}\ \bibnamefont
  {Lattimer}}\ and\ \bibinfo {author} {\bibfnamefont {M.}~\bibnamefont
  {Prakash}},\ }\bibfield  {title} {\bibinfo {title} {What a two solar mass
  neutron star really means},\ }in\ \href@noop {} {\emph {\bibinfo {booktitle}
  {From Nuclei to Stars: Festschrift in Honor of Gerald E Brown}}}\ (\bibinfo
  {publisher} {World Scientific},\ \bibinfo {year} {2011})\ pp.\ \bibinfo
  {pages} {275--304}\BibitemShut {NoStop}%
\bibitem [{\citenamefont {{Abbott}}\ and\ \citenamefont
  {et~al.}(2017)}]{GW170817-detection}%
  \BibitemOpen
  \bibfield  {author} {\bibinfo {author} {\bibfnamefont {B.~P.}\ \bibnamefont
  {{Abbott}}}\ and\ \bibinfo {author} {\bibnamefont {et~al.}},\ }\bibfield
  {title} {\bibinfo {title} {{GW170817: Observation of Gravitational Waves from
  a Binary Neutron Star Inspiral}},\ }\href
  {https://doi.org/10.1103/PhysRevLett.119.161101} {\bibfield  {journal}
  {\bibinfo  {journal} {Physical Review Letters}\ }\textbf {\bibinfo {volume}
  {119}},\ \bibinfo {eid} {161101} (\bibinfo {year} {2017})},\ \Eprint
  {https://arxiv.org/abs/1710.05832} {arXiv:1710.05832 [gr-qc]} \BibitemShut
  {NoStop}%
\bibitem [{\citenamefont {{Abbott}}\ and\ \citenamefont {et.
  al}(2017)}]{GW170817-em}%
  \BibitemOpen
  \bibfield  {author} {\bibinfo {author} {\bibfnamefont {B.~P.}\ \bibnamefont
  {{Abbott}}}\ and\ \bibinfo {author} {\bibnamefont {et. al}},\ }\bibfield
  {title} {\bibinfo {title} {{Multi-messenger Observations of a Binary Neutron
  Star Merger}},\ }\href {https://doi.org/10.3847/2041-8213/aa91c9} {\bibfield
  {journal} {\bibinfo  {journal} {\apjl}\ }\textbf {\bibinfo {volume} {848}},\
  \bibinfo {eid} {L12} (\bibinfo {year} {2017})},\ \Eprint
  {https://arxiv.org/abs/1710.05833} {arXiv:1710.05833 [astro-ph.HE]}
  \BibitemShut {NoStop}%
\bibitem [{\citenamefont {Raithel}\ \emph {et~al.}(2018)\citenamefont
  {Raithel}, \citenamefont {Özel},\ and\ \citenamefont
  {Psaltis}}]{Raithel2018}%
  \BibitemOpen
  \bibfield  {author} {\bibinfo {author} {\bibfnamefont {C.}~\bibnamefont
  {Raithel}}, \bibinfo {author} {\bibfnamefont {F.}~\bibnamefont {Özel}},\
  and\ \bibinfo {author} {\bibfnamefont {D.}~\bibnamefont {Psaltis}},\
  }\bibfield  {title} {\bibinfo {title} {{Tidal deformability from GW170817 as
  a direct probe of the neutron star radius}},\ }\href
  {https://doi.org/10.3847/2041-8213/aabcbf} {\bibfield  {journal} {\bibinfo
  {journal} {Astrophys. J. Lett.}\ }\textbf {\bibinfo {volume} {857}},\
  \bibinfo {pages} {L23} (\bibinfo {year} {2018})},\ \Eprint
  {https://arxiv.org/abs/1803.07687} {arXiv:1803.07687 [astro-ph.HE]}
  \BibitemShut {NoStop}%
\bibitem [{\citenamefont {Abbott}\ \emph {et~al.}(2018)\citenamefont {Abbott}
  \emph {et~al.}}]{Abbott:2018}%
  \BibitemOpen
  \bibfield  {author} {\bibinfo {author} {\bibfnamefont {B.~P.}\ \bibnamefont
  {Abbott}} \emph {et~al.} (\bibinfo {collaboration} {LIGO Scientific,
  Virgo}),\ }\bibfield  {title} {\bibinfo {title} {{GW170817: Measurements of
  neutron star radii and equation of state}},\ }\href
  {https://doi.org/10.1103/PhysRevLett.121.161101} {\bibfield  {journal}
  {\bibinfo  {journal} {Phys. Rev. Lett.}\ }\textbf {\bibinfo {volume} {121}},\
  \bibinfo {pages} {161101} (\bibinfo {year} {2018})},\ \Eprint
  {https://arxiv.org/abs/1805.11581} {arXiv:1805.11581 [gr-qc]} \BibitemShut
  {NoStop}%
\bibitem [{\citenamefont {{Abbott}}\ and\ \citenamefont
  {et~al.}(2020)}]{GW190425-detection}%
  \BibitemOpen
  \bibfield  {author} {\bibinfo {author} {\bibfnamefont {B.~P.}\ \bibnamefont
  {{Abbott}}}\ and\ \bibinfo {author} {\bibnamefont {et~al.}},\ }\bibfield
  {title} {\bibinfo {title} {{GW190425: Observation of a Compact Binary
  Coalescence with Total Mass {\ensuremath{\sim}} 3.4 M$_{\odot}$}},\ }\href
  {https://doi.org/10.3847/2041-8213/ab75f5} {\bibfield  {journal} {\bibinfo
  {journal} {Astrophys. J. Lett.}\ }\textbf {\bibinfo {volume} {892}},\
  \bibinfo {eid} {L3} (\bibinfo {year} {2020})},\ \Eprint
  {https://arxiv.org/abs/2001.01761} {arXiv:2001.01761 [astro-ph.HE]}
  \BibitemShut {NoStop}%
\bibitem [{\citenamefont {Han}\ \emph {et~al.}(2020)\citenamefont {Han},
  \citenamefont {Tang}, \citenamefont {Hu}, \citenamefont {Li}, \citenamefont
  {Jiang}, \citenamefont {Jin}, \citenamefont {Fan},\ and\ \citenamefont
  {Wei}}]{Han:2020qmn}%
  \BibitemOpen
  \bibfield  {author} {\bibinfo {author} {\bibfnamefont {M.-Z.}\ \bibnamefont
  {Han}}, \bibinfo {author} {\bibfnamefont {S.-P.}\ \bibnamefont {Tang}},
  \bibinfo {author} {\bibfnamefont {Y.-M.}\ \bibnamefont {Hu}}, \bibinfo
  {author} {\bibfnamefont {Y.-J.}\ \bibnamefont {Li}}, \bibinfo {author}
  {\bibfnamefont {J.-L.}\ \bibnamefont {Jiang}}, \bibinfo {author}
  {\bibfnamefont {Z.-P.}\ \bibnamefont {Jin}}, \bibinfo {author} {\bibfnamefont
  {Y.-Z.}\ \bibnamefont {Fan}},\ and\ \bibinfo {author} {\bibfnamefont {D.-M.}\
  \bibnamefont {Wei}},\ }\bibfield  {title} {\bibinfo {title} {{Is GW190425
  consistent with being a neutron star$-$black hole merger?}},\ }\href
  {https://doi.org/10.3847/2041-8213/ab745a} {\bibfield  {journal} {\bibinfo
  {journal} {Astrophys. J. Lett.}\ }\textbf {\bibinfo {volume} {891}},\
  \bibinfo {pages} {L5} (\bibinfo {year} {2020})},\ \Eprint
  {https://arxiv.org/abs/2001.07882} {arXiv:2001.07882 [astro-ph.HE]}
  \BibitemShut {NoStop}%
\bibitem [{\citenamefont {Kyutoku}\ \emph {et~al.}(2020)\citenamefont
  {Kyutoku}, \citenamefont {Fujibayashi}, \citenamefont {Hayashi},
  \citenamefont {Kawaguchi}, \citenamefont {Kiuchi}, \citenamefont {Shibata},\
  and\ \citenamefont {Tanaka}}]{Kyutoku:2020xka}%
  \BibitemOpen
  \bibfield  {author} {\bibinfo {author} {\bibfnamefont {K.}~\bibnamefont
  {Kyutoku}}, \bibinfo {author} {\bibfnamefont {S.}~\bibnamefont
  {Fujibayashi}}, \bibinfo {author} {\bibfnamefont {K.}~\bibnamefont
  {Hayashi}}, \bibinfo {author} {\bibfnamefont {K.}~\bibnamefont {Kawaguchi}},
  \bibinfo {author} {\bibfnamefont {K.}~\bibnamefont {Kiuchi}}, \bibinfo
  {author} {\bibfnamefont {M.}~\bibnamefont {Shibata}},\ and\ \bibinfo {author}
  {\bibfnamefont {M.}~\bibnamefont {Tanaka}},\ }\bibfield  {title} {\bibinfo
  {title} {{On the Possibility of GW190425 Being a Black
  Hole\textendash{}Neutron Star Binary Merger}},\ }\href
  {https://doi.org/10.3847/2041-8213/ab6e70} {\bibfield  {journal} {\bibinfo
  {journal} {Astrophys. J. Lett.}\ }\textbf {\bibinfo {volume} {890}},\
  \bibinfo {pages} {L4} (\bibinfo {year} {2020})},\ \Eprint
  {https://arxiv.org/abs/2001.04474} {arXiv:2001.04474 [astro-ph.HE]}
  \BibitemShut {NoStop}%
\bibitem [{\citenamefont {Riley}\ and\ \citenamefont
  {et~al.}(2019)}]{Riley2019}%
  \BibitemOpen
  \bibfield  {author} {\bibinfo {author} {\bibfnamefont {T.~E.}\ \bibnamefont
  {Riley}}\ and\ \bibinfo {author} {\bibnamefont {et~al.}},\ }\bibfield
  {title} {\bibinfo {title} {A {NICER} view of {PSR} j0030+0451: Millisecond
  pulsar parameter estimation},\ }\href
  {https://doi.org/10.3847/2041-8213/ab481c} {\bibfield  {journal} {\bibinfo
  {journal} {Astrophys. J. Lett.}\ }\textbf {\bibinfo {volume} {887}},\
  \bibinfo {pages} {L21} (\bibinfo {year} {2019})}\BibitemShut {NoStop}%
\bibitem [{\citenamefont {Miller}\ and\ \citenamefont
  {et~al.}(2019)}]{Miller2019}%
  \BibitemOpen
  \bibfield  {author} {\bibinfo {author} {\bibfnamefont {M.~C.}\ \bibnamefont
  {Miller}}\ and\ \bibinfo {author} {\bibnamefont {et~al.}},\ }\bibfield
  {title} {\bibinfo {title} {{PSR} j0030+0451 mass and radius from {NICER} data
  and implications for the properties of neutron star matter},\ }\href
  {https://doi.org/10.3847/2041-8213/ab50c5} {\bibfield  {journal} {\bibinfo
  {journal} {Astrophys. J. Lett.}\ }\textbf {\bibinfo {volume} {887}},\
  \bibinfo {pages} {L24} (\bibinfo {year} {2019})}\BibitemShut {NoStop}%
\bibitem [{\citenamefont {Bilous}\ and\ \citenamefont
  {et~al.}(2019)}]{Bilous:2019}%
  \BibitemOpen
  \bibfield  {author} {\bibinfo {author} {\bibfnamefont {A.~V.}\ \bibnamefont
  {Bilous}}\ and\ \bibinfo {author} {\bibnamefont {et~al.}},\ }\bibfield
  {title} {\bibinfo {title} {A {NICER} view of {PSR} j0030+0451: Evidence for a
  global-scale multipolar magnetic field},\ }\href
  {https://doi.org/10.3847/2041-8213/ab53e7} {\bibfield  {journal} {\bibinfo
  {journal} {Astrophys. J. Lett.}\ }\textbf {\bibinfo {volume} {887}},\
  \bibinfo {pages} {L23} (\bibinfo {year} {2019})}\BibitemShut {NoStop}%
\bibitem [{\citenamefont {{Riley}}\ and\ \citenamefont
  {et~al.}(2021)}]{riley2021ApJ-j0740}%
  \BibitemOpen
  \bibfield  {author} {\bibinfo {author} {\bibfnamefont {T.~E.}\ \bibnamefont
  {{Riley}}}\ and\ \bibinfo {author} {\bibnamefont {et~al.}},\ }\bibfield
  {title} {\bibinfo {title} {{A NICER View of the Massive Pulsar PSR J0740+6620
  Informed by Radio Timing and XMM-Newton Spectroscopy}},\ }\href
  {https://doi.org/10.3847/2041-8213/ac0a81} {\bibfield  {journal} {\bibinfo
  {journal} {\apjl}\ }\textbf {\bibinfo {volume} {918}},\ \bibinfo {eid} {L27}
  (\bibinfo {year} {2021})},\ \Eprint {https://arxiv.org/abs/2105.06980}
  {arXiv:2105.06980 [astro-ph.HE]} \BibitemShut {NoStop}%
\bibitem [{\citenamefont {{Miller}}\ and\ \citenamefont
  {et~al.}(2021)}]{miller2021ApJ-j0740}%
  \BibitemOpen
  \bibfield  {author} {\bibinfo {author} {\bibfnamefont {M.~C.}\ \bibnamefont
  {{Miller}}}\ and\ \bibinfo {author} {\bibnamefont {et~al.}},\ }\bibfield
  {title} {\bibinfo {title} {{The Radius of PSR J0740+6620 from NICER and
  XMM-Newton Data}},\ }\href {https://doi.org/10.3847/2041-8213/ac089b}
  {\bibfield  {journal} {\bibinfo  {journal} {\apjl}\ }\textbf {\bibinfo
  {volume} {918}},\ \bibinfo {eid} {L28} (\bibinfo {year} {2021})},\ \Eprint
  {https://arxiv.org/abs/2105.06979} {arXiv:2105.06979 [astro-ph.HE]}
  \BibitemShut {NoStop}%
\bibitem [{\citenamefont {Annala}\ \emph {et~al.}(2020)\citenamefont {Annala},
  \citenamefont {Gorda}, \citenamefont {Kurkela}, \citenamefont {N\"attil\"a},\
  and\ \citenamefont {Vuorinen}}]{Annala:2019puf}%
  \BibitemOpen
  \bibfield  {author} {\bibinfo {author} {\bibfnamefont {E.}~\bibnamefont
  {Annala}}, \bibinfo {author} {\bibfnamefont {T.}~\bibnamefont {Gorda}},
  \bibinfo {author} {\bibfnamefont {A.}~\bibnamefont {Kurkela}}, \bibinfo
  {author} {\bibfnamefont {J.}~\bibnamefont {N\"attil\"a}},\ and\ \bibinfo
  {author} {\bibfnamefont {A.}~\bibnamefont {Vuorinen}},\ }\bibfield  {title}
  {\bibinfo {title} {{Evidence for quark-matter cores in massive neutron
  stars}},\ }\bibfield  {journal} {\bibinfo  {journal} {Nature Phys.}\ }\href
  {https://doi.org/10.1038/s41567-020-0914-9} {10.1038/s41567-020-0914-9}
  (\bibinfo {year} {2020}),\ \Eprint {https://arxiv.org/abs/1903.09121}
  {arXiv:1903.09121 [astro-ph.HE]} \BibitemShut {NoStop}%
\bibitem [{\citenamefont {{Tsue}}\ \emph {et~al.}(2010)\citenamefont {{Tsue}},
  \citenamefont {{Provid{\^e}ncia}}, \citenamefont {{Provid{\^e}ncia}},\ and\
  \citenamefont {{Yamamura}}}]{tsue2010PThPh}%
  \BibitemOpen
  \bibfield  {author} {\bibinfo {author} {\bibfnamefont {Y.}~\bibnamefont
  {{Tsue}}}, \bibinfo {author} {\bibfnamefont {J.~D.}\ \bibnamefont
  {{Provid{\^e}ncia}}}, \bibinfo {author} {\bibfnamefont {C.}~\bibnamefont
  {{Provid{\^e}ncia}}},\ and\ \bibinfo {author} {\bibfnamefont
  {M.}~\bibnamefont {{Yamamura}}},\ }\bibfield  {title} {\bibinfo {title}
  {{First-Order Quark-Hadron Phase-Transition in a NJL-Type Model for Nuclear
  and Quark Matter ---The Case of Symmetric Nuclear Matter---}},\ }\href
  {https://doi.org/10.1143/PTP.123.1013} {\bibfield  {journal} {\bibinfo
  {journal} {Progress of Theoretical Physics}\ }\textbf {\bibinfo {volume}
  {123}},\ \bibinfo {pages} {1013} (\bibinfo {year} {2010})},\ \Eprint
  {https://arxiv.org/abs/1003.2843} {arXiv:1003.2843 [hep-ph]} \BibitemShut
  {NoStop}%
\bibitem [{\citenamefont {{Chamel}}\ \emph {et~al.}(2013)\citenamefont
  {{Chamel}}, \citenamefont {{Fantina}}, \citenamefont {{Pearson}},\ and\
  \citenamefont {{Goriely}}}]{chamel2013A&A}%
  \BibitemOpen
  \bibfield  {author} {\bibinfo {author} {\bibfnamefont {N.}~\bibnamefont
  {{Chamel}}}, \bibinfo {author} {\bibfnamefont {A.~F.}\ \bibnamefont
  {{Fantina}}}, \bibinfo {author} {\bibfnamefont {J.~M.}\ \bibnamefont
  {{Pearson}}},\ and\ \bibinfo {author} {\bibfnamefont {S.}~\bibnamefont
  {{Goriely}}},\ }\bibfield  {title} {\bibinfo {title} {{Phase transitions in
  dense matter and the maximum mass of neutron stars}},\ }\href
  {https://doi.org/10.1051/0004-6361/201220986} {\bibfield  {journal} {\bibinfo
   {journal} {\aap}\ }\textbf {\bibinfo {volume} {553}},\ \bibinfo {eid} {A22}
  (\bibinfo {year} {2013})},\ \Eprint {https://arxiv.org/abs/1205.0983}
  {arXiv:1205.0983 [nucl-th]} \BibitemShut {NoStop}%
\bibitem [{\citenamefont {{Dexheimer}}\ \emph {et~al.}(2018)\citenamefont
  {{Dexheimer}}, \citenamefont {{Soethe}}, \citenamefont {{Roark}},
  \citenamefont {{Gomes}}, \citenamefont {{Kepler}},\ and\ \citenamefont
  {{Schramm}}}]{dexheimer2018IJMPE}%
  \BibitemOpen
  \bibfield  {author} {\bibinfo {author} {\bibfnamefont {V.}~\bibnamefont
  {{Dexheimer}}}, \bibinfo {author} {\bibfnamefont {L.~T.~T.}\ \bibnamefont
  {{Soethe}}}, \bibinfo {author} {\bibfnamefont {J.}~\bibnamefont {{Roark}}},
  \bibinfo {author} {\bibfnamefont {R.~O.}\ \bibnamefont {{Gomes}}}, \bibinfo
  {author} {\bibfnamefont {S.~O.}\ \bibnamefont {{Kepler}}},\ and\ \bibinfo
  {author} {\bibfnamefont {S.}~\bibnamefont {{Schramm}}},\ }\bibfield  {title}
  {\bibinfo {title} {{Phase transitions in neutron stars}},\ }\href
  {https://doi.org/10.1142/S0218301318300084} {\bibfield  {journal} {\bibinfo
  {journal} {International Journal of Modern Physics E}\ }\textbf {\bibinfo
  {volume} {27}},\ \bibinfo {pages} {1830008} (\bibinfo {year}
  {2018})}\BibitemShut {NoStop}%
\bibitem [{\citenamefont {{Hatsuda}}\ \emph {et~al.}(2006)\citenamefont
  {{Hatsuda}}, \citenamefont {{Tachibana}}, \citenamefont {{Yamamoto}},\ and\
  \citenamefont {{Baym}}}]{hatsuda2006PhRvL}%
  \BibitemOpen
  \bibfield  {author} {\bibinfo {author} {\bibfnamefont {T.}~\bibnamefont
  {{Hatsuda}}}, \bibinfo {author} {\bibfnamefont {M.}~\bibnamefont
  {{Tachibana}}}, \bibinfo {author} {\bibfnamefont {N.}~\bibnamefont
  {{Yamamoto}}},\ and\ \bibinfo {author} {\bibfnamefont {G.}~\bibnamefont
  {{Baym}}},\ }\bibfield  {title} {\bibinfo {title} {{New Critical Point
  Induced By the Axial Anomaly in Dense QCD}},\ }\href
  {https://doi.org/10.1103/PhysRevLett.97.122001} {\bibfield  {journal}
  {\bibinfo  {journal} {\prl}\ }\textbf {\bibinfo {volume} {97}},\ \bibinfo
  {eid} {122001} (\bibinfo {year} {2006})},\ \Eprint
  {https://arxiv.org/abs/hep-ph/0605018} {arXiv:hep-ph/0605018 [hep-ph]}
  \BibitemShut {NoStop}%
\bibitem [{\citenamefont {{Baym}}\ \emph {et~al.}(2019)\citenamefont {{Baym}},
  \citenamefont {{Furusawa}}, \citenamefont {{Hatsuda}}, \citenamefont
  {{Kojo}},\ and\ \citenamefont {{Togashi}}}]{baym2019ApJ}%
  \BibitemOpen
  \bibfield  {author} {\bibinfo {author} {\bibfnamefont {G.}~\bibnamefont
  {{Baym}}}, \bibinfo {author} {\bibfnamefont {S.}~\bibnamefont {{Furusawa}}},
  \bibinfo {author} {\bibfnamefont {T.}~\bibnamefont {{Hatsuda}}}, \bibinfo
  {author} {\bibfnamefont {T.}~\bibnamefont {{Kojo}}},\ and\ \bibinfo {author}
  {\bibfnamefont {H.}~\bibnamefont {{Togashi}}},\ }\bibfield  {title} {\bibinfo
  {title} {{New Neutron Star Equation of State with Quark-Hadron Crossover}},\
  }\href {https://doi.org/10.3847/1538-4357/ab441e} {\bibfield  {journal}
  {\bibinfo  {journal} {\apj}\ }\textbf {\bibinfo {volume} {885}},\ \bibinfo
  {eid} {42} (\bibinfo {year} {2019})},\ \Eprint
  {https://arxiv.org/abs/1903.08963} {arXiv:1903.08963 [astro-ph.HE]}
  \BibitemShut {NoStop}%
\bibitem [{\citenamefont {{Lugones}}\ and\ \citenamefont
  {{Grunfeld}}(2021{\natexlab{a}})}]{LugGrunf-universe:2021}%
  \BibitemOpen
  \bibfield  {author} {\bibinfo {author} {\bibfnamefont {G.}~\bibnamefont
  {{Lugones}}}\ and\ \bibinfo {author} {\bibfnamefont {A.~G.}\ \bibnamefont
  {{Grunfeld}}},\ }\bibfield  {title} {\bibinfo {title} {{Phase Conversions in
  Neutron Stars: Implications for Stellar Stability and Gravitational Wave
  Astrophysics}},\ }\href {https://doi.org/10.3390/universe7120493} {\bibfield
  {journal} {\bibinfo  {journal} {Universe}\ }\textbf {\bibinfo {volume} {7}},\
  \bibinfo {pages} {493} (\bibinfo {year} {2021}{\natexlab{a}})}\BibitemShut
  {NoStop}%
\bibitem [{\citenamefont {{Maruyama}}\ \emph {et~al.}(2007)\citenamefont
  {{Maruyama}}, \citenamefont {{Chiba}}, \citenamefont {{Schulze}},\ and\
  \citenamefont {{Tatsumi}}}]{Maruyama:2007ey}%
  \BibitemOpen
  \bibfield  {author} {\bibinfo {author} {\bibfnamefont {T.}~\bibnamefont
  {{Maruyama}}}, \bibinfo {author} {\bibfnamefont {S.}~\bibnamefont {{Chiba}}},
  \bibinfo {author} {\bibfnamefont {H.-J.}\ \bibnamefont {{Schulze}}},\ and\
  \bibinfo {author} {\bibfnamefont {T.}~\bibnamefont {{Tatsumi}}},\ }\bibfield
  {title} {\bibinfo {title} {{Hadron-quark mixed phase in hyperon stars}},\
  }\href {https://doi.org/10.1103/PhysRevD.76.123015} {\bibfield  {journal}
  {\bibinfo  {journal} {\prd}\ }\textbf {\bibinfo {volume} {76}},\ \bibinfo
  {eid} {123015} (\bibinfo {year} {2007})},\ \Eprint
  {https://arxiv.org/abs/0708.3277} {arXiv:0708.3277 [nucl-th]} \BibitemShut
  {NoStop}%
\bibitem [{\citenamefont {Lugones}\ \emph {et~al.}(2013)\citenamefont
  {Lugones}, \citenamefont {Grunfeld},\ and\ \citenamefont
  {Ajmi}}]{Lugones:2013ema}%
  \BibitemOpen
  \bibfield  {author} {\bibinfo {author} {\bibfnamefont {G.}~\bibnamefont
  {Lugones}}, \bibinfo {author} {\bibfnamefont {A.~G.}\ \bibnamefont
  {Grunfeld}},\ and\ \bibinfo {author} {\bibfnamefont {M.~A.}\ \bibnamefont
  {Ajmi}},\ }\bibfield  {title} {\bibinfo {title} {Surface tension and
  curvature energy of quark matter in the nambu--jona-lasinio model},\ }\href
  {https://doi.org/10.1103/PhysRevC.88.045803} {\bibfield  {journal} {\bibinfo
  {journal} {Phys. Rev. C}\ }\textbf {\bibinfo {volume} {88}},\ \bibinfo
  {pages} {045803} (\bibinfo {year} {2013})}\BibitemShut {NoStop}%
\bibitem [{\citenamefont {{Lugones}}\ and\ \citenamefont
  {{Grunfeld}}(2021{\natexlab{b}})}]{LugGrunf.2021PhRvDL}%
  \BibitemOpen
  \bibfield  {author} {\bibinfo {author} {\bibfnamefont {G.}~\bibnamefont
  {{Lugones}}}\ and\ \bibinfo {author} {\bibfnamefont {A.~G.}\ \bibnamefont
  {{Grunfeld}}},\ }\bibfield  {title} {\bibinfo {title} {{Vector interactions
  inhibit quark-hadron mixed phases in neutron stars}},\ }\href
  {https://doi.org/10.1103/PhysRevD.104.L101301} {\bibfield  {journal}
  {\bibinfo  {journal} {\prd}\ }\textbf {\bibinfo {volume} {104}},\ \bibinfo
  {eid} {L101301} (\bibinfo {year} {2021}{\natexlab{b}})},\ \Eprint
  {https://arxiv.org/abs/2109.01749} {arXiv:2109.01749 [nucl-th]} \BibitemShut
  {NoStop}%
\bibitem [{\citenamefont {{Pereira}}\ \emph {et~al.}(2018)\citenamefont
  {{Pereira}}, \citenamefont {{Flores}},\ and\ \citenamefont
  {{Lugones}}}]{Pereira:2017rmp}%
  \BibitemOpen
  \bibfield  {author} {\bibinfo {author} {\bibfnamefont {J.~P.}\ \bibnamefont
  {{Pereira}}}, \bibinfo {author} {\bibfnamefont {C.~V.}\ \bibnamefont
  {{Flores}}},\ and\ \bibinfo {author} {\bibfnamefont {G.}~\bibnamefont
  {{Lugones}}},\ }\bibfield  {title} {\bibinfo {title} {{Phase Transition
  Effects on the Dynamical Stability of Hybrid Neutron Stars}},\ }\href
  {https://doi.org/10.3847/1538-4357/aabfbf} {\bibfield  {journal} {\bibinfo
  {journal} {\apj}\ }\textbf {\bibinfo {volume} {860}},\ \bibinfo {eid} {12}
  (\bibinfo {year} {2018})},\ \Eprint {https://arxiv.org/abs/1706.09371}
  {arXiv:1706.09371 [gr-qc]} \BibitemShut {NoStop}%
\bibitem [{\citenamefont {Baym}\ \emph {et~al.}(2018)\citenamefont {Baym},
  \citenamefont {Hatsuda}, \citenamefont {Kojo}, \citenamefont {Powell},
  \citenamefont {Song},\ and\ \citenamefont {Takatsuka}}]{Baym-review}%
  \BibitemOpen
  \bibfield  {author} {\bibinfo {author} {\bibfnamefont {G.}~\bibnamefont
  {Baym}}, \bibinfo {author} {\bibfnamefont {T.}~\bibnamefont {Hatsuda}},
  \bibinfo {author} {\bibfnamefont {T.}~\bibnamefont {Kojo}}, \bibinfo {author}
  {\bibfnamefont {P.~D.}\ \bibnamefont {Powell}}, \bibinfo {author}
  {\bibfnamefont {Y.}~\bibnamefont {Song}},\ and\ \bibinfo {author}
  {\bibfnamefont {T.}~\bibnamefont {Takatsuka}},\ }\bibfield  {title} {\bibinfo
  {title} {{From hadrons to quarks in neutron stars: a review}},\ }\href
  {https://doi.org/10.1088/1361-6633/aaae14} {\bibfield  {journal} {\bibinfo
  {journal} {Rept. Prog. Phys.}\ }\textbf {\bibinfo {volume} {81}},\ \bibinfo
  {pages} {056902} (\bibinfo {year} {2018})},\ \Eprint
  {https://arxiv.org/abs/1707.04966} {arXiv:1707.04966 [astro-ph.HE]}
  \BibitemShut {NoStop}%
\bibitem [{\citenamefont {{Andersson}}(2021)}]{andersson2021Univ}%
  \BibitemOpen
  \bibfield  {author} {\bibinfo {author} {\bibfnamefont {N.}~\bibnamefont
  {{Andersson}}},\ }\bibfield  {title} {\bibinfo {title} {{A Gravitational-Wave
  Perspective on Neutron-Star Seismology}},\ }\href
  {https://doi.org/10.3390/universe7040097} {\bibfield  {journal} {\bibinfo
  {journal} {Universe}\ }\textbf {\bibinfo {volume} {7}},\ \bibinfo {pages}
  {97} (\bibinfo {year} {2021})},\ \Eprint {https://arxiv.org/abs/2103.10223}
  {arXiv:2103.10223 [gr-qc]} \BibitemShut {NoStop}%
\bibitem [{\citenamefont {{Punturo}}\ and\ \citenamefont
  {et~al.}(2010)}]{Punturo:2010zz}%
  \BibitemOpen
  \bibfield  {author} {\bibinfo {author} {\bibfnamefont {M.}~\bibnamefont
  {{Punturo}}}\ and\ \bibinfo {author} {\bibnamefont {et~al.}},\ }\bibfield
  {title} {\bibinfo {title} {{The Einstein Telescope: a third-generation
  gravitational wave observatory}},\ }\href
  {https://doi.org/10.1088/0264-9381/27/19/194002} {\bibfield  {journal}
  {\bibinfo  {journal} {Classical and Quantum Gravity}\ }\textbf {\bibinfo
  {volume} {27}},\ \bibinfo {eid} {194002} (\bibinfo {year}
  {2010})}\BibitemShut {NoStop}%
\bibitem [{\citenamefont {{Andersson}}\ and\ \citenamefont
  {{Kokkotas}}(1998)}]{AK}%
  \BibitemOpen
  \bibfield  {author} {\bibinfo {author} {\bibfnamefont {N.}~\bibnamefont
  {{Andersson}}}\ and\ \bibinfo {author} {\bibfnamefont {K.~D.}\ \bibnamefont
  {{Kokkotas}}},\ }\bibfield  {title} {\bibinfo {title} {{Towards gravitational
  wave asteroseismology}},\ }\href
  {https://doi.org/10.1046/j.1365-8711.1998.01840.x} {\bibfield  {journal}
  {\bibinfo  {journal} {\mnras}\ }\textbf {\bibinfo {volume} {299}},\ \bibinfo
  {pages} {1059} (\bibinfo {year} {1998})},\ \Eprint
  {https://arxiv.org/abs/gr-qc/9711088} {arXiv:gr-qc/9711088 [gr-qc]}
  \BibitemShut {NoStop}%
\bibitem [{\citenamefont {{Rosofsky}}\ \emph {et~al.}(2019)\citenamefont
  {{Rosofsky}}, \citenamefont {{Gold}}, \citenamefont {{Chirenti}},
  \citenamefont {{Huerta}},\ and\ \citenamefont {{Miller}}}]{Rosofsky:2018}%
  \BibitemOpen
  \bibfield  {author} {\bibinfo {author} {\bibfnamefont {S.~G.}\ \bibnamefont
  {{Rosofsky}}}, \bibinfo {author} {\bibfnamefont {R.}~\bibnamefont {{Gold}}},
  \bibinfo {author} {\bibfnamefont {C.}~\bibnamefont {{Chirenti}}}, \bibinfo
  {author} {\bibfnamefont {E.~A.}\ \bibnamefont {{Huerta}}},\ and\ \bibinfo
  {author} {\bibfnamefont {M.~C.}\ \bibnamefont {{Miller}}},\ }\bibfield
  {title} {\bibinfo {title} {{Probing neutron star structure via f -mode
  oscillations and damping in dynamical spacetime models}},\ }\href
  {https://doi.org/10.1103/PhysRevD.99.084024} {\bibfield  {journal} {\bibinfo
  {journal} {\prd}\ }\textbf {\bibinfo {volume} {99}},\ \bibinfo {eid} {084024}
  (\bibinfo {year} {2019})},\ \Eprint {https://arxiv.org/abs/1812.06126}
  {arXiv:1812.06126 [gr-qc]} \BibitemShut {NoStop}%
\bibitem [{\citenamefont {{V{\'a}squez Flores}}\ and\ \citenamefont
  {{Lugones}}(2018)}]{Flores:2018pnn}%
  \BibitemOpen
  \bibfield  {author} {\bibinfo {author} {\bibfnamefont {C.}~\bibnamefont
  {{V{\'a}squez Flores}}}\ and\ \bibinfo {author} {\bibfnamefont
  {G.}~\bibnamefont {{Lugones}}},\ }\bibfield  {title} {\bibinfo {title}
  {{Gravitational wave asteroseismology limits from low density nuclear matter
  and perturbative QCD}},\ }\href
  {https://doi.org/10.1088/1475-7516/2018/08/046} {\bibfield  {journal}
  {\bibinfo  {journal} {\jcap}\ }\textbf {\bibinfo {volume} {2018}},\ \bibinfo
  {eid} {046} (\bibinfo {year} {2018})},\ \Eprint
  {https://arxiv.org/abs/1804.05155} {arXiv:1804.05155 [astro-ph.HE]}
  \BibitemShut {NoStop}%
\bibitem [{\citenamefont {{Ranea-Sandoval}}\ \emph {et~al.}(2018)\citenamefont
  {{Ranea-Sandoval}}, \citenamefont {{Guilera}}, \citenamefont {{Mariani}},\
  and\ \citenamefont {{Orsaria}}}]{RSetalJCAP}%
  \BibitemOpen
  \bibfield  {author} {\bibinfo {author} {\bibfnamefont {I.~F.}\ \bibnamefont
  {{Ranea-Sandoval}}}, \bibinfo {author} {\bibfnamefont {O.~M.}\ \bibnamefont
  {{Guilera}}}, \bibinfo {author} {\bibfnamefont {M.}~\bibnamefont
  {{Mariani}}},\ and\ \bibinfo {author} {\bibfnamefont {M.~G.}\ \bibnamefont
  {{Orsaria}}},\ }\bibfield  {title} {\bibinfo {title} {{Oscillation modes of
  hybrid stars within the relativistic Cowling approximation}},\ }\href
  {https://doi.org/10.1088/1475-7516/2018/12/031} {\bibfield  {journal}
  {\bibinfo  {journal} {\jcap}\ }\textbf {\bibinfo {volume} {2018}},\ \bibinfo
  {eid} {031} (\bibinfo {year} {2018})},\ \Eprint
  {https://arxiv.org/abs/1807.02166} {arXiv:1807.02166 [astro-ph.HE]}
  \BibitemShut {NoStop}%
\bibitem [{\citenamefont {{Rodr{\'\i}guez}}\ \emph {et~al.}(2021)\citenamefont
  {{Rodr{\'\i}guez}}, \citenamefont {{Ranea-Sandoval}}, \citenamefont
  {{Mariani}}, \citenamefont {{Orsaria}}, \citenamefont {{Malfatti}},\ and\
  \citenamefont {{Guilera}}}]{Rodriguez-etalg2}%
  \BibitemOpen
  \bibfield  {author} {\bibinfo {author} {\bibfnamefont {M.~C.}\ \bibnamefont
  {{Rodr{\'\i}guez}}}, \bibinfo {author} {\bibfnamefont {I.~F.}\ \bibnamefont
  {{Ranea-Sandoval}}}, \bibinfo {author} {\bibfnamefont {M.}~\bibnamefont
  {{Mariani}}}, \bibinfo {author} {\bibfnamefont {M.~G.}\ \bibnamefont
  {{Orsaria}}}, \bibinfo {author} {\bibfnamefont {G.}~\bibnamefont
  {{Malfatti}}},\ and\ \bibinfo {author} {\bibfnamefont {O.~M.}\ \bibnamefont
  {{Guilera}}},\ }\bibfield  {title} {\bibinfo {title} {{Hybrid stars with
  sequential phase transitions: the emergence of the g$_{2}$ mode}},\ }\href
  {https://doi.org/10.1088/1475-7516/2021/02/009} {\bibfield  {journal}
  {\bibinfo  {journal} {\jcap}\ }\textbf {\bibinfo {volume} {2021}},\ \bibinfo
  {eid} {009} (\bibinfo {year} {2021})},\ \Eprint
  {https://arxiv.org/abs/2009.03769} {arXiv:2009.03769 [astro-ph.HE]}
  \BibitemShut {NoStop}%
\bibitem [{\citenamefont {{Mena-Fern{\'a}ndez}}\ and\ \citenamefont
  {{Gonz{\'a}lez-Romero}}(2019)}]{Mena-Fernandez2019}%
  \BibitemOpen
  \bibfield  {author} {\bibinfo {author} {\bibfnamefont {J.}~\bibnamefont
  {{Mena-Fern{\'a}ndez}}}\ and\ \bibinfo {author} {\bibfnamefont {L.~M.}\
  \bibnamefont {{Gonz{\'a}lez-Romero}}},\ }\bibfield  {title} {\bibinfo {title}
  {{Reconstruction of the neutron star equation of state from w -quasinormal
  modes spectra with a piecewise polytropic meshing and refinement method}},\
  }\href {https://doi.org/10.1103/PhysRevD.99.104005} {\bibfield  {journal}
  {\bibinfo  {journal} {\prd}\ }\textbf {\bibinfo {volume} {99}},\ \bibinfo
  {eid} {104005} (\bibinfo {year} {2019})},\ \Eprint
  {https://arxiv.org/abs/1901.10851} {arXiv:1901.10851 [gr-qc]} \BibitemShut
  {NoStop}%
\bibitem [{\citenamefont {{Benitez}}\ \emph {et~al.}(2021)\citenamefont
  {{Benitez}}, \citenamefont {{Weller}}, \citenamefont {{Guedes}},
  \citenamefont {{Chirenti}},\ and\ \citenamefont
  {{Miller}}}]{chirenti2020-wmode}%
  \BibitemOpen
  \bibfield  {author} {\bibinfo {author} {\bibfnamefont {E.}~\bibnamefont
  {{Benitez}}}, \bibinfo {author} {\bibfnamefont {J.}~\bibnamefont {{Weller}}},
  \bibinfo {author} {\bibfnamefont {V.}~\bibnamefont {{Guedes}}}, \bibinfo
  {author} {\bibfnamefont {C.}~\bibnamefont {{Chirenti}}},\ and\ \bibinfo
  {author} {\bibfnamefont {M.~C.}\ \bibnamefont {{Miller}}},\ }\bibfield
  {title} {\bibinfo {title} {{Investigating the I-Love-Q and w -mode universal
  relations using piecewise polytropes}},\ }\href
  {https://doi.org/10.1103/PhysRevD.103.023007} {\bibfield  {journal} {\bibinfo
   {journal} {\prd}\ }\textbf {\bibinfo {volume} {103}},\ \bibinfo {eid}
  {023007} (\bibinfo {year} {2021})},\ \Eprint
  {https://arxiv.org/abs/2010.02619} {arXiv:2010.02619 [astro-ph.HE]}
  \BibitemShut {NoStop}%
\bibitem [{\citenamefont {{Chandrasekhar}}(1970)}]{chandra-CFS}%
  \BibitemOpen
  \bibfield  {author} {\bibinfo {author} {\bibfnamefont {S.}~\bibnamefont
  {{Chandrasekhar}}},\ }\bibfield  {title} {\bibinfo {title} {{Solutions of Two
  Problems in the Theory of Gravitational Radiation}},\ }\href
  {https://doi.org/10.1103/PhysRevLett.24.611} {\bibfield  {journal} {\bibinfo
  {journal} {Physical Review Letters}\ }\textbf {\bibinfo {volume} {24}},\
  \bibinfo {pages} {611} (\bibinfo {year} {1970})}\BibitemShut {NoStop}%
\bibitem [{\citenamefont {{Friedman}}\ and\ \citenamefont
  {{Schutz}}(1978)}]{FS-CFS}%
  \BibitemOpen
  \bibfield  {author} {\bibinfo {author} {\bibfnamefont {J.~L.}\ \bibnamefont
  {{Friedman}}}\ and\ \bibinfo {author} {\bibfnamefont {B.~F.}\ \bibnamefont
  {{Schutz}}},\ }\bibfield  {title} {\bibinfo {title} {{Secular instability of
  rotating Newtonian stars.}},\ }\href {https://doi.org/10.1086/156143}
  {\bibfield  {journal} {\bibinfo  {journal} {\apj}\ }\textbf {\bibinfo
  {volume} {222}},\ \bibinfo {pages} {281} (\bibinfo {year}
  {1978})}\BibitemShut {NoStop}%
\bibitem [{\citenamefont {Bl\'azquez-Salcedo}\ \emph
  {et~al.}(2013)\citenamefont {Bl\'azquez-Salcedo}, \citenamefont
  {Gonz\'alez-Romero},\ and\ \citenamefont
  {Navarro-L\'erida}}]{w-modes_universal}%
  \BibitemOpen
  \bibfield  {author} {\bibinfo {author} {\bibfnamefont {J.~L.}\ \bibnamefont
  {Bl\'azquez-Salcedo}}, \bibinfo {author} {\bibfnamefont {L.~M.}\ \bibnamefont
  {Gonz\'alez-Romero}},\ and\ \bibinfo {author} {\bibfnamefont
  {F.}~\bibnamefont {Navarro-L\'erida}},\ }\bibfield  {title} {\bibinfo {title}
  {Phenomenological relations for axial quasinormal modes of neutron stars with
  realistic equations of state},\ }\href
  {https://doi.org/10.1103/PhysRevD.87.104042} {\bibfield  {journal} {\bibinfo
  {journal} {Phys. Rev. D}\ }\textbf {\bibinfo {volume} {87}},\ \bibinfo
  {pages} {104042} (\bibinfo {year} {2013})}\BibitemShut {NoStop}%
\bibitem [{\citenamefont {{Benhar}}\ \emph {et~al.}(2004)\citenamefont
  {{Benhar}}, \citenamefont {{Ferrari}},\ and\ \citenamefont
  {{Gualtieri}}}]{Benhar2004PRD}%
  \BibitemOpen
  \bibfield  {author} {\bibinfo {author} {\bibfnamefont {O.}~\bibnamefont
  {{Benhar}}}, \bibinfo {author} {\bibfnamefont {V.}~\bibnamefont
  {{Ferrari}}},\ and\ \bibinfo {author} {\bibfnamefont {L.}~\bibnamefont
  {{Gualtieri}}},\ }\bibfield  {title} {\bibinfo {title} {{Gravitational wave
  asteroseismology reexamined}},\ }\href
  {https://doi.org/10.1103/PhysRevD.70.124015} {\bibfield  {journal} {\bibinfo
  {journal} {\prd}\ }\textbf {\bibinfo {volume} {70}},\ \bibinfo {eid} {124015}
  (\bibinfo {year} {2004})},\ \Eprint {https://arxiv.org/abs/astro-ph/0407529}
  {arXiv:astro-ph/0407529 [astro-ph]} \BibitemShut {NoStop}%
\bibitem [{\citenamefont {{Tsui}}\ and\ \citenamefont
  {{Leung}}(2005)}]{Tsui2005MNRAS}%
  \BibitemOpen
  \bibfield  {author} {\bibinfo {author} {\bibfnamefont {L.~K.}\ \bibnamefont
  {{Tsui}}}\ and\ \bibinfo {author} {\bibfnamefont {P.~T.}\ \bibnamefont
  {{Leung}}},\ }\bibfield  {title} {\bibinfo {title} {{Universality in
  quasi-normal modes of neutron stars}},\ }\href
  {https://doi.org/10.1111/j.1365-2966.2005.08710.x} {\bibfield  {journal}
  {\bibinfo  {journal} {\mnras}\ }\textbf {\bibinfo {volume} {357}},\ \bibinfo
  {pages} {1029} (\bibinfo {year} {2005})},\ \Eprint
  {https://arxiv.org/abs/gr-qc/0412024} {arXiv:gr-qc/0412024 [gr-qc]}
  \BibitemShut {NoStop}%
\bibitem [{\citenamefont {{Detar}}\ and\ \citenamefont
  {{Heller}}(2009)}]{qcd01}%
  \BibitemOpen
  \bibfield  {author} {\bibinfo {author} {\bibfnamefont {C.~E.}\ \bibnamefont
  {{Detar}}}\ and\ \bibinfo {author} {\bibfnamefont {U.~M.}\ \bibnamefont
  {{Heller}}},\ }\bibfield  {title} {\bibinfo {title} {{QCD thermodynamics from
  the lattice}},\ }\href {https://doi.org/10.1140/epja/i2009-10825-3}
  {\bibfield  {journal} {\bibinfo  {journal} {European Physical Journal A}\
  }\textbf {\bibinfo {volume} {41}},\ \bibinfo {pages} {405} (\bibinfo {year}
  {2009})},\ \Eprint {https://arxiv.org/abs/0905.2949} {arXiv:0905.2949
  [hep-lat]} \BibitemShut {NoStop}%
\bibitem [{\citenamefont {{Muroya}}\ \emph {et~al.}(2003)\citenamefont
  {{Muroya}}, \citenamefont {{Nakamura}}, \citenamefont {{Nonaka}},\ and\
  \citenamefont {{Takaishi}}}]{qcd02}%
  \BibitemOpen
  \bibfield  {author} {\bibinfo {author} {\bibfnamefont {S.}~\bibnamefont
  {{Muroya}}}, \bibinfo {author} {\bibfnamefont {A.}~\bibnamefont
  {{Nakamura}}}, \bibinfo {author} {\bibfnamefont {C.}~\bibnamefont
  {{Nonaka}}},\ and\ \bibinfo {author} {\bibfnamefont {T.}~\bibnamefont
  {{Takaishi}}},\ }\bibfield  {title} {\bibinfo {title} {{Lattice QCD at Finite
  Density ---An Introductory Review}},\ }\href
  {https://doi.org/10.1143/PTP.110.615} {\bibfield  {journal} {\bibinfo
  {journal} {Progress of Theoretical Physics}\ }\textbf {\bibinfo {volume}
  {110}},\ \bibinfo {pages} {615} (\bibinfo {year} {2003})},\ \Eprint
  {https://arxiv.org/abs/hep-lat/0306031} {arXiv:hep-lat/0306031 [hep-lat]}
  \BibitemShut {NoStop}%
\bibitem [{\citenamefont {{Braun-Munzinger}}\ and\ \citenamefont
  {{Wambach}}(2009)}]{BraunMunzinger:2009}%
  \BibitemOpen
  \bibfield  {author} {\bibinfo {author} {\bibfnamefont {P.}~\bibnamefont
  {{Braun-Munzinger}}}\ and\ \bibinfo {author} {\bibfnamefont {J.}~\bibnamefont
  {{Wambach}}},\ }\bibfield  {title} {\bibinfo {title} {{Colloquium: Phase
  diagram of strongly interacting matter}},\ }\href
  {https://doi.org/10.1103/RevModPhys.81.1031} {\bibfield  {journal} {\bibinfo
  {journal} {Reviews of Modern Physics}\ }\textbf {\bibinfo {volume} {81}},\
  \bibinfo {pages} {1031} (\bibinfo {year} {2009})}\BibitemShut {NoStop}%
\bibitem [{\citenamefont {{Goy}}\ \emph {et~al.}(2017)\citenamefont {{Goy}},
  \citenamefont {{Bornyakov}}, \citenamefont {{Boyda}}, \citenamefont
  {{Molochkov}}, \citenamefont {{Nakamura}}, \citenamefont {{Nikolaev}},\ and\
  \citenamefont {{Zakharov}}}]{2017PTEP.2017c1D01G}%
  \BibitemOpen
  \bibfield  {author} {\bibinfo {author} {\bibfnamefont {V.~A.}\ \bibnamefont
  {{Goy}}}, \bibinfo {author} {\bibfnamefont {V.}~\bibnamefont {{Bornyakov}}},
  \bibinfo {author} {\bibfnamefont {D.}~\bibnamefont {{Boyda}}}, \bibinfo
  {author} {\bibfnamefont {A.}~\bibnamefont {{Molochkov}}}, \bibinfo {author}
  {\bibfnamefont {A.}~\bibnamefont {{Nakamura}}}, \bibinfo {author}
  {\bibfnamefont {A.}~\bibnamefont {{Nikolaev}}},\ and\ \bibinfo {author}
  {\bibfnamefont {V.}~\bibnamefont {{Zakharov}}},\ }\bibfield  {title}
  {\bibinfo {title} {{Sign problem in finite density lattice QCD}},\ }\href
  {https://doi.org/10.1093/ptep/ptx018} {\bibfield  {journal} {\bibinfo
  {journal} {Progress of Theoretical and Experimental Physics}\ }\textbf
  {\bibinfo {volume} {2017}},\ \bibinfo {eid} {031D01} (\bibinfo {year}
  {2017})},\ \Eprint {https://arxiv.org/abs/1611.08093} {arXiv:1611.08093
  [hep-lat]} \BibitemShut {NoStop}%
\bibitem [{\citenamefont {{Mariani}}\ \emph {et~al.}(2017)\citenamefont
  {{Mariani}}, \citenamefont {{Orsaria}},\ and\ \citenamefont
  {{Vucetich}}}]{mariani-nestor}%
  \BibitemOpen
  \bibfield  {author} {\bibinfo {author} {\bibfnamefont {M.}~\bibnamefont
  {{Mariani}}}, \bibinfo {author} {\bibfnamefont {M.}~\bibnamefont
  {{Orsaria}}},\ and\ \bibinfo {author} {\bibfnamefont {H.}~\bibnamefont
  {{Vucetich}}},\ }\bibfield  {title} {\bibinfo {title} {{Constant entropy
  hybrid stars: a first approximation of cooling evolution}},\ }\href
  {https://doi.org/10.1051/0004-6361/201629315} {\bibfield  {journal} {\bibinfo
   {journal} {\aap}\ }\textbf {\bibinfo {volume} {601}},\ \bibinfo {eid} {A21}
  (\bibinfo {year} {2017})},\ \Eprint {https://arxiv.org/abs/1607.05200}
  {arXiv:1607.05200 [astro-ph.HE]} \BibitemShut {NoStop}%
\bibitem [{\citenamefont {{Chodos}}\ \emph {et~al.}(1974)\citenamefont
  {{Chodos}}, \citenamefont {{Jaffe}}, \citenamefont {{Johnson}}, \citenamefont
  {{Thorn}},\ and\ \citenamefont {{Weisskopf}}}]{Chodos1974}%
  \BibitemOpen
  \bibfield  {author} {\bibinfo {author} {\bibfnamefont {A.}~\bibnamefont
  {{Chodos}}}, \bibinfo {author} {\bibfnamefont {R.~L.}\ \bibnamefont
  {{Jaffe}}}, \bibinfo {author} {\bibfnamefont {K.}~\bibnamefont {{Johnson}}},
  \bibinfo {author} {\bibfnamefont {C.~B.}\ \bibnamefont {{Thorn}}},\ and\
  \bibinfo {author} {\bibfnamefont {V.~F.}\ \bibnamefont {{Weisskopf}}},\
  }\bibfield  {title} {\bibinfo {title} {{New extended model of hadrons}},\
  }\href {https://doi.org/10.1103/PhysRevD.9.3471} {\bibfield  {journal}
  {\bibinfo  {journal} {\prd}\ }\textbf {\bibinfo {volume} {9}},\ \bibinfo
  {pages} {3471} (\bibinfo {year} {1974})}\BibitemShut {NoStop}%
\bibitem [{\citenamefont {Nefediev}\ \emph {et~al.}(2009)\citenamefont
  {Nefediev}, \citenamefont {Simonov},\ and\ \citenamefont
  {Trusov}}]{Nefediev2009}%
  \BibitemOpen
  \bibfield  {author} {\bibinfo {author} {\bibfnamefont {A.~V.}\ \bibnamefont
  {Nefediev}}, \bibinfo {author} {\bibfnamefont {Y.~A.}\ \bibnamefont
  {Simonov}},\ and\ \bibinfo {author} {\bibfnamefont {M.~A.}\ \bibnamefont
  {Trusov}},\ }\bibfield  {title} {\bibinfo {title} {Deconfinement and
  quark–gluon plasma},\ }\href {https://doi.org/10.1142/s0218301309012768}
  {\bibfield  {journal} {\bibinfo  {journal} {International Journal of Modern
  Physics E}\ }\textbf {\bibinfo {volume} {18}},\ \bibinfo {pages} {549–599}
  (\bibinfo {year} {2009})}\BibitemShut {NoStop}%
\bibitem [{\citenamefont {{Dexheimer}}\ and\ \citenamefont
  {{Schramm}}(2010)}]{Dexheimer:2009}%
  \BibitemOpen
  \bibfield  {author} {\bibinfo {author} {\bibfnamefont {V.~A.}\ \bibnamefont
  {{Dexheimer}}}\ and\ \bibinfo {author} {\bibfnamefont {S.}~\bibnamefont
  {{Schramm}}},\ }\bibfield  {title} {\bibinfo {title} {{Novel approach to
  modeling hybrid stars}},\ }\href {https://doi.org/10.1103/PhysRevC.81.045201}
  {\bibfield  {journal} {\bibinfo  {journal} {\prc}\ }\textbf {\bibinfo
  {volume} {81}},\ \bibinfo {eid} {045201} (\bibinfo {year} {2010})},\ \Eprint
  {https://arxiv.org/abs/0901.1748} {arXiv:0901.1748 [astro-ph.SR]}
  \BibitemShut {NoStop}%
\bibitem [{\citenamefont {{Contrera, G. A. and Orsaria, M. and Scoccola, N.
  N.}}(2010)}]{Contrera2010}%
  \BibitemOpen
  \bibfield  {author} {\bibinfo {author} {\bibnamefont {{Contrera, G. A. and
  Orsaria, M. and Scoccola, N. N.}}},\ }\bibfield  {title} {\bibinfo {title}
  {{Nonlocal Polyakov-Nambu-Jona-Lasinio model with wave function
  renormalization at finite temperature and chemical potential}},\ }\href
  {https://doi.org/{10.1103/PhysRevD.82.054026}} {\bibfield  {journal}
  {\bibinfo  {journal} {{Phys. Rev. D}}\ }\textbf {\bibinfo {volume} {{82}}},\
  \bibinfo {pages} {{054026}} (\bibinfo {year} {{2010}})}\BibitemShut {NoStop}%
\bibitem [{\citenamefont {{Orsaria, M. and Rodrigues, H. and Weber, F. and
  Contrera, G. A.}}(2014)}]{Orsaria2013}%
  \BibitemOpen
  \bibfield  {author} {\bibinfo {author} {\bibnamefont {{Orsaria, M. and
  Rodrigues, H. and Weber, F. and Contrera, G. A.}}},\ }\bibfield  {title}
  {\bibinfo {title} {{Quark deconfinement in high-mass neutron stars}},\ }\href
  {https://doi.org/{10.1103/PhysRevC.89.015806}} {\bibfield  {journal}
  {\bibinfo  {journal} {{Phys. Rev. C}}\ }\textbf {\bibinfo {volume} {{89}}},\
  \bibinfo {pages} {{015806}} (\bibinfo {year} {{2014}})}\BibitemShut {NoStop}%
\bibitem [{\citenamefont {{Beni{\'c}}}\ \emph {et~al.}(2015)\citenamefont
  {{Beni{\'c}}}, \citenamefont {{Blaschke}}, \citenamefont
  {{Alvarez-Castillo}}, \citenamefont {{Fischer}},\ and\ \citenamefont
  {{Typel}}}]{benic2014}%
  \BibitemOpen
  \bibfield  {author} {\bibinfo {author} {\bibfnamefont {S.}~\bibnamefont
  {{Beni{\'c}}}}, \bibinfo {author} {\bibfnamefont {D.}~\bibnamefont
  {{Blaschke}}}, \bibinfo {author} {\bibfnamefont {D.~E.}\ \bibnamefont
  {{Alvarez-Castillo}}}, \bibinfo {author} {\bibfnamefont {T.}~\bibnamefont
  {{Fischer}}},\ and\ \bibinfo {author} {\bibfnamefont {S.}~\bibnamefont
  {{Typel}}},\ }\bibfield  {title} {\bibinfo {title} {{A new quark-hadron
  hybrid equation of state for astrophysics. I. High-mass twin compact
  stars}},\ }\href {https://doi.org/10.1051/0004-6361/201425318} {\bibfield
  {journal} {\bibinfo  {journal} {Astron. Astrophys.}\ }\textbf {\bibinfo
  {volume} {577}},\ \bibinfo {eid} {A40} (\bibinfo {year} {2015})},\ \Eprint
  {https://arxiv.org/abs/1411.2856} {arXiv:1411.2856 [astro-ph.HE]}
  \BibitemShut {NoStop}%
\bibitem [{\citenamefont {Ranea-Sandoval}\ \emph {et~al.}(2017)\citenamefont
  {Ranea-Sandoval}, \citenamefont {Orsaria}, \citenamefont {Han}, \citenamefont
  {Weber},\ and\ \citenamefont {Spinella}}]{Ranea-Sandoval:2017}%
  \BibitemOpen
  \bibfield  {author} {\bibinfo {author} {\bibfnamefont {I.~F.}\ \bibnamefont
  {Ranea-Sandoval}}, \bibinfo {author} {\bibfnamefont {M.~G.}\ \bibnamefont
  {Orsaria}}, \bibinfo {author} {\bibfnamefont {S.}~\bibnamefont {Han}},
  \bibinfo {author} {\bibfnamefont {F.}~\bibnamefont {Weber}},\ and\ \bibinfo
  {author} {\bibfnamefont {W.~M.}\ \bibnamefont {Spinella}},\ }\bibfield
  {title} {\bibinfo {title} {Color superconductivity in compact stellar hybrid
  configurations},\ }\href {https://doi.org/10.1103/PhysRevC.96.065807}
  {\bibfield  {journal} {\bibinfo  {journal} {Phys. Rev. C}\ }\textbf {\bibinfo
  {volume} {96}},\ \bibinfo {pages} {065807} (\bibinfo {year}
  {2017})}\BibitemShut {NoStop}%
\bibitem [{\citenamefont {Malfatti}\ \emph {et~al.}(2019)\citenamefont
  {Malfatti}, \citenamefont {Orsaria}, \citenamefont {Contrera}, \citenamefont
  {Weber},\ and\ \citenamefont {Ranea-Sandoval}}]{Malfatti:2019}%
  \BibitemOpen
  \bibfield  {author} {\bibinfo {author} {\bibfnamefont {G.}~\bibnamefont
  {Malfatti}}, \bibinfo {author} {\bibfnamefont {M.~G.}\ \bibnamefont
  {Orsaria}}, \bibinfo {author} {\bibfnamefont {G.~A.}\ \bibnamefont
  {Contrera}}, \bibinfo {author} {\bibfnamefont {F.}~\bibnamefont {Weber}},\
  and\ \bibinfo {author} {\bibfnamefont {I.~F.}\ \bibnamefont
  {Ranea-Sandoval}},\ }\bibfield  {title} {\bibinfo {title} {Hot quark matter
  and (proto-) neutron stars},\ }\href
  {https://doi.org/10.1103/PhysRevC.100.015803} {\bibfield  {journal} {\bibinfo
   {journal} {Phys. Rev. C}\ }\textbf {\bibinfo {volume} {100}},\ \bibinfo
  {pages} {015803} (\bibinfo {year} {2019})}\BibitemShut {NoStop}%
\bibitem [{\citenamefont {{Baym}}\ \emph
  {et~al.}(1971{\natexlab{a}})\citenamefont {{Baym}}, \citenamefont
  {{Pethick}},\ and\ \citenamefont {{Sutherland}}}]{BPS1971}%
  \BibitemOpen
  \bibfield  {author} {\bibinfo {author} {\bibfnamefont {G.}~\bibnamefont
  {{Baym}}}, \bibinfo {author} {\bibfnamefont {C.}~\bibnamefont {{Pethick}}},\
  and\ \bibinfo {author} {\bibfnamefont {P.}~\bibnamefont {{Sutherland}}},\
  }\bibfield  {title} {\bibinfo {title} {{The Ground State of Matter at High
  Densities: Equation of State and Stellar Models}},\ }\href
  {https://doi.org/10.1086/151216} {\bibfield  {journal} {\bibinfo  {journal}
  {Astrophys. J.}\ }\textbf {\bibinfo {volume} {170}},\ \bibinfo {pages} {299}
  (\bibinfo {year} {1971}{\natexlab{a}})}\BibitemShut {NoStop}%
\bibitem [{\citenamefont {{Baym}}\ \emph
  {et~al.}(1971{\natexlab{b}})\citenamefont {{Baym}}, \citenamefont {{Bethe}},\
  and\ \citenamefont {{Pethick}}}]{BBP1971}%
  \BibitemOpen
  \bibfield  {author} {\bibinfo {author} {\bibfnamefont {G.}~\bibnamefont
  {{Baym}}}, \bibinfo {author} {\bibfnamefont {H.~A.}\ \bibnamefont
  {{Bethe}}},\ and\ \bibinfo {author} {\bibfnamefont {C.~J.}\ \bibnamefont
  {{Pethick}}},\ }\bibfield  {title} {\bibinfo {title} {{Neutron star
  matter}},\ }\href {https://doi.org/10.1016/0375-9474(71)90281-8} {\bibfield
  {journal} {\bibinfo  {journal} {Nuclear Physics A}\ }\textbf {\bibinfo
  {volume} {175}},\ \bibinfo {pages} {225} (\bibinfo {year}
  {1971}{\natexlab{b}})}\BibitemShut {NoStop}%
\bibitem [{\citenamefont {Kippenhahn}\ \emph {et~al.}(2012)\citenamefont
  {Kippenhahn}, \citenamefont {Weigert},\ and\ \citenamefont {Weiss}}]{kipp}%
  \BibitemOpen
  \bibfield  {author} {\bibinfo {author} {\bibfnamefont {R.}~\bibnamefont
  {Kippenhahn}}, \bibinfo {author} {\bibfnamefont {A.}~\bibnamefont
  {Weigert}},\ and\ \bibinfo {author} {\bibfnamefont {A.}~\bibnamefont
  {Weiss}},\ }\href@noop {} {\emph {\bibinfo {title} {Stellar Structure and
  Evolution}}}\ (\bibinfo  {publisher} {Springer Science \& Business Media},\
  \bibinfo {year} {2012})\BibitemShut {NoStop}%
\bibitem [{\citenamefont {{Walecka}}(1974)}]{walecka1974}%
  \BibitemOpen
  \bibfield  {author} {\bibinfo {author} {\bibfnamefont {J.~D.}\ \bibnamefont
  {{Walecka}}},\ }\bibfield  {title} {\bibinfo {title} {{A theory of highly
  condensed matter.}},\ }\href {https://doi.org/10.1016/0003-4916(74)90208-5}
  {\bibfield  {journal} {\bibinfo  {journal} {Annals of Physics}\ }\textbf
  {\bibinfo {volume} {83}},\ \bibinfo {pages} {491} (\bibinfo {year}
  {1974})}\BibitemShut {NoStop}%
\bibitem [{\citenamefont {{Lalazissis}}\ \emph {et~al.}(1997)\citenamefont
  {{Lalazissis}}, \citenamefont {{K{\"o}nig}},\ and\ \citenamefont
  {{Ring}}}]{Lalazissis:1996}%
  \BibitemOpen
  \bibfield  {author} {\bibinfo {author} {\bibfnamefont {G.~A.}\ \bibnamefont
  {{Lalazissis}}}, \bibinfo {author} {\bibfnamefont {J.}~\bibnamefont
  {{K{\"o}nig}}},\ and\ \bibinfo {author} {\bibfnamefont {P.}~\bibnamefont
  {{Ring}}},\ }\bibfield  {title} {\bibinfo {title} {{New parametrization for
  the Lagrangian density of relativistic mean field theory}},\ }\href
  {https://doi.org/10.1103/PhysRevC.55.540} {\bibfield  {journal} {\bibinfo
  {journal} {\prc}\ }\textbf {\bibinfo {volume} {55}},\ \bibinfo {pages} {540}
  (\bibinfo {year} {1997})},\ \Eprint {https://arxiv.org/abs/nucl-th/9607039}
  {arXiv:nucl-th/9607039 [nucl-th]} \BibitemShut {NoStop}%
\bibitem [{\citenamefont {{Spinella}}(2017)}]{spinellaTH}%
  \BibitemOpen
  \bibfield  {author} {\bibinfo {author} {\bibfnamefont {W.~M.}\ \bibnamefont
  {{Spinella}}},\ }\emph {\bibinfo {title} {{A Systematic Investigation of
  Exotic Matter in Neutron Stars}}},\ \href@noop {} {Ph.D. thesis},\ \bibinfo
  {school} {The Claremont Graduate University} (\bibinfo {year}
  {2017})\BibitemShut {NoStop}%
\bibitem [{\citenamefont {{Typel}}\ \emph {et~al.}(2010)\citenamefont
  {{Typel}}, \citenamefont {{R{\"o}pke}}, \citenamefont {{Kl{\"a}hn}},
  \citenamefont {{Blaschke}},\ and\ \citenamefont {{Wolter}}}]{DD2}%
  \BibitemOpen
  \bibfield  {author} {\bibinfo {author} {\bibfnamefont {S.}~\bibnamefont
  {{Typel}}}, \bibinfo {author} {\bibfnamefont {G.}~\bibnamefont
  {{R{\"o}pke}}}, \bibinfo {author} {\bibfnamefont {T.}~\bibnamefont
  {{Kl{\"a}hn}}}, \bibinfo {author} {\bibfnamefont {D.}~\bibnamefont
  {{Blaschke}}},\ and\ \bibinfo {author} {\bibfnamefont {H.~H.}\ \bibnamefont
  {{Wolter}}},\ }\bibfield  {title} {\bibinfo {title} {{Composition and
  thermodynamics of nuclear matter with light clusters}},\ }\href
  {https://doi.org/10.1103/PhysRevC.81.015803} {\bibfield  {journal} {\bibinfo
  {journal} {\prc}\ }\textbf {\bibinfo {volume} {81}},\ \bibinfo {eid} {015803}
  (\bibinfo {year} {2010})},\ \Eprint {https://arxiv.org/abs/0908.2344}
  {arXiv:0908.2344 [nucl-th]} \BibitemShut {NoStop}%
\bibitem [{\citenamefont {{Spinella}}\ and\ \citenamefont
  {{Weber}}(2019)}]{swl4}%
  \BibitemOpen
  \bibfield  {author} {\bibinfo {author} {\bibfnamefont {W.~M.}\ \bibnamefont
  {{Spinella}}}\ and\ \bibinfo {author} {\bibfnamefont {F.}~\bibnamefont
  {{Weber}}},\ }\bibfield  {title} {\bibinfo {title} {{Hyperonic neutron star
  matter in light of GW170817}},\ }\href
  {https://doi.org/10.1002/asna.201913579} {\bibfield  {journal} {\bibinfo
  {journal} {Astronomische Nachrichten}\ }\textbf {\bibinfo {volume} {340}},\
  \bibinfo {pages} {145} (\bibinfo {year} {2019})},\ \Eprint
  {https://arxiv.org/abs/1812.03600} {arXiv:1812.03600 [nucl-th]} \BibitemShut
  {NoStop}%
\bibitem [{\citenamefont {Orsaria}\ \emph {et~al.}(2019)\citenamefont
  {Orsaria}, \citenamefont {Malfatti}, \citenamefont {Mariani}, \citenamefont
  {Ranea-Sandoval}, \citenamefont {García}, \citenamefont {Spinella},
  \citenamefont {Contrera}, \citenamefont {Lugones},\ and\ \citenamefont
  {Weber}}]{orsaria-rev}%
  \BibitemOpen
  \bibfield  {author} {\bibinfo {author} {\bibfnamefont {M.~G.}\ \bibnamefont
  {Orsaria}}, \bibinfo {author} {\bibfnamefont {G.}~\bibnamefont {Malfatti}},
  \bibinfo {author} {\bibfnamefont {M.}~\bibnamefont {Mariani}}, \bibinfo
  {author} {\bibfnamefont {I.~F.}\ \bibnamefont {Ranea-Sandoval}}, \bibinfo
  {author} {\bibfnamefont {F.}~\bibnamefont {García}}, \bibinfo {author}
  {\bibfnamefont {W.~M.}\ \bibnamefont {Spinella}}, \bibinfo {author}
  {\bibfnamefont {G.~A.}\ \bibnamefont {Contrera}}, \bibinfo {author}
  {\bibfnamefont {G.}~\bibnamefont {Lugones}},\ and\ \bibinfo {author}
  {\bibfnamefont {F.}~\bibnamefont {Weber}},\ }\bibfield  {title} {\bibinfo
  {title} {{Phase transitions in neutron stars and their links to gravitational
  waves}},\ }\href {https://doi.org/10.1088/1361-6471/ab1d81} {\bibfield
  {journal} {\bibinfo  {journal} {J. Phys. G}\ }\textbf {\bibinfo {volume}
  {46}},\ \bibinfo {pages} {073002} (\bibinfo {year} {2019})},\ \Eprint
  {https://arxiv.org/abs/1907.04654} {arXiv:1907.04654 [astro-ph.HE]}
  \BibitemShut {NoStop}%
\bibitem [{\citenamefont {{Lattimer}}(2012)}]{Lattimer:2012ARNPS}%
  \BibitemOpen
  \bibfield  {author} {\bibinfo {author} {\bibfnamefont {J.~M.}\ \bibnamefont
  {{Lattimer}}},\ }\bibfield  {title} {\bibinfo {title} {{The Nuclear Equation
  of State and Neutron Star Masses}},\ }\href
  {https://doi.org/10.1146/annurev-nucl-102711-095018} {\bibfield  {journal}
  {\bibinfo  {journal} {Annual Review of Nuclear and Particle Science}\
  }\textbf {\bibinfo {volume} {62}},\ \bibinfo {pages} {485} (\bibinfo {year}
  {2012})},\ \Eprint {https://arxiv.org/abs/1305.3510} {arXiv:1305.3510
  [nucl-th]} \BibitemShut {NoStop}%
\bibitem [{\citenamefont {{Burgio}}\ \emph {et~al.}(2021)\citenamefont
  {{Burgio}}, \citenamefont {{Schulze}}, \citenamefont {{Vida{\~n}a}},\ and\
  \citenamefont {{Wei}}}]{Burgio:2021PrPNP}%
  \BibitemOpen
  \bibfield  {author} {\bibinfo {author} {\bibfnamefont {G.~F.}\ \bibnamefont
  {{Burgio}}}, \bibinfo {author} {\bibfnamefont {H.~J.}\ \bibnamefont
  {{Schulze}}}, \bibinfo {author} {\bibfnamefont {I.}~\bibnamefont
  {{Vida{\~n}a}}},\ and\ \bibinfo {author} {\bibfnamefont {J.~B.}\ \bibnamefont
  {{Wei}}},\ }\bibfield  {title} {\bibinfo {title} {{Neutron stars and the
  nuclear equation of state}},\ }\href
  {https://doi.org/10.1016/j.ppnp.2021.103879} {\bibfield  {journal} {\bibinfo
  {journal} {Progress in Particle and Nuclear Physics}\ }\textbf {\bibinfo
  {volume} {120}},\ \bibinfo {eid} {103879} (\bibinfo {year} {2021})},\ \Eprint
  {https://arxiv.org/abs/2105.03747} {arXiv:2105.03747 [nucl-th]} \BibitemShut
  {NoStop}%
\bibitem [{\citenamefont {{Alcock}}\ \emph {et~al.}(1986)\citenamefont
  {{Alcock}}, \citenamefont {{Farhi}},\ and\ \citenamefont
  {{Olinto}}}]{Alcock:1986hz}%
  \BibitemOpen
  \bibfield  {author} {\bibinfo {author} {\bibfnamefont {C.}~\bibnamefont
  {{Alcock}}}, \bibinfo {author} {\bibfnamefont {E.}~\bibnamefont {{Farhi}}},\
  and\ \bibinfo {author} {\bibfnamefont {A.}~\bibnamefont {{Olinto}}},\
  }\bibfield  {title} {\bibinfo {title} {{Strange Stars}},\ }\href
  {https://doi.org/10.1086/164679} {\bibfield  {journal} {\bibinfo  {journal}
  {\apj}\ }\textbf {\bibinfo {volume} {310}},\ \bibinfo {pages} {261} (\bibinfo
  {year} {1986})}\BibitemShut {NoStop}%
\bibitem [{\citenamefont {{Lattimer}}\ and\ \citenamefont
  {{Prakash}}(2004)}]{Lattimer:2004}%
  \BibitemOpen
  \bibfield  {author} {\bibinfo {author} {\bibfnamefont {J.~M.}\ \bibnamefont
  {{Lattimer}}}\ and\ \bibinfo {author} {\bibfnamefont {M.}~\bibnamefont
  {{Prakash}}},\ }\bibfield  {title} {\bibinfo {title} {{The Physics of Neutron
  Stars}},\ }\href {https://doi.org/10.1126/science.1090720} {\bibfield
  {journal} {\bibinfo  {journal} {Science}\ }\textbf {\bibinfo {volume}
  {304}},\ \bibinfo {pages} {536} (\bibinfo {year} {2004})},\ \Eprint
  {https://arxiv.org/abs/astro-ph/0405262} {arXiv:astro-ph/0405262 [astro-ph]}
  \BibitemShut {NoStop}%
\bibitem [{\citenamefont {{Weber}}(2005)}]{Weber:2004}%
  \BibitemOpen
  \bibfield  {author} {\bibinfo {author} {\bibfnamefont {F.}~\bibnamefont
  {{Weber}}},\ }\bibfield  {title} {\bibinfo {title} {{Strange quark matter and
  compact stars}},\ }\href {https://doi.org/10.1016/j.ppnp.2004.07.001}
  {\bibfield  {journal} {\bibinfo  {journal} {Progress in Particle and Nuclear
  Physics}\ }\textbf {\bibinfo {volume} {54}},\ \bibinfo {pages} {193}
  (\bibinfo {year} {2005})},\ \Eprint {https://arxiv.org/abs/astro-ph/0407155}
  {arXiv:astro-ph/0407155 [astro-ph]} \BibitemShut {NoStop}%
\bibitem [{\citenamefont {{Weissenborn}}\ \emph {et~al.}(2011)\citenamefont
  {{Weissenborn}}, \citenamefont {{Sagert}}, \citenamefont {{Pagliara}},
  \citenamefont {{Hempel}},\ and\ \citenamefont
  {{Schaffner-Bielich}}}]{Weissenborn:2011}%
  \BibitemOpen
  \bibfield  {author} {\bibinfo {author} {\bibfnamefont {S.}~\bibnamefont
  {{Weissenborn}}}, \bibinfo {author} {\bibfnamefont {I.}~\bibnamefont
  {{Sagert}}}, \bibinfo {author} {\bibfnamefont {G.}~\bibnamefont
  {{Pagliara}}}, \bibinfo {author} {\bibfnamefont {M.}~\bibnamefont
  {{Hempel}}},\ and\ \bibinfo {author} {\bibfnamefont {J.}~\bibnamefont
  {{Schaffner-Bielich}}},\ }\bibfield  {title} {\bibinfo {title} {{Quark Matter
  in Massive Compact Stars}},\ }\href
  {https://doi.org/10.1088/2041-8205/740/1/L14} {\bibfield  {journal} {\bibinfo
   {journal} {\apjl}\ }\textbf {\bibinfo {volume} {740}},\ \bibinfo {eid} {L14}
  (\bibinfo {year} {2011})},\ \Eprint {https://arxiv.org/abs/1102.2869}
  {arXiv:1102.2869 [astro-ph.HE]} \BibitemShut {NoStop}%
\bibitem [{\citenamefont {{Baldo}}\ \emph {et~al.}(2003)\citenamefont
  {{Baldo}}, \citenamefont {{Buballa}}, \citenamefont {{Burgio}}, \citenamefont
  {{Neumann}}, \citenamefont {{Oertel}},\ and\ \citenamefont
  {{Schulze}}}]{Baldo:2002}%
  \BibitemOpen
  \bibfield  {author} {\bibinfo {author} {\bibfnamefont {M.}~\bibnamefont
  {{Baldo}}}, \bibinfo {author} {\bibfnamefont {M.}~\bibnamefont {{Buballa}}},
  \bibinfo {author} {\bibfnamefont {G.~F.}\ \bibnamefont {{Burgio}}}, \bibinfo
  {author} {\bibfnamefont {F.}~\bibnamefont {{Neumann}}}, \bibinfo {author}
  {\bibfnamefont {M.}~\bibnamefont {{Oertel}}},\ and\ \bibinfo {author}
  {\bibfnamefont {H.~J.}\ \bibnamefont {{Schulze}}},\ }\bibfield  {title}
  {\bibinfo {title} {{Neutron stars and the transition to color superconducting
  quark matter}},\ }\href {https://doi.org/10.1016/S0370-2693(03)00556-2}
  {\bibfield  {journal} {\bibinfo  {journal} {Physics Letters B}\ }\textbf
  {\bibinfo {volume} {562}},\ \bibinfo {pages} {153} (\bibinfo {year}
  {2003})},\ \Eprint {https://arxiv.org/abs/nucl-th/0212096}
  {arXiv:nucl-th/0212096 [nucl-th]} \BibitemShut {NoStop}%
\bibitem [{\citenamefont {{Alford}}\ \emph {et~al.}(2008)\citenamefont
  {{Alford}}, \citenamefont {{Schmitt}}, \citenamefont {{Rajagopal}},\ and\
  \citenamefont {{Sch{\"a}fer}}}]{Alford:2007}%
  \BibitemOpen
  \bibfield  {author} {\bibinfo {author} {\bibfnamefont {M.~G.}\ \bibnamefont
  {{Alford}}}, \bibinfo {author} {\bibfnamefont {A.}~\bibnamefont {{Schmitt}}},
  \bibinfo {author} {\bibfnamefont {K.}~\bibnamefont {{Rajagopal}}},\ and\
  \bibinfo {author} {\bibfnamefont {T.}~\bibnamefont {{Sch{\"a}fer}}},\
  }\bibfield  {title} {\bibinfo {title} {{Color superconductivity in dense
  quark matter}},\ }\href {https://doi.org/10.1103/RevModPhys.80.1455}
  {\bibfield  {journal} {\bibinfo  {journal} {Reviews of Modern Physics}\
  }\textbf {\bibinfo {volume} {80}},\ \bibinfo {pages} {1455} (\bibinfo {year}
  {2008})},\ \Eprint {https://arxiv.org/abs/0709.4635} {arXiv:0709.4635
  [hep-ph]} \BibitemShut {NoStop}%
\bibitem [{\citenamefont {Alford}\ \emph {et~al.}(2001)\citenamefont {Alford},
  \citenamefont {Rajagopal}, \citenamefont {Reddy},\ and\ \citenamefont
  {Wilczek}}]{Alford:2001zr}%
  \BibitemOpen
  \bibfield  {author} {\bibinfo {author} {\bibfnamefont {M.~G.}\ \bibnamefont
  {Alford}}, \bibinfo {author} {\bibfnamefont {K.}~\bibnamefont {Rajagopal}},
  \bibinfo {author} {\bibfnamefont {S.}~\bibnamefont {Reddy}},\ and\ \bibinfo
  {author} {\bibfnamefont {F.}~\bibnamefont {Wilczek}},\ }\bibfield  {title}
  {\bibinfo {title} {{The Minimal CFL nuclear interface}},\ }\href
  {https://doi.org/10.1103/PhysRevD.64.074017} {\bibfield  {journal} {\bibinfo
  {journal} {Phys. Rev. D}\ }\textbf {\bibinfo {volume} {64}},\ \bibinfo
  {pages} {074017} (\bibinfo {year} {2001})},\ \Eprint
  {https://arxiv.org/abs/hep-ph/0105009} {arXiv:hep-ph/0105009} \BibitemShut
  {NoStop}%
\bibitem [{\citenamefont {Olesen}\ and\ \citenamefont
  {Madsen}(1994)}]{Olesen:1993ek}%
  \BibitemOpen
  \bibfield  {author} {\bibinfo {author} {\bibfnamefont {M.~L.}\ \bibnamefont
  {Olesen}}\ and\ \bibinfo {author} {\bibfnamefont {J.}~\bibnamefont
  {Madsen}},\ }\bibfield  {title} {\bibinfo {title} {{Nucleation of quark
  matter bubbles in neutron stars}},\ }\href
  {https://doi.org/10.1103/PhysRevD.49.2698} {\bibfield  {journal} {\bibinfo
  {journal} {Phys. Rev. D}\ }\textbf {\bibinfo {volume} {49}},\ \bibinfo
  {pages} {2698} (\bibinfo {year} {1994})},\ \Eprint
  {https://arxiv.org/abs/astro-ph/9401002} {arXiv:astro-ph/9401002}
  \BibitemShut {NoStop}%
\bibitem [{\citenamefont {Lugones}\ and\ \citenamefont
  {Benvenuto}(1998)}]{Lugones:1997gg}%
  \BibitemOpen
  \bibfield  {author} {\bibinfo {author} {\bibfnamefont {G.}~\bibnamefont
  {Lugones}}\ and\ \bibinfo {author} {\bibfnamefont {O.}~\bibnamefont
  {Benvenuto}},\ }\bibfield  {title} {\bibinfo {title} {{Effect of trapped
  neutrinos in the hadron matter to quark matter transition}},\ }\href
  {https://doi.org/10.1103/PhysRevD.58.083001} {\bibfield  {journal} {\bibinfo
  {journal} {Phys. Rev. D}\ }\textbf {\bibinfo {volume} {58}},\ \bibinfo
  {pages} {083001} (\bibinfo {year} {1998})}\BibitemShut {NoStop}%
\bibitem [{\citenamefont {Iida}\ and\ \citenamefont
  {Sato}(1998)}]{Iida:1998pi}%
  \BibitemOpen
  \bibfield  {author} {\bibinfo {author} {\bibfnamefont {K.}~\bibnamefont
  {Iida}}\ and\ \bibinfo {author} {\bibfnamefont {K.}~\bibnamefont {Sato}},\
  }\bibfield  {title} {\bibinfo {title} {{Effects of hyperons on the dynamical
  deconfinement transition in cold neutron star matter}},\ }\href
  {https://doi.org/10.1103/PhysRevC.58.2538} {\bibfield  {journal} {\bibinfo
  {journal} {Phys. Rev. C}\ }\textbf {\bibinfo {volume} {58}},\ \bibinfo
  {pages} {2538} (\bibinfo {year} {1998})},\ \Eprint
  {https://arxiv.org/abs/nucl-th/9808056} {arXiv:nucl-th/9808056} \BibitemShut
  {NoStop}%
\bibitem [{\citenamefont {Bombaci}\ \emph {et~al.}(2004)\citenamefont
  {Bombaci}, \citenamefont {Parenti},\ and\ \citenamefont
  {Vidana}}]{Bombaci:2004mt}%
  \BibitemOpen
  \bibfield  {author} {\bibinfo {author} {\bibfnamefont {I.}~\bibnamefont
  {Bombaci}}, \bibinfo {author} {\bibfnamefont {I.}~\bibnamefont {Parenti}},\
  and\ \bibinfo {author} {\bibfnamefont {I.}~\bibnamefont {Vidana}},\
  }\bibfield  {title} {\bibinfo {title} {{Quark deconfinement and implications
  for the radius and the limiting mass of compact stars}},\ }\href
  {https://doi.org/10.1086/423658} {\bibfield  {journal} {\bibinfo  {journal}
  {Astrophys. J.}\ }\textbf {\bibinfo {volume} {614}},\ \bibinfo {pages} {314}
  (\bibinfo {year} {2004})},\ \Eprint {https://arxiv.org/abs/astro-ph/0402404}
  {arXiv:astro-ph/0402404} \BibitemShut {NoStop}%
\bibitem [{\citenamefont {Lugones}(2016)}]{Lugones:2015bya}%
  \BibitemOpen
  \bibfield  {author} {\bibinfo {author} {\bibfnamefont {G.}~\bibnamefont
  {Lugones}},\ }\bibfield  {title} {\bibinfo {title} {{From quark drops to
  quark stars: some aspects of the role of quark matter in compact stars}},\
  }\href {https://doi.org/10.1140/epja/i2016-16053-x} {\bibfield  {journal}
  {\bibinfo  {journal} {Eur. Phys. J. A}\ }\textbf {\bibinfo {volume} {52}},\
  \bibinfo {pages} {53} (\bibinfo {year} {2016})},\ \Eprint
  {https://arxiv.org/abs/1508.05548} {arXiv:1508.05548 [astro-ph.HE]}
  \BibitemShut {NoStop}%
\bibitem [{\citenamefont {Bombaci}\ \emph {et~al.}(2016)\citenamefont
  {Bombaci}, \citenamefont {Logoteta}, \citenamefont {Vida\~na},\ and\
  \citenamefont {Provid\^encia}}]{Bombaci:2016xuj}%
  \BibitemOpen
  \bibfield  {author} {\bibinfo {author} {\bibfnamefont {I.}~\bibnamefont
  {Bombaci}}, \bibinfo {author} {\bibfnamefont {D.}~\bibnamefont {Logoteta}},
  \bibinfo {author} {\bibfnamefont {I.}~\bibnamefont {Vida\~na}},\ and\
  \bibinfo {author} {\bibfnamefont {C.}~\bibnamefont {Provid\^encia}},\
  }\bibfield  {title} {\bibinfo {title} {{Quark matter nucleation in neutron
  stars and astrophysical implications}},\ }\href
  {https://doi.org/10.1140/epja/i2016-16058-5} {\bibfield  {journal} {\bibinfo
  {journal} {Eur. Phys. J. A}\ }\textbf {\bibinfo {volume} {52}},\ \bibinfo
  {pages} {58} (\bibinfo {year} {2016})},\ \Eprint
  {https://arxiv.org/abs/1601.04559} {arXiv:1601.04559 [astro-ph.HE]}
  \BibitemShut {NoStop}%
\bibitem [{\citenamefont {Vasquez~Flores}\ \emph {et~al.}(2012)\citenamefont
  {Vasquez~Flores}, \citenamefont {Lenzi},\ and\ \citenamefont
  {Lugones}}]{VasquezFlores:2012vf}%
  \BibitemOpen
  \bibfield  {author} {\bibinfo {author} {\bibfnamefont {C.}~\bibnamefont
  {Vasquez~Flores}}, \bibinfo {author} {\bibfnamefont {C.}~\bibnamefont
  {Lenzi}},\ and\ \bibinfo {author} {\bibfnamefont {G.}~\bibnamefont
  {Lugones}},\ }\bibfield  {title} {\bibinfo {title} {{Radial pulsations of
  hybrid neutron stars}},\ }\href {https://doi.org/10.1142/S201019451200829X}
  {\bibfield  {journal} {\bibinfo  {journal} {Int. J. Mod. Phys. Conf. Ser.}\
  }\textbf {\bibinfo {volume} {18}},\ \bibinfo {pages} {105} (\bibinfo {year}
  {2012})}\BibitemShut {NoStop}%
\bibitem [{\citenamefont {Mariani}\ \emph {et~al.}(2019)\citenamefont
  {Mariani}, \citenamefont {Orsaria}, \citenamefont {Ranea-Sandoval},\ and\
  \citenamefont {Lugones}}]{Mariani:2019vve}%
  \BibitemOpen
  \bibfield  {author} {\bibinfo {author} {\bibfnamefont {M.}~\bibnamefont
  {Mariani}}, \bibinfo {author} {\bibfnamefont {M.~G.}\ \bibnamefont
  {Orsaria}}, \bibinfo {author} {\bibfnamefont {I.~F.}\ \bibnamefont
  {Ranea-Sandoval}},\ and\ \bibinfo {author} {\bibfnamefont {G.}~\bibnamefont
  {Lugones}},\ }\bibfield  {title} {\bibinfo {title} {{Magnetized hybrid stars:
  effects of slow and rapid phase transitions at the quark\textendash{}hadron
  interface}},\ }\href {https://doi.org/10.1093/mnras/stz2392} {\bibfield
  {journal} {\bibinfo  {journal} {Mon. Not. Roy. Astron. Soc.}\ }\textbf
  {\bibinfo {volume} {489}},\ \bibinfo {pages} {4261} (\bibinfo {year}
  {2019})},\ \Eprint {https://arxiv.org/abs/1909.08661} {arXiv:1909.08661
  [astro-ph.HE]} \BibitemShut {NoStop}%
\bibitem [{\citenamefont {{Malfatti}}\ \emph {et~al.}(2020)\citenamefont
  {{Malfatti}}, \citenamefont {{Orsaria}}, \citenamefont {{Ranea-Sandoval}},
  \citenamefont {{Contrera}},\ and\ \citenamefont {{Weber}}}]{Malfatti2020PRD}%
  \BibitemOpen
  \bibfield  {author} {\bibinfo {author} {\bibfnamefont {G.}~\bibnamefont
  {{Malfatti}}}, \bibinfo {author} {\bibfnamefont {M.~G.}\ \bibnamefont
  {{Orsaria}}}, \bibinfo {author} {\bibfnamefont {I.~F.}\ \bibnamefont
  {{Ranea-Sandoval}}}, \bibinfo {author} {\bibfnamefont {G.~A.}\ \bibnamefont
  {{Contrera}}},\ and\ \bibinfo {author} {\bibfnamefont {F.}~\bibnamefont
  {{Weber}}},\ }\bibfield  {title} {\bibinfo {title} {{Delta baryons and
  diquark formation in the cores of neutron stars}},\ }\href
  {https://doi.org/10.1103/PhysRevD.102.063008} {\bibfield  {journal} {\bibinfo
   {journal} {\prd}\ }\textbf {\bibinfo {volume} {102}},\ \bibinfo {eid}
  {063008} (\bibinfo {year} {2020})},\ \Eprint
  {https://arxiv.org/abs/2008.06459} {arXiv:2008.06459 [astro-ph.HE]}
  \BibitemShut {NoStop}%
\bibitem [{\citenamefont {{O'Boyle}}\ \emph {et~al.}(2020)\citenamefont
  {{O'Boyle}}, \citenamefont {{Markakis}}, \citenamefont {{Stergioulas}},\ and\
  \citenamefont {{Read}}}]{OBoyle-etal-2020}%
  \BibitemOpen
  \bibfield  {author} {\bibinfo {author} {\bibfnamefont {M.~F.}\ \bibnamefont
  {{O'Boyle}}}, \bibinfo {author} {\bibfnamefont {C.}~\bibnamefont
  {{Markakis}}}, \bibinfo {author} {\bibfnamefont {N.}~\bibnamefont
  {{Stergioulas}}},\ and\ \bibinfo {author} {\bibfnamefont {J.~S.}\
  \bibnamefont {{Read}}},\ }\bibfield  {title} {\bibinfo {title} {{Parametrized
  equation of state for neutron star matter with continuous sound speed}},\
  }\href {https://doi.org/10.1103/PhysRevD.102.083027} {\bibfield  {journal}
  {\bibinfo  {journal} {\prd}\ }\textbf {\bibinfo {volume} {102}},\ \bibinfo
  {eid} {083027} (\bibinfo {year} {2020})},\ \Eprint
  {https://arxiv.org/abs/2008.03342} {arXiv:2008.03342 [astro-ph.HE]}
  \BibitemShut {NoStop}%
\bibitem [{\citenamefont {{Alford}}\ \emph {et~al.}(2013)\citenamefont
  {{Alford}}, \citenamefont {{Han}},\ and\ \citenamefont
  {{Prakash}}}]{css-original}%
  \BibitemOpen
  \bibfield  {author} {\bibinfo {author} {\bibfnamefont {M.~G.}\ \bibnamefont
  {{Alford}}}, \bibinfo {author} {\bibfnamefont {S.}~\bibnamefont {{Han}}},\
  and\ \bibinfo {author} {\bibfnamefont {M.}~\bibnamefont {{Prakash}}},\
  }\bibfield  {title} {\bibinfo {title} {{Generic conditions for stable hybrid
  stars}},\ }\href {https://doi.org/10.1103/PhysRevD.88.083013} {\bibfield
  {journal} {\bibinfo  {journal} {\prd}\ }\textbf {\bibinfo {volume} {88}},\
  \bibinfo {eid} {083013} (\bibinfo {year} {2013})},\ \Eprint
  {https://arxiv.org/abs/1302.4732} {arXiv:1302.4732 [astro-ph.SR]}
  \BibitemShut {NoStop}%
\bibitem [{\citenamefont {Rezzolla}\ \emph {et~al.}(2018)\citenamefont
  {Rezzolla}, \citenamefont {Most},\ and\ \citenamefont
  {Weih}}]{Rezzolla:2017aly}%
  \BibitemOpen
  \bibfield  {author} {\bibinfo {author} {\bibfnamefont {L.}~\bibnamefont
  {Rezzolla}}, \bibinfo {author} {\bibfnamefont {E.~R.}\ \bibnamefont {Most}},\
  and\ \bibinfo {author} {\bibfnamefont {L.~R.}\ \bibnamefont {Weih}},\
  }\bibfield  {title} {\bibinfo {title} {{Using gravitational-wave observations
  and quasi-universal relations to constrain the maximum mass of neutron
  stars}},\ }\href {https://doi.org/10.3847/2041-8213/aaa401} {\bibfield
  {journal} {\bibinfo  {journal} {Astrophys. J.}\ }\textbf {\bibinfo {volume}
  {852}},\ \bibinfo {pages} {L25} (\bibinfo {year} {2018})},\ \Eprint
  {https://arxiv.org/abs/1711.00314} {arXiv:1711.00314 [astro-ph.HE]}
  \BibitemShut {NoStop}%
\bibitem [{\citenamefont {{Regge}}\ and\ \citenamefont
  {{Wheeler}}(1957)}]{regge1957}%
  \BibitemOpen
  \bibfield  {author} {\bibinfo {author} {\bibfnamefont {T.}~\bibnamefont
  {{Regge}}}\ and\ \bibinfo {author} {\bibfnamefont {J.~A.}\ \bibnamefont
  {{Wheeler}}},\ }\bibfield  {title} {\bibinfo {title} {{Stability of a
  Schwarzschild Singularity}},\ }\href
  {https://doi.org/10.1103/PhysRev.108.1063} {\bibfield  {journal} {\bibinfo
  {journal} {Physical Review}\ }\textbf {\bibinfo {volume} {108}},\ \bibinfo
  {pages} {1063} (\bibinfo {year} {1957})}\BibitemShut {NoStop}%
\bibitem [{\citenamefont {{Kokkotas}}\ and\ \citenamefont
  {{Schutz}}(1992)}]{kokkotas:1992}%
  \BibitemOpen
  \bibfield  {author} {\bibinfo {author} {\bibfnamefont {K.~D.}\ \bibnamefont
  {{Kokkotas}}}\ and\ \bibinfo {author} {\bibfnamefont {B.~F.}\ \bibnamefont
  {{Schutz}}},\ }\bibfield  {title} {\bibinfo {title} {{W-modes - A new family
  of normal modes of pulsating relativistic stars}},\ }\href
  {https://doi.org/10.1093/mnras/255.1.119} {\bibfield  {journal} {\bibinfo
  {journal} {\mnras}\ }\textbf {\bibinfo {volume} {255}},\ \bibinfo {pages}
  {119} (\bibinfo {year} {1992})}\BibitemShut {NoStop}%
\bibitem [{\citenamefont {{Benhar}}\ \emph {et~al.}(1999)\citenamefont
  {{Benhar}}, \citenamefont {{Berti}},\ and\ \citenamefont
  {{Ferrari}}}]{Benhar:1998}%
  \BibitemOpen
  \bibfield  {author} {\bibinfo {author} {\bibfnamefont {O.}~\bibnamefont
  {{Benhar}}}, \bibinfo {author} {\bibfnamefont {E.}~\bibnamefont {{Berti}}},\
  and\ \bibinfo {author} {\bibfnamefont {V.}~\bibnamefont {{Ferrari}}},\
  }\bibfield  {title} {\bibinfo {title} {{The imprint of the equation of state
  on the axial w-modes of oscillating neutron stars}},\ }\href
  {https://doi.org/10.1046/j.1365-8711.1999.02983.x} {\bibfield  {journal}
  {\bibinfo  {journal} {\mnras}\ }\textbf {\bibinfo {volume} {310}},\ \bibinfo
  {pages} {797} (\bibinfo {year} {1999})},\ \Eprint
  {https://arxiv.org/abs/gr-qc/9901037} {arXiv:gr-qc/9901037 [gr-qc]}
  \BibitemShut {NoStop}%
\bibitem [{\citenamefont {{Kokkotas}}\ and\ \citenamefont
  {{Schmidt}}(1999)}]{kokkotas:1999}%
  \BibitemOpen
  \bibfield  {author} {\bibinfo {author} {\bibfnamefont {K.~D.}\ \bibnamefont
  {{Kokkotas}}}\ and\ \bibinfo {author} {\bibfnamefont {B.~G.}\ \bibnamefont
  {{Schmidt}}},\ }\bibfield  {title} {\bibinfo {title} {{Quasi-Normal Modes of
  Stars and Black Holes}},\ }\href {https://doi.org/10.12942/lrr-1999-2}
  {\bibfield  {journal} {\bibinfo  {journal} {Living Reviews in Relativity}\
  }\textbf {\bibinfo {volume} {2}},\ \bibinfo {eid} {2} (\bibinfo {year}
  {1999})},\ \Eprint {https://arxiv.org/abs/gr-qc/9909058} {arXiv:gr-qc/9909058
  [gr-qc]} \BibitemShut {NoStop}%
\bibitem [{\citenamefont {Alvarez-Castillo}\ \emph {et~al.}(2019)\citenamefont
  {Alvarez-Castillo}, \citenamefont {Blaschke}, \citenamefont {Grunfeld},\ and\
  \citenamefont {Pagura}}]{Alvarez-Castillo:2018pve}%
  \BibitemOpen
  \bibfield  {author} {\bibinfo {author} {\bibfnamefont {D.~E.}\ \bibnamefont
  {Alvarez-Castillo}}, \bibinfo {author} {\bibfnamefont {D.~B.}\ \bibnamefont
  {Blaschke}}, \bibinfo {author} {\bibfnamefont {A.~G.}\ \bibnamefont
  {Grunfeld}},\ and\ \bibinfo {author} {\bibfnamefont {V.~P.}\ \bibnamefont
  {Pagura}},\ }\bibfield  {title} {\bibinfo {title} {{Third family of compact
  stars within a nonlocal chiral quark model equation of state}},\ }\href
  {https://doi.org/10.1103/PhysRevD.99.063010} {\bibfield  {journal} {\bibinfo
  {journal} {Phys. Rev.}\ }\textbf {\bibinfo {volume} {D99}},\ \bibinfo {pages}
  {063010} (\bibinfo {year} {2019})},\ \Eprint
  {https://arxiv.org/abs/1805.04105} {arXiv:1805.04105 [hep-ph]} \BibitemShut
  {NoStop}%
\bibitem [{\citenamefont {{Jakobus}}\ \emph {et~al.}(2021)\citenamefont
  {{Jakobus}}, \citenamefont {{Motornenko}}, \citenamefont {{Gomes}},
  \citenamefont {{Steinheimer}},\ and\ \citenamefont
  {{Stoecker}}}]{jakobus2021EPJC}%
  \BibitemOpen
  \bibfield  {author} {\bibinfo {author} {\bibfnamefont {P.}~\bibnamefont
  {{Jakobus}}}, \bibinfo {author} {\bibfnamefont {A.}~\bibnamefont
  {{Motornenko}}}, \bibinfo {author} {\bibfnamefont {R.~O.}\ \bibnamefont
  {{Gomes}}}, \bibinfo {author} {\bibfnamefont {J.}~\bibnamefont
  {{Steinheimer}}},\ and\ \bibinfo {author} {\bibfnamefont {H.}~\bibnamefont
  {{Stoecker}}},\ }\bibfield  {title} {\bibinfo {title} {{The possibility of
  twin star solutions in a model based on lattice QCD thermodynamics}},\ }\href
  {https://doi.org/10.1140/epjc/s10052-020-08779-x} {\bibfield  {journal}
  {\bibinfo  {journal} {European Physical Journal C}\ }\textbf {\bibinfo
  {volume} {81}},\ \bibinfo {eid} {41} (\bibinfo {year} {2021})},\ \Eprint
  {https://arxiv.org/abs/2004.07026} {arXiv:2004.07026 [nucl-th]} \BibitemShut
  {NoStop}%
\bibitem [{\citenamefont {{Gon{\c{c}}alves}}\ and\ \citenamefont
  {{Lazzari}}(2022)}]{2022EPJC}%
  \BibitemOpen
  \bibfield  {author} {\bibinfo {author} {\bibfnamefont {V.~P.}\ \bibnamefont
  {{Gon{\c{c}}alves}}}\ and\ \bibinfo {author} {\bibfnamefont {L.}~\bibnamefont
  {{Lazzari}}},\ }\bibfield  {title} {\bibinfo {title} {{Impact of slow
  conversions on hybrid stars with sequential QCD phase transitions}},\ }\href
  {https://doi.org/10.1140/epjc/s10052-022-10273-5} {\bibfield  {journal}
  {\bibinfo  {journal} {European Physical Journal C}\ }\textbf {\bibinfo
  {volume} {82}},\ \bibinfo {eid} {288} (\bibinfo {year} {2022})},\ \Eprint
  {https://arxiv.org/abs/2201.03304} {arXiv:2201.03304 [nucl-th]} \BibitemShut
  {NoStop}%
\bibitem [{\citenamefont {{Lugones}}\ \emph {et~al.}(2021)\citenamefont
  {{Lugones}}, \citenamefont {{Mariani}},\ and\ \citenamefont
  {{Ranea-Sandoval}}}]{lugones2021arXiv}%
  \BibitemOpen
  \bibfield  {author} {\bibinfo {author} {\bibfnamefont {G.}~\bibnamefont
  {{Lugones}}}, \bibinfo {author} {\bibfnamefont {M.}~\bibnamefont
  {{Mariani}}},\ and\ \bibinfo {author} {\bibfnamefont {I.~F.}\ \bibnamefont
  {{Ranea-Sandoval}}},\ }\bibfield  {title} {\bibinfo {title} {{Slow stable
  hybrid stars: a new class of compact stars that fulfills all current
  observational constraints}},\ }\href@noop {} {\bibfield  {journal} {\bibinfo
  {journal} {arXiv e-prints}\ ,\ \bibinfo {eid} {arXiv:2106.10380}} (\bibinfo
  {year} {2021})},\ \Eprint {https://arxiv.org/abs/2106.10380}
  {arXiv:2106.10380 [nucl-th]} \BibitemShut {NoStop}%
\bibitem [{\citenamefont {{Curin}}\ \emph {et~al.}(2021)\citenamefont
  {{Curin}}, \citenamefont {{Ranea-Sandoval}}, \citenamefont {{Mariani}},
  \citenamefont {{Orsaria}},\ and\ \citenamefont {{Weber}}}]{curin:2021hsw}%
  \BibitemOpen
  \bibfield  {author} {\bibinfo {author} {\bibfnamefont {D.}~\bibnamefont
  {{Curin}}}, \bibinfo {author} {\bibfnamefont {I.~F.}\ \bibnamefont
  {{Ranea-Sandoval}}}, \bibinfo {author} {\bibfnamefont {M.}~\bibnamefont
  {{Mariani}}}, \bibinfo {author} {\bibfnamefont {M.~G.}\ \bibnamefont
  {{Orsaria}}},\ and\ \bibinfo {author} {\bibfnamefont {F.}~\bibnamefont
  {{Weber}}},\ }\bibfield  {title} {\bibinfo {title} {{Hybrid Stars with Color
  Superconducting Cores in an Extended FCM Model}},\ }\href
  {https://doi.org/10.3390/universe7100370} {\bibfield  {journal} {\bibinfo
  {journal} {Universe}\ }\textbf {\bibinfo {volume} {7}},\ \bibinfo {pages}
  {370} (\bibinfo {year} {2021})},\ \Eprint {https://arxiv.org/abs/2110.08892}
  {arXiv:2110.08892 [nucl-th]} \BibitemShut {NoStop}%
\bibitem [{\citenamefont {{Mariani}}\ \emph {et~al.}(2022)\citenamefont
  {{Mariani}}, \citenamefont {{Tonetto}}, \citenamefont {{Rodr{\'\i}guez}},
  \citenamefont {{Celi}}, \citenamefont {{Ranea-Sandoval}}, \citenamefont
  {{Orsaria}},\ and\ \citenamefont {{P{\'e}rez
  Mart{\'\i}nez}}}]{mariani2022MNRAS}%
  \BibitemOpen
  \bibfield  {author} {\bibinfo {author} {\bibfnamefont {M.}~\bibnamefont
  {{Mariani}}}, \bibinfo {author} {\bibfnamefont {L.}~\bibnamefont
  {{Tonetto}}}, \bibinfo {author} {\bibfnamefont {M.~C.}\ \bibnamefont
  {{Rodr{\'\i}guez}}}, \bibinfo {author} {\bibfnamefont {M.~O.}\ \bibnamefont
  {{Celi}}}, \bibinfo {author} {\bibfnamefont {I.~F.}\ \bibnamefont
  {{Ranea-Sandoval}}}, \bibinfo {author} {\bibfnamefont {M.~G.}\ \bibnamefont
  {{Orsaria}}},\ and\ \bibinfo {author} {\bibfnamefont {A.}~\bibnamefont
  {{P{\'e}rez Mart{\'\i}nez}}},\ }\bibfield  {title} {\bibinfo {title}
  {{Oscillating magnetized hybrid stars under the magnifying glass of
  multimessenger observations}},\ }\href
  {https://doi.org/10.1093/mnras/stac546} {\bibfield  {journal} {\bibinfo
  {journal} {\mnras}\ }\textbf {\bibinfo {volume} {512}},\ \bibinfo {pages}
  {517} (\bibinfo {year} {2022})},\ \Eprint {https://arxiv.org/abs/2202.12222}
  {arXiv:2202.12222 [astro-ph.HE]} \BibitemShut {NoStop}%
\bibitem [{\citenamefont {{Tonetto}}\ and\ \citenamefont
  {{Lugones}}(2020)}]{lucasygerman}%
  \BibitemOpen
  \bibfield  {author} {\bibinfo {author} {\bibfnamefont {L.}~\bibnamefont
  {{Tonetto}}}\ and\ \bibinfo {author} {\bibfnamefont {G.}~\bibnamefont
  {{Lugones}}},\ }\bibfield  {title} {\bibinfo {title} {{Discontinuity gravity
  modes in hybrid stars: Assessing the role of rapid and slow phase
  conversions}},\ }\href {https://doi.org/10.1103/PhysRevD.101.123029}
  {\bibfield  {journal} {\bibinfo  {journal} {\prd}\ }\textbf {\bibinfo
  {volume} {101}},\ \bibinfo {eid} {123029} (\bibinfo {year} {2020})},\ \Eprint
  {https://arxiv.org/abs/2003.01259} {arXiv:2003.01259 [astro-ph.HE]}
  \BibitemShut {NoStop}%
\bibitem [{\citenamefont {Pratten}\ \emph {et~al.}(2020)\citenamefont
  {Pratten}, \citenamefont {Schmidt},\ and\ \citenamefont
  {Hinderer}}]{Pratten:2019sed}%
  \BibitemOpen
  \bibfield  {author} {\bibinfo {author} {\bibfnamefont {G.}~\bibnamefont
  {Pratten}}, \bibinfo {author} {\bibfnamefont {P.}~\bibnamefont {Schmidt}},\
  and\ \bibinfo {author} {\bibfnamefont {T.}~\bibnamefont {Hinderer}},\
  }\bibfield  {title} {\bibinfo {title} {{Gravitational-Wave Asteroseismology
  with Fundamental Modes from Compact Binary Inspirals}},\ }\href
  {https://doi.org/10.1038/s41467-020-15984-5} {\bibfield  {journal} {\bibinfo
  {journal} {Nature Commun.}\ }\textbf {\bibinfo {volume} {11}},\ \bibinfo
  {pages} {2553} (\bibinfo {year} {2020})},\ \Eprint
  {https://arxiv.org/abs/1905.00817} {arXiv:1905.00817 [gr-qc]} \BibitemShut
  {NoStop}%
\end{thebibliography}%

\end{document}